\documentclass[traditabstract,longauth]{aa}

\def\setsymbol#1#2{\expandafter\def\csname #1\endcsname{#2}}
\def\getsymbol#1{\csname #1\endcsname}

\def\Planck{\textit{Planck}}

\newbox\tablebox    \newdimen\tablewidth
\def\leaderfil{\leaders\hbox to 5pt{\hss.\hss}\hfil}
\def\endPlancktable{\tablewidth=\columnwidth 
    $$\hss\copy\tablebox\hss$$
    \vskip-\lastskip\vskip -2pt}
\def\endPlancktablewide{\tablewidth=\textwidth 
    $$\hss\copy\tablebox\hss$$
    \vskip-\lastskip\vskip -2pt}
\def\tablenote#1 #2\par{\begingroup \parindent=0.8em
    \abovedisplayshortskip=0pt\belowdisplayshortskip=0pt
    \noindent
    $$\hss\vbox{\hsize\tablewidth \hangindent=\parindent \hangafter=1 \noindent
    \hbox to \parindent{$^#1$\hss}\strut#2\strut\par}\hss$$
    \endgroup}
\def\doubleline{\vskip 3pt\hrule \vskip 1.5pt \hrule \vskip 5pt}

\def\L2{\ifmmode L_2\else $L_2$\fi}

\def\DeltaT{\ifmmode \Delta T\else $\Delta T$\fi}
\def\deltat{\ifmmode \Delta t\else $\Delta t$\fi}
\def\fknee{\ifmmode f_{\rm knee}\else $f_{\rm knee}$\fi}
\def\Fmax{\ifmmode F_{\rm max}\else $F_{\rm max}$\fi}
\def\solar{\ifmmode{\rm M}_{\mathord\odot}\else${\rm M}_{\mathord\odot}$\fi}
\def\Msolar{\ifmmode{\rm M}_{\mathord\odot}\else${\rm M}_{\mathord\odot}$\fi}
\def\Lsolar{\ifmmode{\rm L}_{\mathord\odot}\else${\rm L}_{\mathord\odot}$\fi}
\def\inv{\ifmmode^{-1}\else$^{-1}$\fi}
\def\mo{\ifmmode^{-1}\else$^{-1}$\fi}
\def\sup#1{\ifmmode ^{\rm #1}\else $^{\rm #1}$\fi}
\def\expo#1{\ifmmode \times 10^{#1}\else $\times 10^{#1}$\fi}
\def\,{\thinspace}
\def\lsim{\mathrel{\raise .4ex\hbox{\rlap{$<$}\lower 1.2ex\hbox{$\sim$}}}}
\def\gsim{\mathrel{\raise .4ex\hbox{\rlap{$>$}\lower 1.2ex\hbox{$\sim$}}}}

\def\simprop{\mathrel{\raise .4ex\hbox{\rlap{$\propto$}\lower 1.2ex\hbox{$\sim$}}}}
\def\deg{\ifmmode^\circ\else$^\circ$\fi}
\def\pdeg{\ifmmode $\setbox0=\hbox{$^{\circ}$}\rlap{\hskip.11\wd0 .}$^{\circ}
          \else \setbox0=\hbox{$^{\circ}$}\rlap{\hskip.11\wd0 .}$^{\circ}$\fi}
\def\arcs{\ifmmode {^{\scriptstyle\prime\prime}}
          \else $^{\scriptstyle\prime\prime}$\fi}
\def\arcm{\ifmmode {^{\scriptstyle\prime}}
          \else $^{\scriptstyle\prime}$\fi}
\newdimen\sa  \newdimen\sb
\def\parcs{\sa=.07em \sb=.03em
     \ifmmode \hbox{\rlap{.}}^{\scriptstyle\prime\kern -\sb\prime}\hbox{\kern -\sa}
     \else \rlap{.}$^{\scriptstyle\prime\kern -\sb\prime}$\kern -\sa\fi}
\def\parcm{\sa=.08em \sb=.03em
     \ifmmode \hbox{\rlap{.}\kern\sa}^{\scriptstyle\prime}\hbox{\kern-\sb}
     \else \rlap{.}\kern\sa$^{\scriptstyle\prime}$\kern-\sb\fi}
\def\ra[#1 #2 #3.#4]{#1\sup{h}#2\sup{m}#3\sup{s}\llap.#4}
\def\dec[#1 #2 #3.#4]{#1\deg#2\arcm#3\arcs\llap.#4}
\def\deco[#1 #2 #3]{#1\deg#2\arcm#3\arcs}
\def\rra[#1 #2]{#1\sup{h}#2\sup{m}}

\def\dots{\relax\ifmmode \ldots\else $\ldots$\fi}
\def\WHzsr{\ifmmode $W\,Hz\mo\,sr\mo$\else W\,Hz\mo\,sr\mo\fi}
\def\mHz{\ifmmode $\,mHz$\else \,mHz\fi}
\def\GHz{\ifmmode $\,GHz$\else \,GHz\fi}
\def\mKs{\ifmmode $\,mK\,s$^{1/2}\else \,mK\,s$^{1/2}$\fi}
\def\muKs{\ifmmode \,\mu$K\,s$^{1/2}\else \,$\mu$K\,s$^{1/2}$\fi}
\def\muKRJs{\ifmmode \,\mu$K$_{\rm RJ}$\,s$^{1/2}\else \,$\mu$K$_{\rm RJ}$\,s$^{1/2}$\fi}
\def\muKHz{\ifmmode \,\mu$K\,Hz$^{-1/2}\else \,$\mu$K\,Hz$^{-1/2}$\fi}
\def\MJysr{\ifmmode \,$MJy\,sr\mo$\else \,MJy\,sr\mo\fi}
\def\MJysrmK{\ifmmode \,$MJy\,sr\mo$\,mK$_{\rm CMB}\mo\else \,MJy\,sr\mo\,mK$_{\rm CMB}\mo$\fi}
\def\microns{\ifmmode \,\mu$m$\else \,$\mu$m\fi}
\def\micron{\microns}
\def\muK{\ifmmode \,\mu$K$\else \,$\mu$\hbox{K}\fi}
\def\microK{\ifmmode \,\mu$K$\else \,$\mu$\hbox{K}\fi}
\def\muW{\ifmmode \,\mu$W$\else \,$\mu$\hbox{W}\fi}
\def\kms{\ifmmode $\,km\,s$^{-1}\else \,km\,s$^{-1}$\fi}
\def\kmsMpc{\ifmmode $\,\kms\,Mpc\mo$\else \,\kms\,Mpc\mo\fi}
\providecommand{\sorthelp}[1]{}

\usepackage{graphicx}
\usepackage{epstopdf}
\usepackage{natbib}
\bibpunct{(}{)}{;}{a}{}{,}
\usepackage{amssymb,,amsmath}
\usepackage{longtable}
\usepackage{multirow}
\usepackage{txfonts}
\usepackage[switch]{lineno}
\usepackage{ifthen}
\usepackage[breaklinks,colorlinks,citecolor=blue]{hyperref}
\usepackage{fixltx2e}

\def\lsim{\mathrel{\raise .4ex\hbox{\rlap{$<$}\lower 1.2ex\hbox{$\sim$}}}}
\def\gsim{\mathrel{\raise .4ex\hbox{\rlap{$>$}\lower 1.2ex\hbox{$\sim$}}}}
\def\muKRJ{\ifmmode \,\mu$K$_{\rm RJ}$\else \,$\mu$\hbox{K}$_{\rm RJ}$\fi}
\def\muKCMB{\ifmmode \,\mu$K$_{\rm CMB}$\else \,$\mu$\hbox{K}$_{\rm CMB}$\fi}

\def\leaderfil{\leaders\hbox to 5pt{\hss.\hss}\hfil}

\DeclareGraphicsExtensions{.eps,.pdf,.png,.jpg}

\pdfmapfile{+txfonts.map}

\newcommand{\be}{\begin{equation}}
\newcommand{\ee}{\end{equation}}

\newcommand{\hatn}{\hat{\vec{n}}}

\newcommand{\Nside}{\ensuremath{N_{\mathrm{side}}}} 

\newcommand{\yslm}[3]{ {}_{#1}\!Y_{#2 #3} }

\newcommand{\elt}{\ell}

\newcommand{\healpix}{{\tt HEALPix}}
\newcommand{\toast}{{\tt TOAST}}
\newcommand{\madam}{{\tt Madam}}
\newcommand{\polkapix}{{\tt Polkapix}}
\newcommand{\madamtoast}{{\tt Madam/TOAST}}
\newcommand{\camb}{{\tt camb}}
\newcommand{\conviqt}{{\tt conviqt}}
\newcommand{\iscalmtod}{{\tt IScalm2TOD}}
\newcommand{\levels}{{\tt LevelS}}
\newcommand{\multimod}{{\tt multimod}}
\newcommand{\febecop}{{\tt FEBeCoP}}
\newcommand{\commander}{{\tt Commander}}
\newcommand{\biposh}{{\tt BipoSH}}

\makeatletter
\renewcommand*{\@fnsymbol}[1]{\ensuremath{\ifcase#1\or \dagger\or \ddagger\or *\or
   \mathsection\or \mathparagraph\or \|\or **\or \dagger\dagger
   \or \ddagger\ddagger \else\@ctrerr\fi}}
\makeatother

\begin{document}

\title{\Planck\ 2015 results. XII. Full Focal Plane simulations}

\titlerunning{Full Focal Plane Simulations}
\authorrunning{Planck Collaboration}
\author{\small
Planck Collaboration: P.~A.~R.~Ade\inst{87}
\and
N.~Aghanim\inst{60}
\and
M.~Arnaud\inst{75}
\and
M.~Ashdown\inst{71, 5}
\and
J.~Aumont\inst{60}
\and
C.~Baccigalupi\inst{86}
\and
A.~J.~Banday\inst{95, 9}
\and
R.~B.~Barreiro\inst{66}
\and
J.~G.~Bartlett\inst{1, 68}
\and
N.~Bartolo\inst{31, 67}
% ...
}
\institute{\small
APC, AstroParticule et Cosmologie, Universit\'{e} Paris Diderot, CNRS/IN2P3, CEA/lrfu, Observatoire de Paris, Sorbonne Paris Cit\'{e}, 10, rue Alice Domon et L\'{e}onie Duquet, 75205 Paris Cedex 13, France\goodbreak
\and
African Institute for Mathematical Sciences, 6-8 Melrose Road, Muizenberg, Cape Town, South Africa\goodbreak
\and
Agenzia Spaziale Italiana Science Data Center, Via del Politecnico snc, 00133, Roma, Italy\goodbreak
\and
Aix Marseille Universit\'{e}, CNRS, LAM (Laboratoire d'Astrophysique de Marseille) UMR 7326, 13388, Marseille, France\goodbreak
\and
Astrophysics Group, Cavendish Laboratory, University of Cambridge, J J Thomson Avenue, Cambridge CB3 0HE, U.K.\goodbreak
\and
Astrophysics \& Cosmology Research Unit, School of Mathematics, Statistics \& Computer Science, University of KwaZulu-Natal, Westville Campus, Private Bag X54001, Durban 4000, South Africa\goodbreak
\and
Atacama Large Millimeter/submillimeter Array, ALMA Santiago Central Offices, Alonso de Cordova 3107, Vitacura, Casilla 763 0355, Santiago, Chile\goodbreak
\and
CITA, University of Toronto, 60 St. George St., Toronto, ON M5S 3H8, Canada\goodbreak
\and
CNRS, IRAP, 9 Av. colonel Roche, BP 44346, F-31028 Toulouse cedex 4, France\goodbreak
\and
CRANN, Trinity College, Dublin, Ireland\goodbreak
\and
California Institute of Technology, Pasadena, California, U.S.A.\goodbreak
\and
Centre for Theoretical Cosmology, DAMTP, University of Cambridge, Wilberforce Road, Cambridge CB3 0WA, U.K.\goodbreak
\and
Centro de Estudios de F\'{i}sica del Cosmos de Arag\'{o}n (CEFCA), Plaza San Juan, 1, planta 2, E-44001, Teruel, Spain\goodbreak
\and
Computational Cosmology Center, Lawrence Berkeley National Laboratory, Berkeley, California, U.S.A.\goodbreak
\and
Consejo Superior de Investigaciones Cient\'{\i}ficas (CSIC), Madrid, Spain\goodbreak
\and
DSM/Irfu/SPP, CEA-Saclay, F-91191 Gif-sur-Yvette Cedex, France\goodbreak
\and
DTU Space, National Space Institute, Technical University of Denmark, Elektrovej 327, DK-2800 Kgs. Lyngby, Denmark\goodbreak
\and
D\'{e}partement de Physique Th\'{e}orique, Universit\'{e} de Gen\`{e}ve, 24, Quai E. Ansermet,1211 Gen\`{e}ve 4, Switzerland\goodbreak
\and
Departamento de Astrof\'{i}sica, Universidad de La Laguna (ULL), E-38206 La Laguna, Tenerife, Spain\goodbreak
\and
Departamento de F\'{\i}sica, Universidad de Oviedo, Avda. Calvo Sotelo s/n, Oviedo, Spain\goodbreak
\and
Department of Astronomy and Astrophysics, University of Toronto, 50 Saint George Street, Toronto, Ontario, Canada\goodbreak
\and
Department of Astrophysics/IMAPP, Radboud University Nijmegen, P.O. Box 9010, 6500 GL Nijmegen, The Netherlands\goodbreak
\and
Department of Physics \& Astronomy, University of British Columbia, 6224 Agricultural Road, Vancouver, British Columbia, Canada\goodbreak
\and
Department of Physics and Astronomy, Dana and David Dornsife College of Letter, Arts and Sciences, University of Southern California, Los Angeles, CA 90089, U.S.A.\goodbreak
\and
Department of Physics and Astronomy, University College London, London WC1E 6BT, U.K.\goodbreak
\and
Department of Physics, Florida State University, Keen Physics Building, 77 Chieftan Way, Tallahassee, Florida, U.S.A.\goodbreak
\and
Department of Physics, Gustaf H\"{a}llstr\"{o}min katu 2a, University of Helsinki, Helsinki, Finland\goodbreak
\and
Department of Physics, Princeton University, Princeton, New Jersey, U.S.A.\goodbreak
\and
Department of Physics, University of California, Santa Barbara, California, U.S.A.\goodbreak
\and
Department of Physics, University of Illinois at Urbana-Champaign, 1110 West Green Street, Urbana, Illinois, U.S.A.\goodbreak
\and
Dipartimento di Fisica e Astronomia G. Galilei, Universit\`{a} degli Studi di Padova, via Marzolo 8, 35131 Padova, Italy\goodbreak
\and
Dipartimento di Fisica e Scienze della Terra, Universit\`{a} di Ferrara, Via Saragat 1, 44122 Ferrara, Italy\goodbreak
\and
Dipartimento di Fisica, Universit\`{a} La Sapienza, P. le A. Moro 2, Roma, Italy\goodbreak
\and
Dipartimento di Fisica, Universit\`{a} degli Studi di Milano, Via Celoria, 16, Milano, Italy\goodbreak
\and
Dipartimento di Fisica, Universit\`{a} degli Studi di Trieste, via A. Valerio 2, Trieste, Italy\goodbreak
\and
Dipartimento di Matematica, Universit\`{a} di Roma Tor Vergata, Via della Ricerca Scientifica, 1, Roma, Italy\goodbreak
\and
Discovery Center, Niels Bohr Institute, Blegdamsvej 17, Copenhagen, Denmark\goodbreak
\and
European Southern Observatory, ESO Vitacura, Alonso de Cordova 3107, Vitacura, Casilla 19001, Santiago, Chile\goodbreak
\and
European Space Agency, ESAC, Planck Science Office, Camino bajo del Castillo, s/n, Urbanizaci\'{o}n Villafranca del Castillo, Villanueva de la Ca\~{n}ada, Madrid, Spain\goodbreak
\and
European Space Agency, ESTEC, Keplerlaan 1, 2201 AZ Noordwijk, The Netherlands\goodbreak
\and
Facolt\`{a} di Ingegneria, Universit\`{a} degli Studi e-Campus, Via Isimbardi 10, Novedrate (CO), 22060, Italy\goodbreak
\and
Gran Sasso Science Institute, INFN, viale F. Crispi 7, 67100 L'Aquila, Italy\goodbreak
\and
HGSFP and University of Heidelberg, Theoretical Physics Department, Philosophenweg 16, 69120, Heidelberg, Germany\goodbreak
\and
Helsinki Institute of Physics, Gustaf H\"{a}llstr\"{o}min katu 2, University of Helsinki, Helsinki, Finland\goodbreak
\and
INAF - Osservatorio Astronomico di Padova, Vicolo dell'Osservatorio 5, Padova, Italy\goodbreak
\and
INAF - Osservatorio Astronomico di Roma, via di Frascati 33, Monte Porzio Catone, Italy\goodbreak
\and
INAF - Osservatorio Astronomico di Trieste, Via G.B. Tiepolo 11, Trieste, Italy\goodbreak
\and
INAF/IASF Bologna, Via Gobetti 101, Bologna, Italy\goodbreak
\and
INAF/IASF Milano, Via E. Bassini 15, Milano, Italy\goodbreak
\and
INFN, Sezione di Bologna, Via Irnerio 46, I-40126, Bologna, Italy\goodbreak
\and
INFN, Sezione di Roma 1, Universit\`{a} di Roma Sapienza, Piazzale Aldo Moro 2, 00185, Roma, Italy\goodbreak
\and
INFN, Sezione di Roma 2, Universit\`{a} di Roma Tor Vergata, Via della Ricerca Scientifica, 1, Roma, Italy\goodbreak
\and
INFN/National Institute for Nuclear Physics, Via Valerio 2, I-34127 Trieste, Italy\goodbreak
\and
IPAG: Institut de Plan\'{e}tologie et d'Astrophysique de Grenoble, Universit\'{e} Grenoble Alpes, IPAG, F-38000 Grenoble, France, CNRS, IPAG, F-38000 Grenoble, France\goodbreak
\and
IUCAA, Post Bag 4, Ganeshkhind, Pune University Campus, Pune 411 007, India\goodbreak
\and
Imperial College London, Astrophysics group, Blackett Laboratory, Prince Consort Road, London, SW7 2AZ, U.K.\goodbreak
\and
Infrared Processing and Analysis Center, California Institute of Technology, Pasadena, CA 91125, U.S.A.\goodbreak
\and
Institut N\'{e}el, CNRS, Universit\'{e} Joseph Fourier Grenoble I, 25 rue des Martyrs, Grenoble, France\goodbreak
\and
Institut Universitaire de France, 103, bd Saint-Michel, 75005, Paris, France\goodbreak
\and
Institut d'Astrophysique Spatiale, CNRS (UMR8617) Universit\'{e} Paris-Sud 11, B\^{a}timent 121, Orsay, France\goodbreak
\and
Institut d'Astrophysique de Paris, CNRS (UMR7095), 98 bis Boulevard Arago, F-75014, Paris, France\goodbreak
\and
Institut f\"ur Theoretische Teilchenphysik und Kosmologie, RWTH Aachen University, D-52056 Aachen, Germany\goodbreak
\and
Institute of Astronomy, University of Cambridge, Madingley Road, Cambridge CB3 0HA, U.K.\goodbreak
\and
Institute of Theoretical Astrophysics, University of Oslo, Blindern, Oslo, Norway\goodbreak
\and
Instituto de Astrof\'{\i}sica de Canarias, C/V\'{\i}a L\'{a}ctea s/n, La Laguna, Tenerife, Spain\goodbreak
\and
Instituto de F\'{\i}sica de Cantabria (CSIC-Universidad de Cantabria), Avda. de los Castros s/n, Santander, Spain\goodbreak
\and
Istituto Nazionale di Fisica Nucleare, Sezione di Padova, via Marzolo 8, I-35131 Padova, Italy\goodbreak
\and
Jet Propulsion Laboratory, California Institute of Technology, 4800 Oak Grove Drive, Pasadena, California, U.S.A.\goodbreak
\and
Jodrell Bank Centre for Astrophysics, Alan Turing Building, School of Physics and Astronomy, The University of Manchester, Oxford Road, Manchester, M13 9PL, U.K.\goodbreak
\and
Kavli Institute for Cosmological Physics, University of Chicago, Chicago, IL 60637, USA\goodbreak
\and
Kavli Institute for Cosmology Cambridge, Madingley Road, Cambridge, CB3 0HA, U.K.\goodbreak
\and
Kazan Federal University, 18 Kremlyovskaya St., Kazan, 420008, Russia\goodbreak
\and
LAL, Universit\'{e} Paris-Sud, CNRS/IN2P3, Orsay, France\goodbreak
\and
LERMA, CNRS, Observatoire de Paris, 61 Avenue de l'Observatoire, Paris, France\goodbreak
\and
Laboratoire AIM, IRFU/Service d'Astrophysique - CEA/DSM - CNRS - Universit\'{e} Paris Diderot, B\^{a}t. 709, CEA-Saclay, F-91191 Gif-sur-Yvette Cedex, France\goodbreak
\and
Laboratoire Traitement et Communication de l'Information, CNRS (UMR 5141) and T\'{e}l\'{e}com ParisTech, 46 rue Barrault F-75634 Paris Cedex 13, France\goodbreak
\and
Laboratoire de Physique Subatomique et Cosmologie, Universit\'{e} Grenoble-Alpes, CNRS/IN2P3, 53, rue des Martyrs, 38026 Grenoble Cedex, France\goodbreak
\and
Laboratoire de Physique Th\'{e}orique, Universit\'{e} Paris-Sud 11 \& CNRS, B\^{a}timent 210, 91405 Orsay, France\goodbreak
\and
Lawrence Berkeley National Laboratory, Berkeley, California, U.S.A.\goodbreak
\and
Lebedev Physical Institute of the Russian Academy of Sciences, Astro Space Centre, 84/32 Profsoyuznaya st., Moscow, GSP-7, 117997, Russia\goodbreak
\and
Max-Planck-Institut f\"{u}r Astrophysik, Karl-Schwarzschild-Str. 1, 85741 Garching, Germany\goodbreak
\and
McGill Physics, Ernest Rutherford Physics Building, McGill University, 3600 rue University, Montr\'{e}al, QC, H3A 2T8, Canada\goodbreak
\and
National University of Ireland, Department of Experimental Physics, Maynooth, Co. Kildare, Ireland\goodbreak
\and
Nicolaus Copernicus Astronomical Center, Bartycka 18, 00-716 Warsaw, Poland\goodbreak
\and
Niels Bohr Institute, Blegdamsvej 17, Copenhagen, Denmark\goodbreak
\and
SISSA, Astrophysics Sector, via Bonomea 265, 34136, Trieste, Italy\goodbreak
\and
School of Physics and Astronomy, Cardiff University, Queens Buildings, The Parade, Cardiff, CF24 3AA, U.K.\goodbreak
\and
School of Physics and Astronomy, University of Nottingham, Nottingham NG7 2RD, U.K.\goodbreak
\and
Sorbonne Universit\'{e}-UPMC, UMR7095, Institut d'Astrophysique de Paris, 98 bis Boulevard Arago, F-75014, Paris, France\goodbreak
\and
Space Sciences Laboratory, University of California, Berkeley, California, U.S.A.\goodbreak
\and
Special Astrophysical Observatory, Russian Academy of Sciences, Nizhnij Arkhyz, Zelenchukskiy region, Karachai-Cherkessian Republic, 369167, Russia\goodbreak
\and
Sub-Department of Astrophysics, University of Oxford, Keble Road, Oxford OX1 3RH, U.K.\goodbreak
\and
Theory Division, PH-TH, CERN, CH-1211, Geneva 23, Switzerland\goodbreak
\and
UPMC Univ Paris 06, UMR7095, 98 bis Boulevard Arago, F-75014, Paris, France\goodbreak
\and
Universit\'{e} de Toulouse, UPS-OMP, IRAP, F-31028 Toulouse cedex 4, France\goodbreak
% ...
}

\abstract{
We present the 8th Full Focal Plane simulation set (FFP8), deployed in support of the \Planck\ 2015 results. FFP8 consists of $10$ fiducial mission realizations reduced to 18\,144 maps, together with the most massive suite of Monte Carlo realizations of instrument noise and CMB ever generated, comprising $10^4$ mission realizations reduced to about $10^6$ maps. The resulting maps incorporate the dominant instrumental, scanning, and data analysis effects; remaining subdominant effects will be included in future updates. Generated at a cost of some 25 million CPU-hours spread across multiple high-performance-computing (HPC) platforms, FFP8 is used for the validation and verification of analysis algorithms, as well as their implementations, and for removing biases from and quantifying uncertainties in the results of analyses of the real data.
}

\maketitle


\section{Introduction}
\label{sec_introduction}

\Planck{\footnote{\Planck\ (\url{http://www.esa.int/Planck}) is a project of the European Space Agency (ESA) with instruments provided by two sci-entific consortia funded by ESA member states and led by Principal Investigators from France and Italy, telescope reflectors provided through a collaboration between ESA and a scientific consortium led and funded by Denmark, and additional contributions from NASA (USA).}} is the third satellite to study the cosmic microwave background (CMB) radiation.

Launched in May 2009, \Planck{} started its science observations from the $L_2$ Lagrange point in August 2009 and completed an all-sky survey \citep{planck2013-p01} approximately every six months until it was decommissioned in October 2013. \Planck{} carried two instruments.  The High Frequency Instrument \citep[HFI;][]{Lamarre2010, planck2011-1.5}, comprising 52 detectors\footnote{The data from two detectors proved to be unusable, so only 50~detectors are included in the analysis.} at six frequencies (100, 143, 217, 353, 545, and 857\GHz), completed its observations when its cryogens were exhausted in January 2012 after almost five surveys, while the Low Frequency Instrument \citep[LFI;][]{Bersanelli2010, planck2011-1.4}, comprising 22 detectors at three frequencies (30, 44, and 70\GHz) continued operating throughout the satellite lifetime, completing more than eight surveys.

The second release of \Planck{} data (hereafter PR2-2015) is based on five HFI and eight LFI surveys (LFI Survey 9 having been reserved for atypical scans designed to assist in the understanding of systematic effects), and for the first time includes results in polarization as well as intensity. PR2-2015 is accompanied by a suite of papers, of which this is one, together with an online explanatory supplement, an ESA legacy data archive (\url{http://pla.esac.esa.int/pla/}) and NASA partial mirror (\url{http://irsa.ipac.caltech.edu/Missions/planck.html}), as well as  resources to access and manipulate the full simulation suite described here (\url{http://crd.lbl.gov/cmb-data}).

Simulations play a number of important roles in the analysis of these \Planck{} data, including:
\begin{enumerate}

\item {\em validating and verifying tools used to measure instrument characteristics} by simulating data with known instrument characteristics, applying the tools used on the real data to measure these, and verifying the accuracy of their recovery;

\item {\em quantifying systematic effect residuals} by simulating data with some particular systematic effect included, applying the treatment used on the real data to ameliorate that effect, and measuring the residuals;

\item {\em validating and verifying data analysis algorithms and their implementations} by simulating data with known science inputs (cosmology and foreground sky) and detector characteristics (beam, bandpass, and noise spectrum), applying the analyses used on the real data to extract this science, and verifying the accuracy of its recovery;

\item {\em debiasing and quantifying uncertainties in the analysis of the real data} by generating massive sets of Monte Carlo (MC) realizations of both the noise and the CMB and passing them separately through the analyses used on the real data to quantify biases and uncertainties.
\end{enumerate}

The first two items are instrument-specific, and distinct pipelines have been developed and employed by the LFI and HFI Data Processing Centres (DPCs); details are provided in the DPC processsing papers \citep{planck2014-a03,planck2014-a04,planck2014-a05,planck2014-a06,planck2014-a07,planck2014-a08,planck2014-a09}. However, the last two items require consistent simulations of LFI and HFI in tandem; such simulations are beyond the scope of either single-instrument pipeline. Furthermore, generating the Monte Carlo simulations is the most computationally intensive part of the \Planck\ data analysis and requires computational capacity and capability far beyond those available at the DPCs. As a result, a massively parallel cross-instrument suite of codes has been developed to provide the collaboration with self-consistent simulations of all detectors at all frequencies - refered to as the Full Focal Plane (FFP) - primarily using the high performance computing (HPC) resources at the National Energy Research Scientific Computing Center\footnote{\url{http://www.nersc.gov}} (NERSC) in the USA and at CSC--IT Center for Science\footnote{\url{http://www.csc.fi}} (CSC) in Finland. Since its first deployment in 2006, this suite of codes has been used to generate a sequence of FFP simulation-sets of increasing veracity, complexity, and volume.  The second \Planck\ data release is supported by FFP8.

This paper is laid out as follows. Section~\ref{sec_specs} outlines the overall specification of FFP8, Sect.~\ref{sec_inputs} details the inputs to FFP8, Sect.~\ref{sec_pipelines} describes the three major pipelines used to generate FFP8, Sect.~\ref{sec_hpc} covers the key HPC aspects of FFP8, and Sect.~\ref{sec_res} summarizes the results.

\section{Specifications}
\label{sec_specs}

The FFP8 simulations include a set of fiducial mission realizations together with separate sets of Monte Carlo realizations of the CMB and the instrument noise. They contain the dominant instrumental (detector beam, bandpass, and correlated noise properties), scanning (pointing and flags), and analysis (mapmaking algorithm and implmentation) effects.

In addition to the baseline maps made from the data from all detectors at a given frequency for the entire mission, there are a number of data cuts that are mapped both for systematics tests and to support cross-spectral analyses. These include:
\begin{itemize}

\item detector subsets (``detsets''), comprising the individual unpolarized detectors and the polarized detector quadruplets corresponding to each leading/trailing horn pair;\footnote{HFI sometimes refers to full channels as detset0; here detset only refers to subsets of detectors.}

\item mission subsets, comprising the surveys, years, and half-missions, with exact boundary definitions given in \citet{planck2014-a03} and \citet{planck2014-a08} for LFI and HFI, respectively; and

\item half-ring subsets, comprising the data from either the first or the second half of each pointing-period ring,
\end{itemize}
The various combinations of these data cuts then define $1\,134$ maps, as enumerated in the top section of Table~\ref{table:maps}. The different types of map are then named according to their included detectors (channel or detset), interval (mission, half-mission, year or survey), and ring-content (full or half-ring); for example the baseline maps are described as channel/mission/full, etc.

\begin{table*}
\renewcommand{\arraystretch}{1.25}
\caption{The numbers of fiducial, MC noise and MC CMB maps at each frequency by detector subset, data interval, and data cut. }
\begin{center}
\begin{tabular}{|c|c|c|r|r|r|r|r|r|r|r|r|r|}
\hline
\multicolumn{13}{|c|}{FFP8 and FFP8.1 Fiducial Maps} \\
\hline
\multirow{2}{*}{Detectors} & \multirow{2}{*}{Interval} & \multirow{2}{*}{Cut} & \multicolumn{9}{|c|}{Frequency} & \multirow{2}{*}{Total} \\
\cline{4-12}
& & & 30 & 44 & 70 & 100 & 143 & 217 & 353 & 545 & 857 & \\
\hline
\multirow{8}{*}{Channel} & \multirow{2}{*}{Mission} & Full & 1 & 1 & 1 & 1 & 1 & 1 & 1 & 1 & 1 & 9 \\
 &  & Half-Ring & 2 & 2 & 2 & 2 & 2 & 2 & 2 & 2 & 2 & 18 \\ \cline{2-13}
 & \multirow{2}{*}{Half-Mission} & Full & 2 & 2 & 2 & 2 & 2 & 2 & 2 & 2 & 2 & 18 \\
 &  & Half-Ring & 4 & 4 & 4 & 4 & 4 & 4 & 4 & 4 & 4 & 36 \\ \cline{2-13}
 & \multirow{2}{*}{Year} & Full & 4 & 4 & 4 & 2 & 2 & 2 & 2 & 2 & 2 & 24 \\
 &  & Half-Ring & 8 & 8 & 8 & 4 & 4 & 4 & 4 & 4 & 4 & 48 \\ \cline{2-13}
 & \multirow{2}{*}{Survey} & Full & 8 & 8 & 8 & 4 & 4 & 4 & 4 & 4 & 4 & 48 \\
 &  & Half-Ring & 16 & 16 & 16 & 8 & 8 & 8 & 8 & 8 & 8 & 96 \\ \hline
\multirow{8}{*}{DetSet} & \multirow{2}{*}{Mission} & Full &\dots&\dots& 3 & 2 & 5 & 6 & 6 & 3 & 4 & 29 \\
 &  & Half-Ring &\dots &\dots & 6 & 4 & 10 & 12 & 12 & 6 & 8 & 58 \\ \cline{2-13}
 & \multirow{2}{*}{Half-Mission} & Full &\dots &\dots & 6 & 4 & 10 & 12 & 12 & 6 & 8 & 58 \\
 &  & Half-Ring &\dots &\dots & 12 & 8 & 20 & 24 & 24 & 12 & 16 & 116 \\ \cline{2-13}
 & \multirow{2}{*}{Year} & Full &\dots &\dots & 12 & 4 & 10 & 12 & 12 & 6 & 8 & 64 \\
 &  & Half-Ring &\dots &\dots & 24 & 8 & 20 & 24 & 24 & 12 & 16 & 128 \\ \cline{2-13}
 & \multirow{2}{*}{Survey} & Full &\dots &\dots & 24 & 8 & 20 & 24 & 24 & 12 & 16 & 128 \\
 &  & Half-Ring &\dots &\dots & 48 & 16 & 40 & 48 & 48 & 24 & 32 & 256 \\
\hline
\multicolumn{3}{|c|}{Total} & 45 & 45 & 180 & 81 & 162 & 189 & 189 & 108 & 135 & 1\,134 \\
\hline
\multicolumn{13}{c}{} \vspace{-0.1in} \\
\hline
\multicolumn{13}{|c|}{FFP8 Noise MC Maps} \\
\hline
\multirow{2}{*}{Detectors} & \multirow{2}{*}{Interval} & \multirow{2}{*}{Cut} & \multicolumn{9}{|c|}{Frequency} & \multirow{2}{*}{Total} \\
\cline{4-12}
 &  &  & 30 & 44 & 70 & 100 & 143 & 217 & 353 & 545 & 857 & \\
\hline
\multirow{8}{*}{Channel} & \multirow{2}{*}{Mission} & Full & 10\,000 & 10\,000 & 10\,000 & 10\,000 & 10\,000 & 10\,000 & 10\,000 & 10\,000 & 10\,000 & 90\,000 \\
 &  & Half-Ring & 20\,000 & 20\,000 & 20\,000 & 200 & 200 & 200 & 200 & 200 & 200 & 61\,200 \\ \cline{2-13}
 & \multirow{2}{*}{Half-Mission} & Full & 2\,000 & 2\,000 & 2\,000 & 2\,000 & 2\,000 & 2\,000 & 2\,000 & 2\,000 & 2\,000 & 18\,000 \\
 &  & Half-Ring & 4\,000 & 4\,000 & 4\,000 & 400 & 400 & 400 & 400 & 400 & 400 & 14\,400 \\ \cline{2-13}
 & \multirow{2}{*}{Year} & Full & 4\,000 & 4\,000 & 4\,000 & 2\,000 & 2\,000 & 2\,000 & 2\,000 & 2\,000 & 2\,000 & 24\,000 \\
 &  & Half-Ring & 8\,000 & 8\,000 & 8\,000 & 400 & 400 & 400 & 400 & 400 & 400 & 26\,400 \\ \cline{2-13}
 & \multirow{2}{*}{Survey} & Full & 8\,000 & 8\,000 & 8\,000 & 4\,000 & 4\,000 & 4\,000 & 4\,000 & 4\,000 & 4\,000 & 48\,000 \\
 &  & Half-Ring & 16\,000 & 16\,000 & 16\,000 & 800 & 800 & 800 & 800 & 800 & 800 & 52\,800 \\ \hline
\multirow{8}{*}{DetSet} & \multirow{2}{*}{Mission} & Full &\dots &\dots & 3\,000 & 2\,000 & 5\,000 & 6\,000 & 6\,000 & 3\,000 & 13\,000 & 38\,000 \\
 &  & Half-Ring &\dots &\dots & 6\,000 & 400 & 1\,000 & 1\,200 & 1\,200 & 600 & 800 & 11\,200 \\ \cline{2-13}
 & \multirow{2}{*}{Half-Mission} & Full &\dots &\dots & 1\,800 & 4\,000 & 10\,000 & 12\,000 & 12\,000 & 6\,000 & 8\,000 & 53\,800 \\
 &  & Half-Ring &\dots &\dots & 3\,600 & 800 & 2\,000 & 2\,400 & 2\,400 & 1\,200 & 1\,600 & 14\,000 \\ \cline{2-13}
 & \multirow{2}{*}{Year} & Full &\dots &\dots & 3\,600 & 4\,000 & 10\,000 & 12\,000 & 12\,000 & 6\,000 & 8\,000 & 55\,600 \\
 &  & Half-Ring &\dots &\dots & 7\,200 & 800 & 2\,000 & 2\,400 & 2\,400 & 1\,200 & 1\,600 & 17\,600 \\ \cline{2-13}
 & \multirow{2}{*}{Survey} & Full &\dots &\dots & 7\,200 & 8\,000 & 20\,000 & 24\,000 & 24\,000 & 12\,000 & 16\,000 & 111\,200 \\
 &  & Half-Ring &\dots &\dots & 14\,400 & 1\,600 & 4\,000 & 4\,800 & 4\,800 & 2\,400 & 3\,200 & 35\,200 \\
\hline
\multicolumn{3}{|c|}{Total} & 72\,000 & 72\,000 & 118\,800 & 41\,400 & 73\,800 & 84\,600 & 84\,600 & 52\,200 & 72\,000 & 671\,400 \\
\hline
\multicolumn{13}{c}{} \vspace{-0.1in} \\
\hline
\multicolumn{13}{|c|}{FFP8 CMB MC Maps} \\
\hline
\multirow{2}{*}{Detectors} & \multirow{2}{*}{Interval} & \multirow{2}{*}{Cut} & \multicolumn{9}{|c|}{Frequency} & \multirow{2}{*}{Total} \\
\cline{4-12}
 &  &  & 30 & 44 & 70 & 100 & 143 & 217 & 353 & 545 & 857 & \\
\hline
\multirow{2}{*}{Channel} & Mission & Full & 10\,000 & 10\,000 & 10\,000 & 10\,000 & 10\,000 & 10\,000 & 10\,000 & 10\,000 & 10\,000 & 90\,000 \\
  & Half-Mission & Full & 20\,000  & 20\,000  & 20\,000  & 20\,000  & 20\,000  & 20\,000  & 20\,000  & 20\,000  & 20\,000  & 180\,000 \\ \hline
Detset & Mission & Full &\dots&\dots& 30\,000 & 20\,000 & 50\,000 & 40\,000 & 20\,000 & 20\,000 & 10\,000 & 190\,000\\
\hline
\multicolumn{3}{|c|}{Total} & 30\,000  & 30\,000  & 60\,000  & 50\,000  & 80\,000  & 70\,000  & 50\,000  & 50\,000  & 40\,000  & 460\,000 \\
\hline
\multicolumn{13}{c}{} \vspace{-0.1in} \\
\hline
\multicolumn{13}{|c|}{FFP8.1 CMB MC Maps} \\
\hline
\multirow{2}{*}{Detectors} & \multirow{2}{*}{Interval} & \multirow{2}{*}{Cut} & \multicolumn{9}{|c|}{Frequency} & \multirow{2}{*}{Total} \\
\cline{4-12}
 &  &  & 30 & 44 & 70 & 100 & 143 & 217 & 353 & 545 & 857 & \\
\hline
\multirow{2}{*}{Channel} & Mission & Full & 1\,000 & 1\,000 & 1\,000 & 1\,000 & 1\,000 & 1\,000 & 1\,000 & 1\,000 & 1\,000 & 9\,000 \\
                                & Half-Mission & Full & 2\,000  & 2\,000  & 2\,000  & 2\,000  & 2\,000  & 2\,000  & 2\,000  & 2\,000  & 2\,000  & 18\,000 \\ 
\hline
\multicolumn{3}{|c|}{Total} & 3\,000  & 3\,000  & 3\,000  & 3\,000  & 3\,000  & 3\,000  & 3\,000  & 3\,000  & 3\,000  & 27\,000 \\
\hline
\end{tabular}
\end{center}
\label{table:maps}
\end{table*}

The fiducial realizations include instrument noise (Sect.~\ref{subsec_noise}), astrophysical foregrounds (Sect.~\ref{subsec:foregrounds}), and the lensed scalar, tensor, and non-Gaussian CMB components (Sect.~\ref{subsec:cmb}), and are primarily designed to support the validation and verification of analysis codes. To test our ability to detect tensor modes and non-Gaussianity, we generate five CMB realizations with various cosmologically interesting -- but undeclared -- values of the tensor-to-scalar ratio $r$ and non-Gaussianity parameter $f_{NL}$ (Table~\ref{table:rfnl}).  To investigate the impact of differences in the bandpasses of the detectors at any given frequency, the foreground sky is simulated using both the individual detector bandpasses and a common average bandpass, to include and exclude the effects of bandpass mismatch. To check that the PR2-2015 results are not sensitive to the exact cosmological parameters used in FFP8 we subsequently generated FFP8.1, exactly matching the PR2-2015 cosmology.

\begin{table}[hbt] 
  \begingroup
  \newdimen\tblskip \tblskip=5pt
  \caption{Values of tensor ($r$) and non-Gaussianity ($f_\mathrm{NL}$) parameters for the FFP8 baseline and FFP8a-d blind challenge cases}
  \label{table:rfnl}
  \nointerlineskip
  \vskip -3mm
  \footnotesize
  \setbox\tablebox=\vbox{
    \newdimen\digitwidth
    \setbox0=\hbox{\rm 0}
    \digitwidth=\wd0
    \catcode`*=\active
    \def*{\kern\digitwidth}
    \newdimen\signwidth
    \setbox0=\hbox{-}
    \signwidth=\wd0
    \catcode`!=\active
    \def!{\kern\signwidth}
\halign{\hbox to 2cm{#\leaderfil}\tabskip 2em&
    \hfil$#$\hfil\tabskip 2em&
    \hfil$#$\hfil\tabskip 0pt\cr
      \noalign{\doubleline}
      \omit\hfil Data set\hfil&r&f_\mathrm{NL}\cr
      \noalign{\vskip 3pt\hrule\vskip 4pt}
      FFP8&\omit\hfil $0$\hfil&\omit\hfil $0$\hfil\cr
      FFP8a&0.222&-4.311\cr
      FFP8b&0.046&!!8.590\cr
      FFP8c&0.088&!!7.140\cr
      FFP8d&0.153&-2.181\cr
      \noalign{\vskip 3pt\hrule\vskip 5pt}
    }
  }
  \endPlancktable
  \endgroup
\end{table}

Since mapmaking is a linear operation, the easiest way to generate all of these different realizations is to build the full set of $1\,134$ maps of each of six components:
\begin{itemize}
\item the lensed scalar CMB ({\tt cmb\_scl});
\item the tensor CMB ({\tt cmb\_ten});
\item the non-Gaussian complement CMB ({\tt cmb\_ngc});
\item the forgreounds including bandpass mismatch ({\tt fg\_bpm});
\item the foregrounds excluding bandpass mismatch ({\tt fg\_nobpm});
\item the noise (noise).
\end{itemize}
We then sum these, weighting the tensor and non-Gaussian complement maps with $\sqrt{r}$ and $f_{\rm NL}$, respectively, and including one of the two foreground maps, to produce 10 total maps of each type. The complete fiducial data set then comprises 18\,144 maps.

While the full set of maps can be generated for the fiducial cases, for the $10^4$-realization MC sets this would result in some $10^7$ maps and require about $6$\,PB of storage. Instead, therefore, the number of realizations generated for each type of map is chosen to balance the improved statistics it supports against the computational cost of its generation and storage. In practice, we generate $10^6$ MC maps in total, including the full $10^4$ realizations for all channel/mission/full maps for both noise and CMB. As detailed in Table~\ref{table:maps}, the remaining noise MCs sample broadly across all data cuts, while the additional CMB MCs are focused on the channel/half-mission/full maps and the subset of the detset/mission/full maps required by the \commander{} component separation code \citep{planck2014-a12}.

\section{Inputs}
\label{sec_inputs}

The inputs to an FFP simulation consist of the characterization of the mission (the satellite pointing and each detector's data flags), the intrument (the focal plane layout and the noise statistics, beam, and bandpass of each of the detectors), and the synthetic sky to be observed.

\subsection{Mission and instrument characteristics}

The goal of FFP8 is to simulate the \Planck{} mission as accurately as possible; however, there are a number of known systematic effects that are not included, either because they are removed in the pre-processing of the time-ordered data (TOD), or because they are insufficiently well-characterized to simulate reliably, or because their inclusion (simulation and removal) would be too computationally expensive. These systematic effects are discussed in detail in \cite{planck2014-a03} and \cite{planck2014-a08} and include:
\begin{itemize}
\item cosmic ray glitches (HFI);
\item spurious spectral lines from the 4-K cooler electronics (HFI);
\item nonlinearity in the analogue-to-digital converter (HFI);
\item imperfect reconstruction of the focal plane geometry.
\end{itemize}
Note that if the residuals from the treatment of any of these effects could be mapped in isolation, then maps of such systematics could simply be added to the existing FFP8 maps to improve their correspondence to the real data.

\subsubsection{Pointing and flags}

The FFP8 detector pointing is calculated by interpolating the satellite attitude to the detector sample times and by applying a fixed rotation from the satellite frame into the detector frame. The fixed rotations are determined by the measured focal plane geometry \citep{planck2014-a05,planck2014-a08}, while the satellite attitude is described in the \Planck\ attitude history files (AHF). The FFP pointing expansion reproduces the DPC pointing to sub-arcsecond accuracy, except for three short and isolated instances during Surveys 6--8 where the LFI sampling frequency was out of specification. Pixelization of the information causes the pointing error to be quantized to either zero (majority of cases) or the distance between pixel centres (3\parcm4 and 1\parcm7 for LFI and HFI, respectively). Since we need a single reconstruction that will serve both instruments efficiently in a massively parallel environment, we use the pointing provided by the Time Ordered Astrophysics Scalable Tools (\toast) package.

\begin{figure}[!ht]
  \centerline{\includegraphics[width=0.49\textwidth]{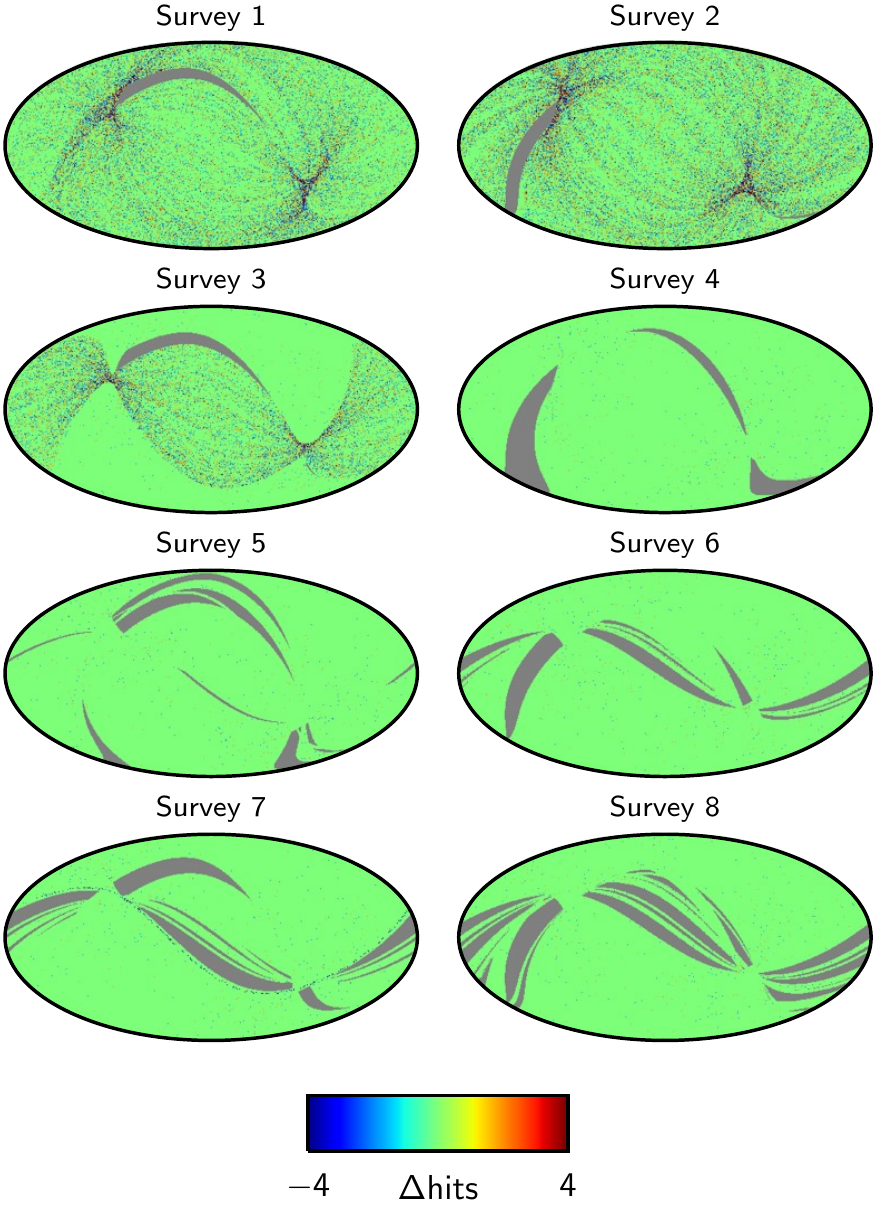}}
  \caption{\label{fig:hit_difference_lfi}
    Survey hit map differences between FFP8 (using \toast\ pointing) and LFI at 70\,GHz. The diffuse error pattern during Surveys 1--3 was caused by differences in pointing interpolation. The matching HFI comparison is presented in Fig.~\ref{fig:hit_difference_hfi}. The FFP pipeline rejected 2.5\,min of integration time on Survey 7, leaving a small deficit of hits on a single scanning ring that is not visible in this projection. A hit is assigned to a pixel where the detector beam centre falls midway through a sample integration period.
  }
\end{figure}

In Fig.~\ref{fig:hit_difference_lfi} we compare LFI DPC and \toast\ ``hit maps,'' which show the number of times a detector beam centre has fallen in a given pixel at the mid-point of a data sample, for the specified time period.  There is a low level, diffuse, difference during the first three surveys, caused by slight differences in the way the pointing correction is interpolated in the two pipelines. The net effect is that some samples near pixel boundaries are assigned to different pixels in the different pipelines. Due to pixels being vastly oversampled, this affects about 25\,\% of the sky pixels in the first three surveys, while the fraction of discrepant pixels after Survey 3 is about 0.5\,\%. In the affected pixels, the relative difference in hit count is $0.2 \pm 0.1\,$\% throughout. For Survey 1, the 70\,GHz median hit count is 812 and 68\,\% of the hit counts fall between 604 and 1\,278.

Except for Survey 7, the channel integration time differs by about a second, with the flight data maps always containing slightly more data. The larger difference during LFI Survey 7 occurs where \toast\ rejected 2.5~min of data due to their irregular sampling rate. All of these differences are entirely negligible compared to the 14\,Ms of overall integration time in each survey. For Survey 7 the relative difference in hits in differing pixels is slightly elevated to 0.3\,\%, with an asymmetric standard deviation of 0.6\,\%.

\begin{figure}[!ht]
  \centerline{\includegraphics[width=0.49\textwidth]{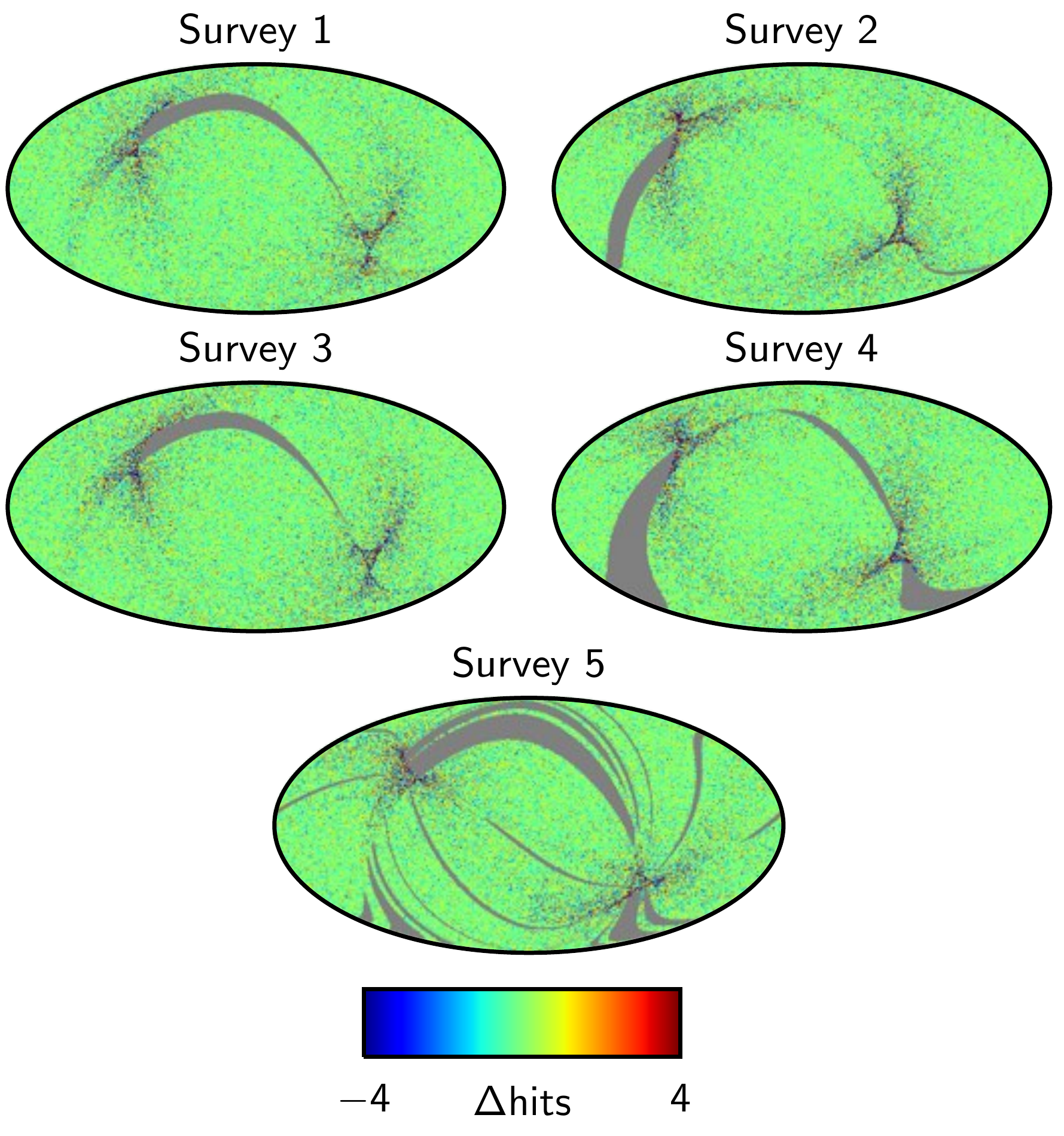}}
  \caption{\label{fig:hit_difference_hfi}
    Survey hit map differences between FFP8 and HFI at 100\,GHz. The diffuse pattern is caused by time-stamp truncation and is not the same as the pointing interpolation difference in the LFI comparison of Surveys 1--3 in Fig.~\ref{fig:hit_difference_lfi}.
  }
\end{figure}

The survey-by-survey HFI comparison in Fig.~\ref{fig:hit_difference_hfi} does not have a diffuse pattern matching Fig.~\ref{fig:hit_difference_lfi} that ends during Survey 3, indicating a better agreement in the pointing correction interpolation between \toast\ and HFI DPC. The more pronounced ecliptic pole caustics are caused by a truncation error in the TOD time stamps that were used as a basis of pointing expansion. In addition, the HFI's smaller pixel size and higher sampling frequency increases the likelihood that small numerical differences can result in samples being assigned across a pixel boundary. On average, about $52\,$\% of the pixels are affected and exhibit hit differences at the level of $0.5 \pm 0.3\,$\%. For $100\,$GHz and Survey 1, the median hit count is $224$ and $68\,$\% of the hit counts fall between $164$ and $361$. The channel integration time between the two pipelines in the HFI case shows negligible differences, of the order of a few tenths of a second for all surveys.

\subsubsection{Noise}
\label{subsec_noise}

We require simulated noise realizations that are representative of the noise in the flight data, including variations in the noise power spectral density (PSD) of each detector over time.  To obtain these we developed a noise estimation pipeline complementary to those of the DPCs. The goal of DPC noise estimation is to monitor instrument health and to derive optimal noise weighting, whereas our estimation is optimized to feed into noise simulation. Key features are the use of full mission maps for signal subtraction, long (about 24 hour) realization length, and the use of autocorrelation functions in place of Fourier transforms to handle flagged and masked data (HFI).

\begin{figure}[!ht]
  \centerline{\includegraphics[width=0.5\textwidth]{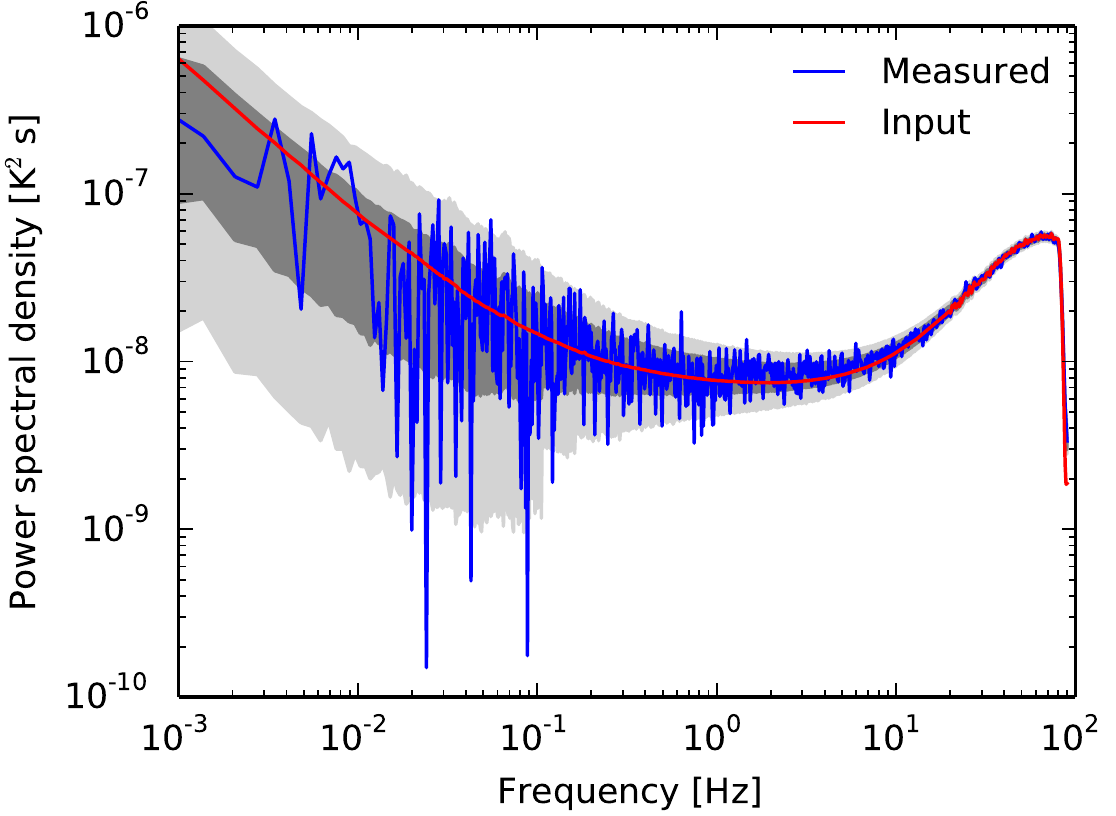}}
  \caption{\label{fig:noise_test}
    Example comparison of the input and recovered noise PSDs for one pointing period of a signal+noise simulation of HFI bolometer 100-1a. The shaded regions around the input model reflect the asymmetric realization scatter of the estimated PSDs at $68\,$\% and $95\,$\% confidence intervals in each of the $707$ logarithmically-placed frequency bins.
  }
\end{figure}

The PSD for each detector and each roughly 45\,min pointing period was estimated as follows.
\begin{enumerate}

\item Read in a 15\% mask for the Galaxy and point sources.

\item Read the calibrated and dipole-subtracted time-ordered data and quality flags for the interval.

\item Expand detector pointing for the interval.

\item Scan and subtract a signal estimated from the full mission, full frequency map using the detector pointing weights.

\item Subtract the mean of the interval.

\item Then either
  \begin{itemize}

  \item Fourier transform the signal-subtracted TOD for a noise PSD estimate, correcting for missing samples by dividing by the unflagged fraction (LFI),
  \end{itemize}
  or
  \begin{itemize}

  \item estimate the autocovariance function from the unflagged and unmasked samples and Fourier transform the autocovariance function into a noise PSD (HFI).
  \end{itemize}

\item Bin the noise PSD estimate into $300$ logarithmically spaced bins.

\item Average the noise estimates into approximately daily noise spectra.

\item Fit an analytic $1/f$ model to the measured spectra.

\end{enumerate}
We show the results of running our pipeline on a signal+noise simulation in Fig.~\ref{fig:noise_test}.

\begin{figure*}[!htb]
  \centerline{\includegraphics[trim=260 0 0 0,clip,width=0.5\textwidth]{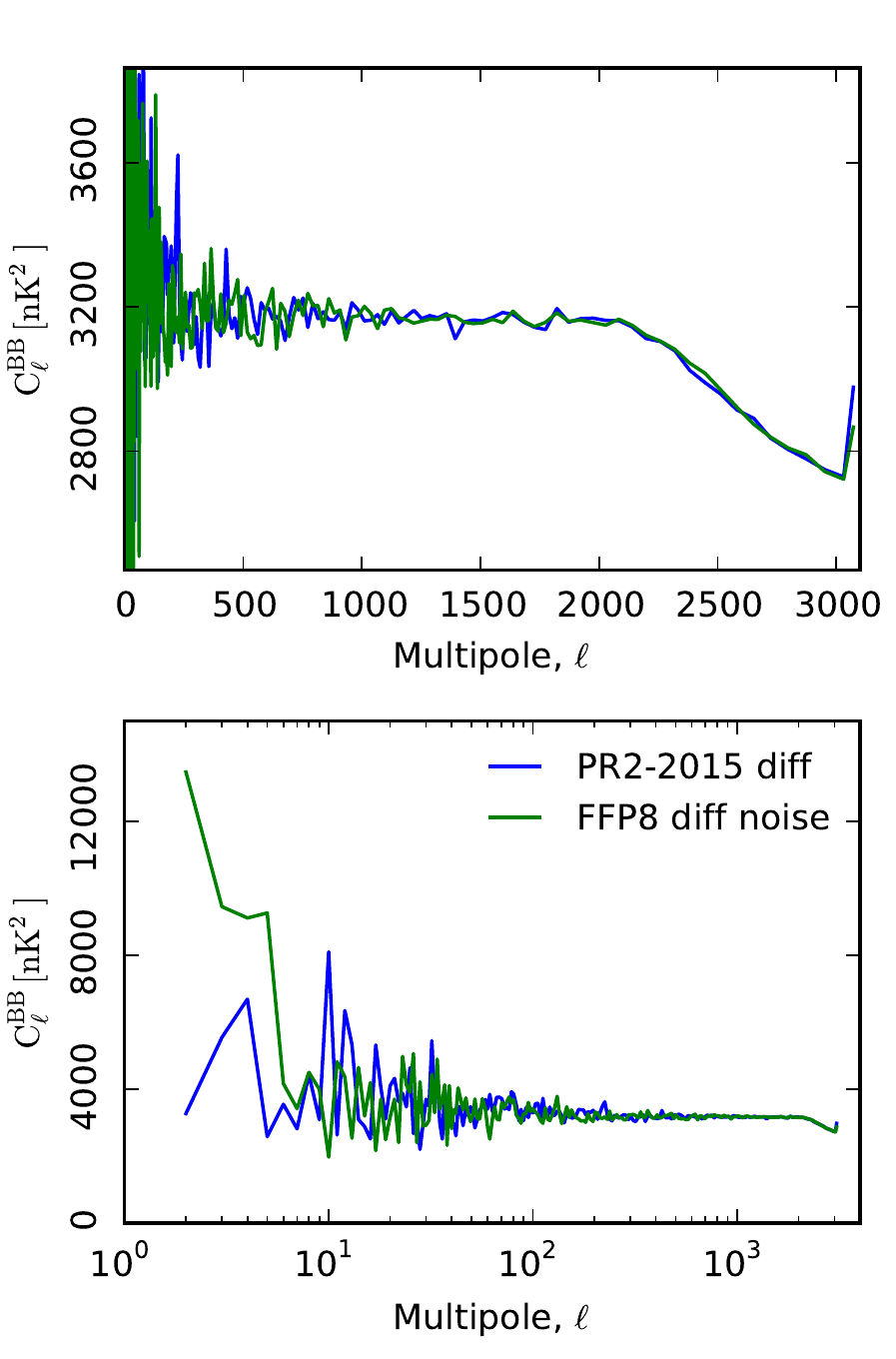}\includegraphics[trim=260 0 0 0,clip,width=0.5\textwidth]{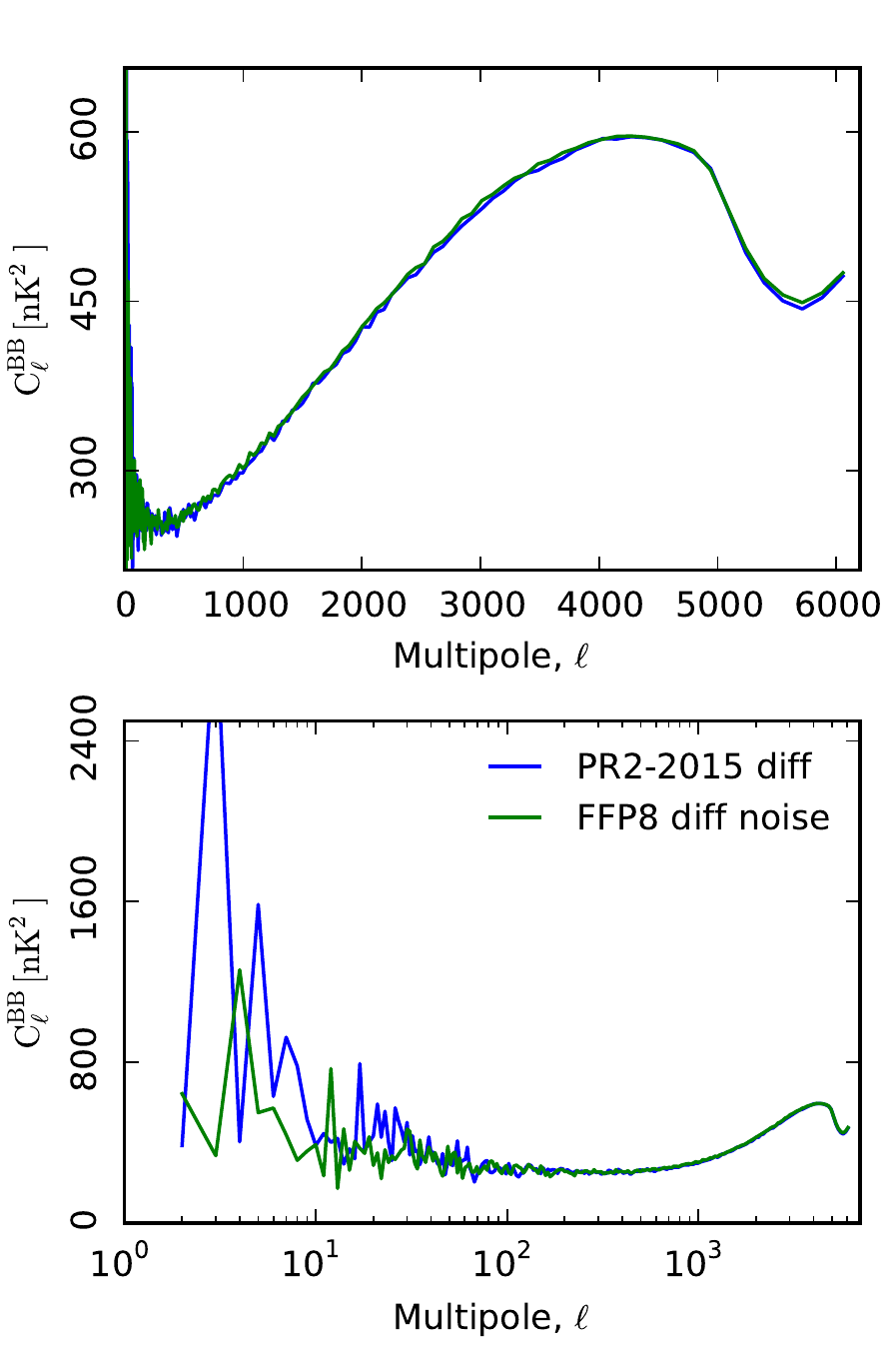}}
  \caption{\label{fig:half_ring_noise}
    $BB$ spectra of 70\,GHz ({\it left}) and 100\,GHz ({\it right}) half-ring half difference noise maps, for both FFP8 simulations ({\it green}) and flight data ({\it black}).  Differences at the low-$\ell$ end are caused by sample variance. These are pseudo-spectra, computed on $75\,$\% of the sky with the Galactic plane and point sources masked. \emph{Top:} Linear horizontal axis to show small-scale behaviour.  \emph{Bottom:} Logarithmic horizontal axis to show large-scale behaviour.
  }
\end{figure*}

As a part of this process we observed significant low frequency ($\sim 1$\,mHz) correlated noise in the flight data between polarized HFI bolometers that share a horn,\footnote{Each of the polarized horns on the \Planck\ focal plane has two orthogonally polarized detectors. Due to the scanning strategy, it takes two polarized horns, i.e., a quadruplet of detectors, to solve for polarization away from the regions close to the ecliptic poles where the scanning rings intersect.} designated $a$ and $b$.  These pairs of detectors form a polarization-sensitive unit, as their relative difference is sensitive to linear polarization of the incident radiation, and the existence of a correlated noise component between them is a known feature of the HFI intrument \cite{planck2013-p03}. When this correlated noise was not included in the simulations, it left a clear signature in the mismatch between our noise simulations and the half ring difference maps. Specifically, $a/b$ correlated noise adds coherently in the $(a+b)$ combination that is sensitive to temperature modulation of the signal, whereas it cancels in the $(a-b)$ combination that detects polarization modulation of the signal. Without simulating the correlated noise component, our simulated noise maps had too little large-scale noise in the $TT$ spectrum, but excess noise in the $EE$ and $BB$ spectra. We estimated the correlated noise spectrum by differencing the $(a+b)$ and $(a-b)$ noise estimates, and added the correlated component into our noise simulation. FFP8 is the first simulation to include realistic $a/b$ noise correlations with the appropriate red PSD.

We also found that our physically motivated HFI noise models contained a mismatch with the measured noise PSDs at the very highest frequencies ($>50$\,Hz). Replacing the models with actual binned noise PSDs in the highest-frequency bins gave better performance in the half-ring map test. Examples of the correlated noise spectrum and the noise model fits to the autospectra can be found in \cite{planck2014-a08}.

To validate the results we considered the angular power spectra of both simulated and flight data half-ring map differences (i.e., splitting each pointing period into two halves, making maps from all of the first halves and second halves separately, and taking the difference of the maps), chosen because they are virtually free of all signal residuals while still supporting most of the same noise power as the full-ring data. Fig.~\ref{fig:half_ring_noise} illustrates the excellent agreement between the simulated and real data half-ring map $BB$ spectra for 70 and 100\,GHz.

\subsubsection{Beams}
\label{subsec_beams}

The simulations use the so-called scanning beams (e.g., \citealt{planck2013-p03}), which give the point-spread function of for a given detector \emph{including} all temporal data processing effects: sample integration, demodulation, ADC nonlinearity residuals, bolometric time constant residuals, etc. In the absence of significant residuals (LFI), the scanning beams may be estimated from the optical beams by ``smearing'' them in the scanning direction to match the finite integration time for each instrument sample. Where there are unknown residuals in the timelines (HFI), the scanning beam must be measured directly from observations of strong point-like sources, namely planets. If the residuals are present but understood, it is possible to simulate the beam measurement and predict the scanning beam shape starting from the optical beam.

For FFP8, the scanning beams are expanded in terms of their spherical harmonic coefficients, $b_{\ell m}$, with the order of the expansion (maximum $\ell$ and $m$ considered) representing a trade-off between the accuracy of the representation and the computational cost of its convolution. The LFI horns have larger beams with larger sidelobes (due to their location on the outside of the focal plane), and we treat them as full $4\pi$ beams divided into main (up to 1\pdeg9, 1\pdeg3, and 0\pdeg9 for 30, 44, and 70\,GHz, respectively), intermediate (up to 5\deg), and sidelobe (above 5\deg) components \citep{planck2014-a05}. This division allows us to tune the expansion orders of the three components separately. HFI horns are limited to the main beam component, measured out to 100\arcm\ \citep{planck2014-a08}. Since detector beams are characterized independently, the simulations naturally include differential beam and pointing systematics.

\subsubsection{Bandpasses}

\begin{figure*}[!hbt]
  \centerline{\includegraphics[width=1.0\textwidth,trim=0 25 0  0,clip=True]{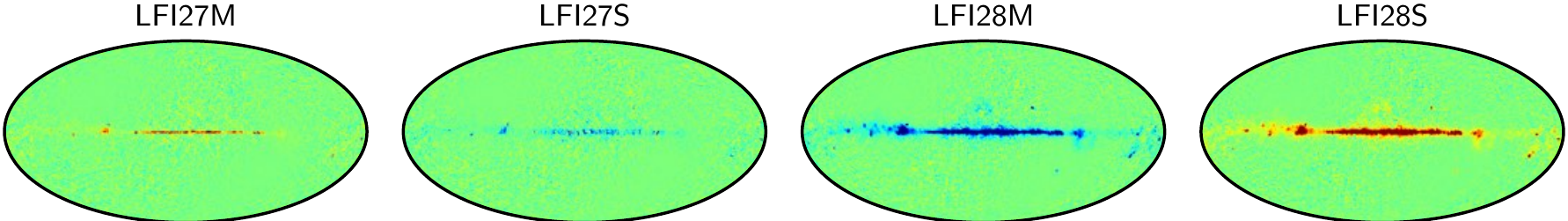}}
  \centerline{\includegraphics[width=1.0\textwidth,trim=0  0 0 10,clip=True]{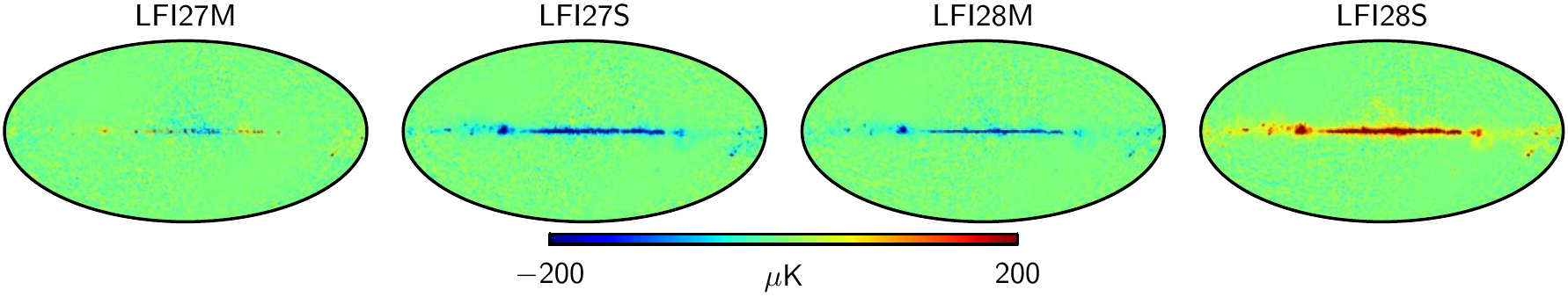}}
  \centerline{\includegraphics[width=1.0\textwidth,trim=0 25 0  0,clip=True]{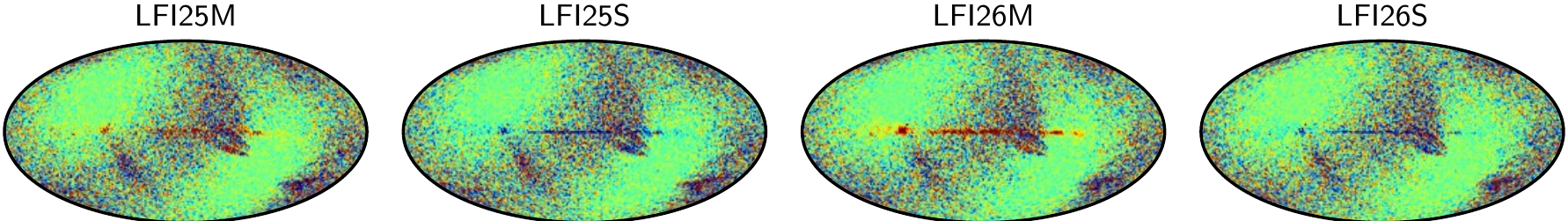}}
  \centerline{\includegraphics[width=1.0\textwidth,trim=0  0 0 10,clip=True]{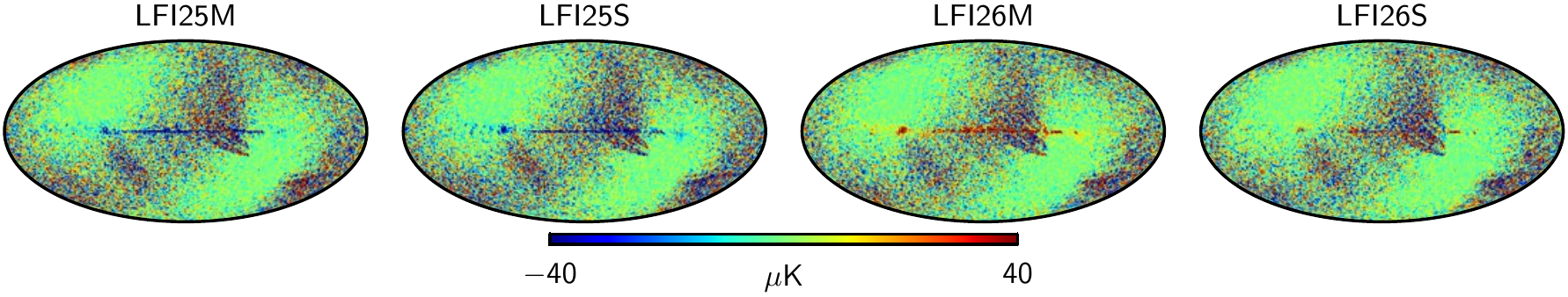}}
  \centerline{\includegraphics[width=1.0\textwidth,trim=0 25 0  0,clip=True]{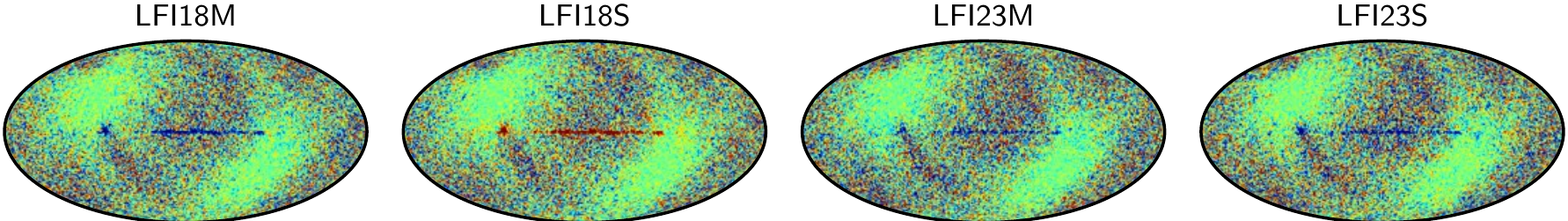}}
  \centerline{\includegraphics[width=1.0\textwidth,trim=0  0 0 10,clip=True]{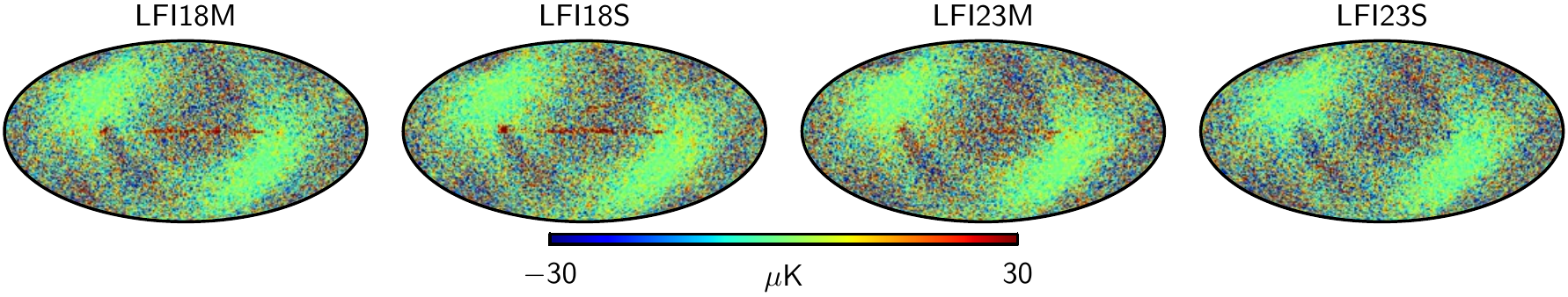}}
  \caption{\label{fig:iqus_ffp8}
    Pairs of measured bandpass mismatch LFI maps, comparing FFP8 simulations (upper) and 2015 \Planck\ (lower) data. Our method of estimating the total bandpass mismatch between each detector and a frequency average is detailed in Appendix~\ref{appendix:iqus}.
  }
\end{figure*}

\begin{figure*}[!ht]
  \centerline{\includegraphics[width=1.0\textwidth,trim=0 25 0  0,clip=True]{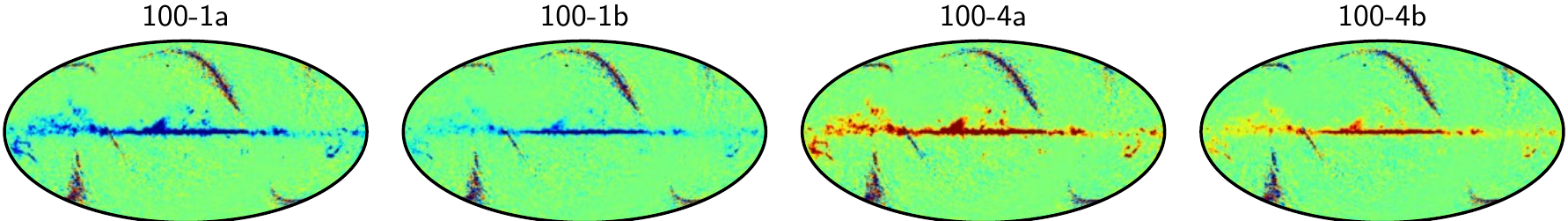}}
  \centerline{\includegraphics[width=1.0\textwidth,trim=0  0 0 10,clip=True]{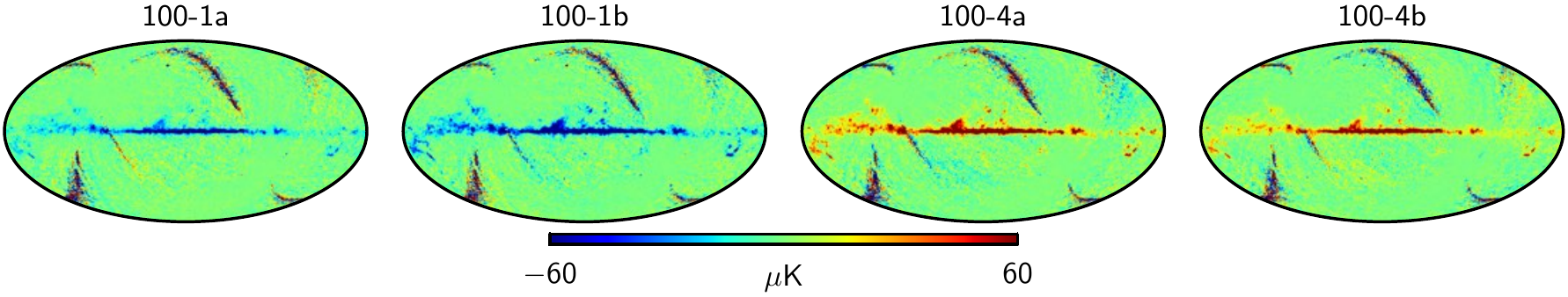}}
  \centerline{\includegraphics[width=1.0\textwidth,trim=0 25 0  0,clip=True]{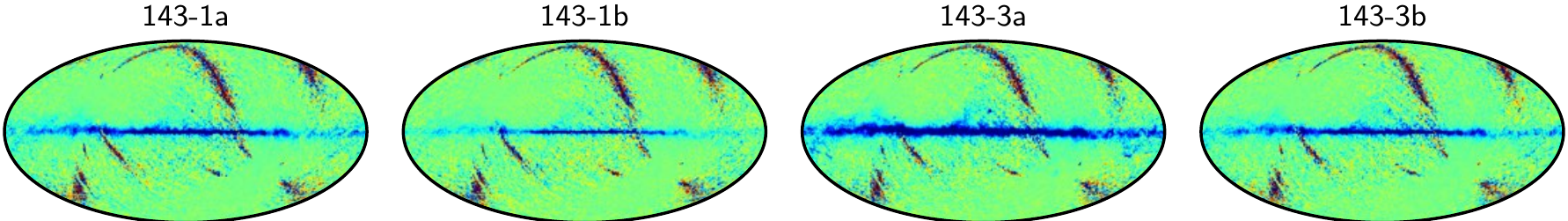}}
  \centerline{\includegraphics[width=1.0\textwidth,trim=0  0 0 10,clip=True]{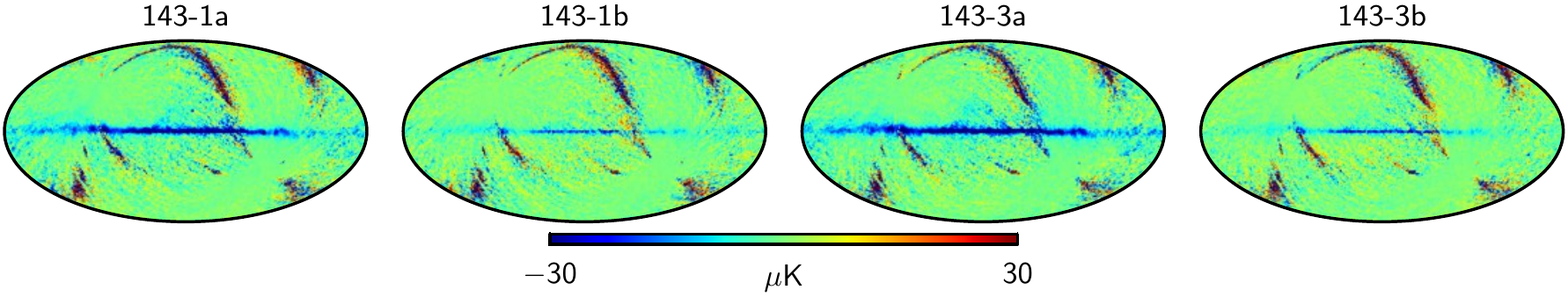}}
  \centerline{\includegraphics[width=1.0\textwidth,trim=0 25 0  0,clip=True]{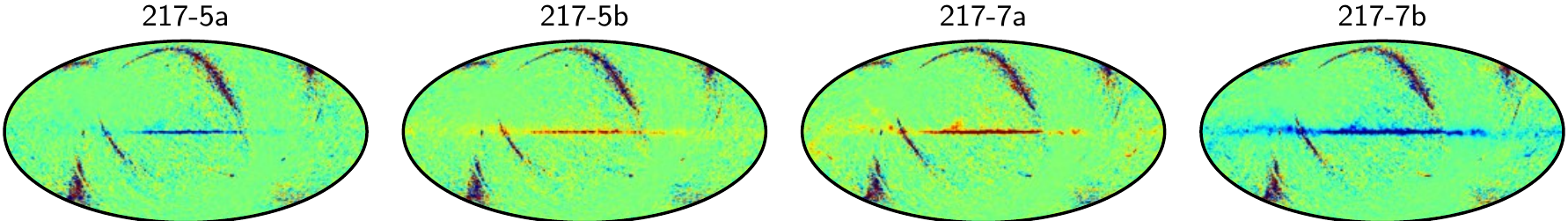}}
  \centerline{\includegraphics[width=1.0\textwidth,trim=0  0 0 10,clip=True]{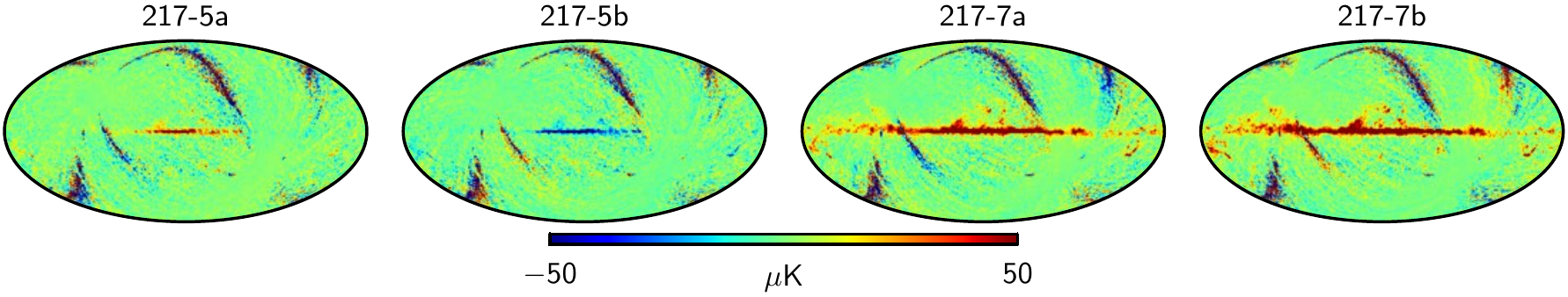}}
  \centerline{\includegraphics[width=1.0\textwidth,trim=0 25 0  0,clip=True]{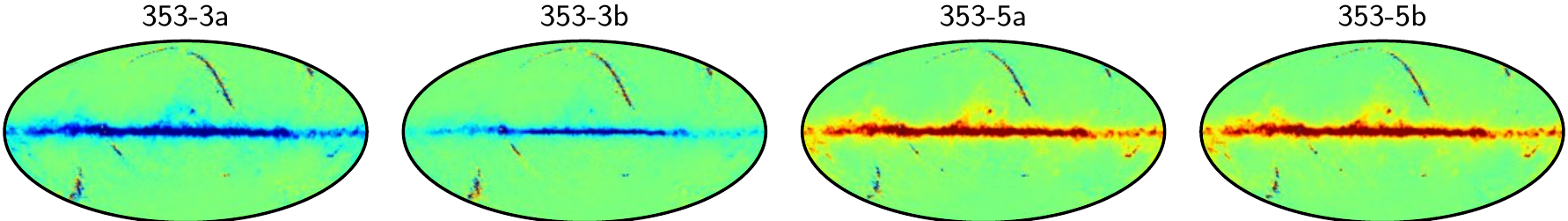}}
  \centerline{\includegraphics[width=1.0\textwidth,trim=0  0 0 10,clip=True]{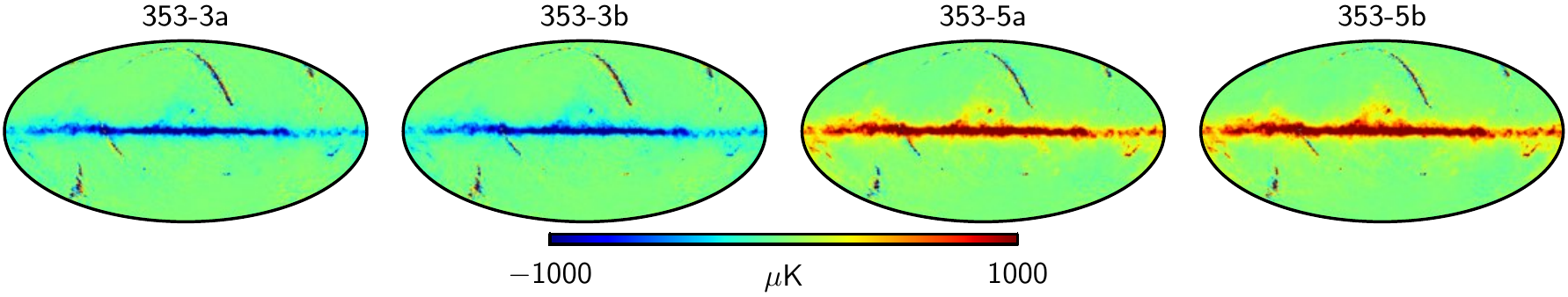}}
  \caption{\label{fig:iqus_dx11}
    Pairs of measured bandpass mismatch HFI maps, comparing FFP8 simulations (upper) and 2015 \Planck\ (lower) data. Our method of estimating the total bandpass mismatch between each detector and a frequency average is detailed in Appendix~\ref{appendix:iqus}.
  }
\end{figure*}

Both the LFI and HFI detector bandpasses are based on ground measurements (see \citealt{Zonca:2010fx} and \citealt{planck2013-p03d}, respectively), although flight data processing for LFI now uses in-flight top-hat approximations rather than the ground measurements that were found to contain systematic errors. Differences in the bandpasses of detectors nominally at the same frequency (the so-called bandpass mismatch) generate spurious signals in the maps, since each detector is seeing a slightly different sky while the mapmaking algorithms assume that the signal in a pixel is the same for all detectors. To quantify the effect of these residuals, in FFP8 we generate detector timelines from foreground maps in two ways, one that incorporates the individual detector bandpasses, the other using an average bandpass for all the detectors at a given frequency.

This effect of the bandpass mismatch can be roughly measured from either flight or simulated data using so-called spurious component mapmaking, which provides noisy all-sky estimates of the observed sky differences (the spurious maps), excluding polarization, between individual detectors and the frequency average. We compare the amount of simulated bandpass mismatch to flight data in Figs.~\ref{fig:iqus_ffp8} (LFI) and \ref{fig:iqus_dx11} (HFI). The spurious component approach is detailed in Appendix~\ref{appendix:iqus}. Mismatch between FFP8 and flight data is driven by inaccurate bandpass description (LFI) and incomplete line emission simulation (HFI). The noisy pixels that align with the \Planck\ scanning rings in the HFI maps are regions where the spurious map solution is degenerate with polarization due to insufficient observation orientations.

\subsection{The foreground sky}
\label{subsec:foregrounds}

The quality of the reconstructed CMB signal depends on the accuracy of the separation of the astrophysical foreground components from the \Planck\ data, which requires building simulations  as close to the complexity of the real sky as possible over a large frequency range. To this end, a complete model of multi-component sky emission, called the \Planck{} sky model (PSM), has been developed. FFP8 simulations make use of version 1.9.0 of the PSM, a major update of the published version 1.7.8 \citep{2013A&A...553A..96D}. The foreground components generated by the PSM are diffuse Galactic emission from thermal dust, spinning dust, relativistic electrons (synchrotron emission), free-free radiation, and CO (the $J$=1$\rightarrow$0, $J$=2$\rightarrow$1, and $J$=3$\rightarrow$2 lines at $115.27$, $230.54$, and $345.80$\,GHz, respectively), plus the cosmic infrared background (CIB), emission from radio sources, and the thermal and kinetic Sunyaev-Zeldovich (SZ) effects.

\subsubsection{Thermal dust}
\label{subsubsec:dust}

The thermal dust emission is modelled using single-frequency template maps of the intensity and polarization, together with a pixel-dependent emission law.

For FFP8 the thermal dust emission templates are derived from the \Planck{} 353\,GHz observations. This update of the original PSM dust model, described in \cite{2013A&A...553A..96D} and based on \cite{1999ApJ...524..867F}, is necessary to provide a better match to the emission observed by \Planck{}. While one option would be simply to use the dust opacity map obtained in \cite{planck2013-p06b}, this map still suffers from significant contamination by CIB anisotropies and infrared point sources. Using it as a $353$\,GHz dust template in simulations would result in an excess of small scale power (from CIB and infrared sources) scaling exactly as thermal dust across frequencies. The resulting component represents correctly neither dust alone (because of an excess of small scale power) nor the sum of dust and infrared sources (because the frequency scaling of the CIB and infrared sources is wrong). For simulation purposes, the main objective is not to have an exact map of the dust, but instead a map that has the right statistical properties. Hence we produce a template dust map at 353\,GHz by removing that fraction of the small-scale power that is due to CIB emission, infra-red sources, CMB, and noise.

The emission of strong point sources is first subtracted from all \Planck{} 2015 maps and from the IRAS 100\micron\ map following \cite{2005ApJS..157..302M}. We mask point sources detected in \Planck{} and IRAS maps with an algorithm designed specifically to avoid false detections in regions of dust emission where localized compact regions of diffuse dust emission can easily be mistaken for point sources. This algorithm detects sources against a background level computed directly on the maps in the immediate vicinity of each pixel.  Specifically, for each pixel $p$, we compute the standard deviation $\sigma(p)$ of the observed map in a neighbourhood $N(p)$ consisting of a small ring centred on the source, between the two radii $r_1(p)$ and $r_2(p)$. We consider as part of point sources those objects for which the flux is above a threshold of 3 to 5 times $\sigma(p)$. The parameters of the filter for each frequency map -- the full-width half-maximum (FWHM) radii defining the disc, and the source detection threshold -- are optimized by trial and error. We then generate a mask from the union of small discs of radius 1~FWHM centred on all pixels above the detection threshold at any frequency. The gaps created by masking the point sources are filled by extrapolation of the neighbouring regions using a minimum-curvature spline gap filler. This procedure avoids most false detections of point sources in filaments of dust emission, unless there is a  compact source significantly brighter than the local emission of the filament. The drawback of this approach is that extended compact sources such as nearby galaxies are not masked.

The dust intensity template is then corrected for CIB, CMB, and noise fluctuations using the multi-dimensional Generalized Needlet Internal Linear Combination (GNILC) of \cite{2011MNRAS.418..467R}. Starting from the 10~de-sourced maps (nine \Planck{} maps and the IRAS 100\micron\ map), we model the total emission at a given frequency $\nu$ as:
\begin{equation}
y_\nu = s_\nu + x_\nu
\end{equation}
where $y_\nu$ is the observed map, $s_\nu$ the map of foreground emission (mostly Galactic emission), and $x_\nu$ the sum of CMB, CIB, and noise.
The total $10 \times 10$ covariance matrix $R_y$ of the maps, in any localized region in pixel and/or harmonic space, is the sum of the foreground term $R_s$ and a contamination term $R_x$ due to CMB, CIB and noise,
\begin{equation}
R_y = R_s + R_x.
\end{equation}
We perform a needlet decomposition of each map as in \cite{2011MNRAS.418..467R}, except that the spectral windows correspond to Gaussian bandpass filters in harmonic space. We use the best fit CMB $C_{\ell}$ from \cite{planck2013-p08}, the CIB power spectrum at each frequency and cross-correlation coefficients from \cite{planck2013-pip56}, and the noise level from \Planck{} half-ring difference maps to obtain a model $\widehat R_x$ of the covariance $R_x$ of CMB, CIB, and noise in a set of needlet domains localized on the sky. We then follow the procedure described in \cite{2011MNRAS.418..467R} to reconstruct maps of $s_\nu$ at up to 5\arcm\ resolution. For each needlet domain, the covariance of the needlet power is whitened locally as
\begin{equation}
\widetilde R_y = \widehat R_x^{-1/2} \widehat R_y \widehat R_x^{-1/2}.
\end{equation}
The matrix $\widetilde R_y$ is diagonalized and the Akaike information criterion (AIC; \citealt{1974ITAC...19..716A}) is used to filter out the CMB, CIB, and noise contamination $x_\nu$ in dimensions of the observation space where the total covariance is compatible with $s_\nu = 0$. The final map of foreground emission $s_\nu$ at each needlet scale is reconstructed by coadding the needlets for which there is significant additional emission from Galactic foregrounds according to the AIC criterion. However, the original source-subtracted 353\,GHz \Planck{} intensity map is used instead of the processed map in the Galactic plane, the LMC, the SMC, regions of very strong localized emission, and the extended galaxies M31, M33, M81, M82, M87, M101, NGC\,55, NGC\,300, and NGC\,2403. The reason for this is that the multi-dimensional GNILC process is not perfectly local in pixel space, so that bright local structures such as the Galactic plane or nearby galaxies are analysed together with regions of significantly lower emission. This results in excessive filtering of some small-scale structure in the bright objects themselves, and insufficient filtering in the pixels located in their immediate vicinity. While the latter is not much of a problem (giving a non-crucial loss of efficiency of the multi-dimensional GNILC filtering), the former is not acceptable since it impacts the brightest objects in the sky. As CIB, CMB, and noise contributions are completely subdominant in such regions of very bright emission, it is better to use the original map than the processed one for the generation of the dust template in these regions.

\begin{figure}
  \begin{center}
\includegraphics[width=0.33\textwidth]{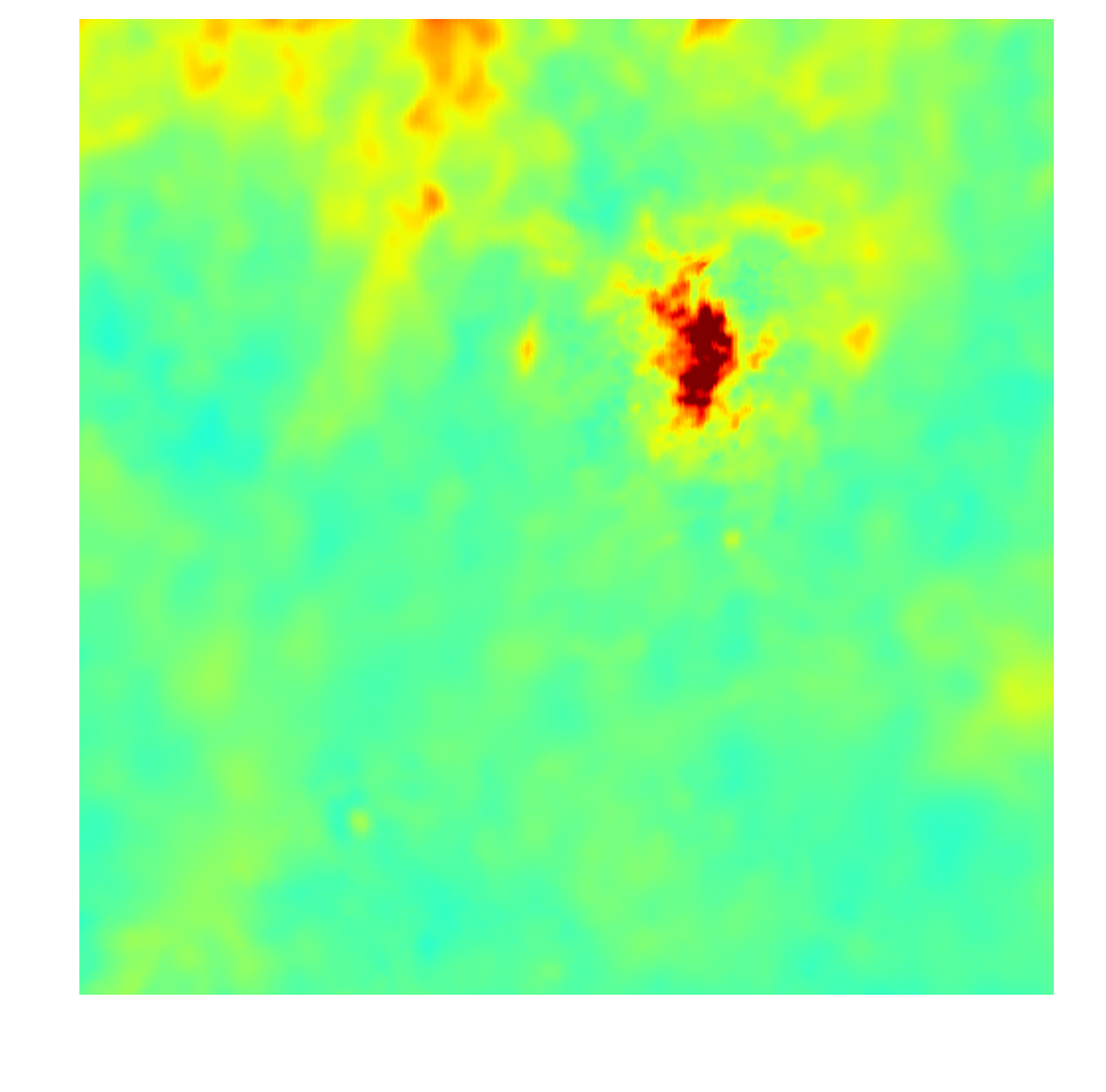} \\
\vspace*{0.05in}
\includegraphics[width=0.33\textwidth]{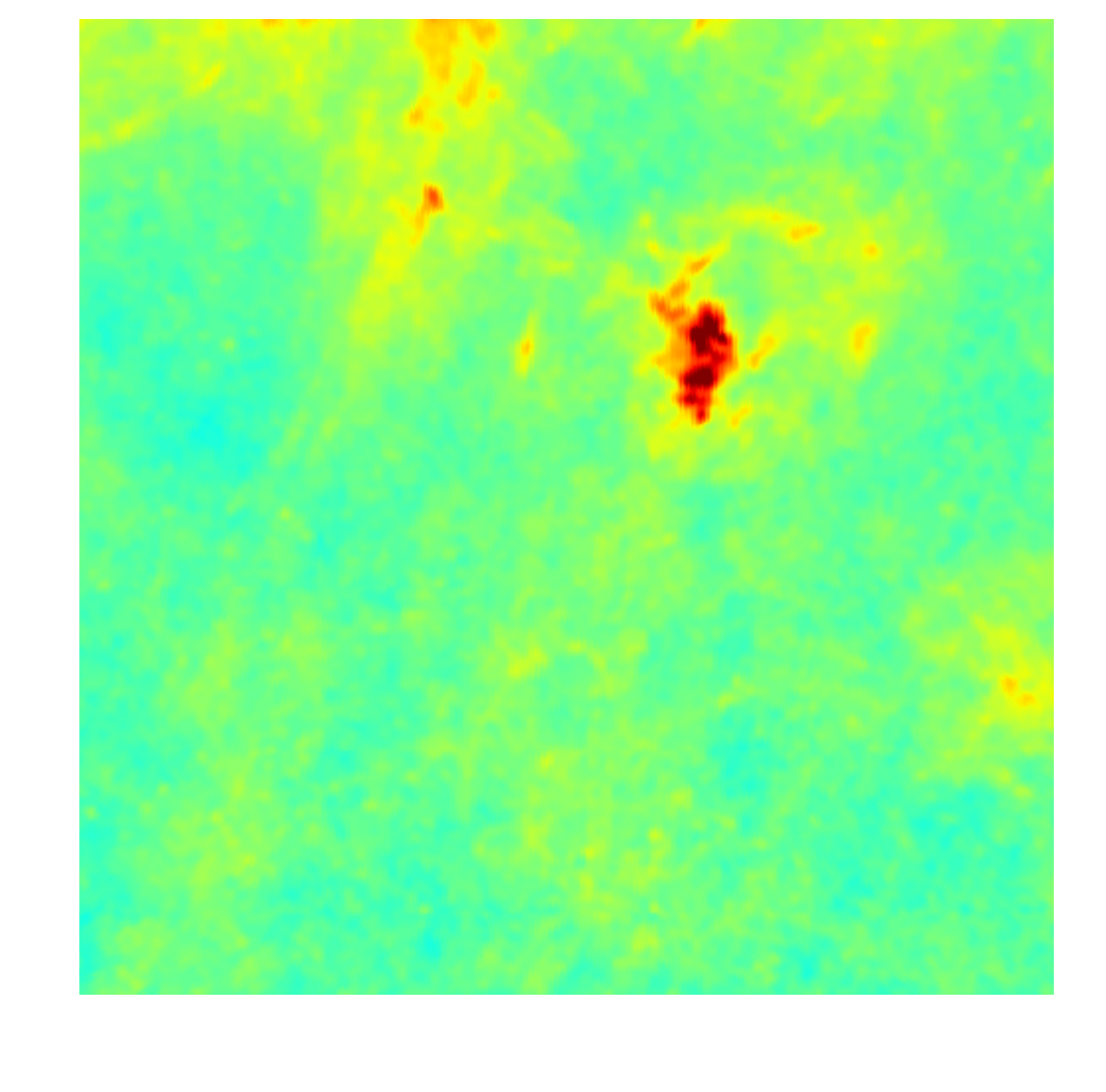} \\
\includegraphics[width=0.38\textwidth]{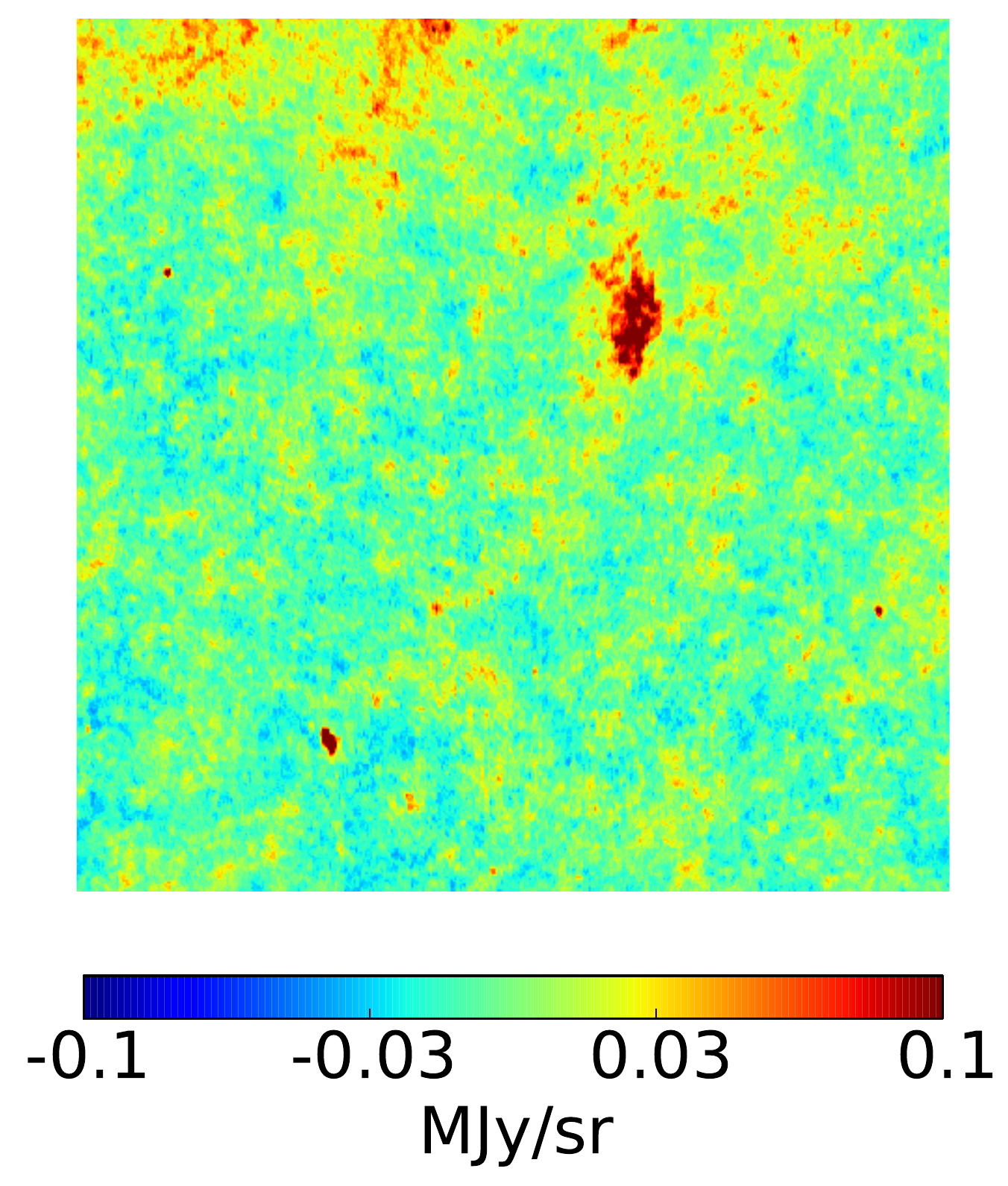}
\end{center}
\caption{$12\pdeg5 \times 12\pdeg5$ gnomonic projection of dust intensity templates centred at high latitude $(l, b) = (90^\circ,-80^\circ)$ at a map resolution of 5\arcm. \emph{Top}: new PSM dust from multi-dimensional GNILC \citep{2011MNRAS.418..467R}. \emph{Middle}: original PSM dust \citep{2013A&A...553A..96D} based on the model of \cite{1999ApJ...524..867F}. \emph{Bottom}: \Planck{} 2013 dust model \citep{planck2013-p06b}. Comparing the top and bottom panels, the improvement in reducing the contamination of the dust by small-scale CIB anisotropies at high Galactic latitude is clearly visible.}
  \label{Fig:dust-intensity}
\end{figure}

Next, a mask is applied to remove from the 353\,GHz dust map residuals of the positive SZ effect towards bright galaxy clusters. The mask consists of discs centred on CLASS1 (i.e., the most reliable) SZ candidates in the PSZ1 catalogue \citep{planck2013-p05a}, with radius $5 \times \theta_{500}$.\footnote{$\theta_{500}$ is the observed radius $R_{500}/D_{\rm A}(z)$, where $D_{\rm A}(z)$ is the angular diameter distance of the cluster.} Gaps generated in the map by this masking are filled in again by a minimum-curvature spline ``inpainting''. Finally, a colour-correction factor is applied to obtain a map of brightness emission at 353\,GHz rather than the map integrated in the \Planck\ 353\,GHz channel. The 353\,GHz dust map obtained in this way, which is subsequently used as a dust template, is illustrated in Fig.~\ref{Fig:dust-intensity}, where it is also compared with the model of \mbox{\cite{1999ApJ...524..867F}} and with the map of optical depth published in \mbox{\cite{planck2013-p06b}}.

The pixel-dependent emission law is derived from \citet{planck2013-p06b}, fitting \Planck{} 353, 545, and 857\,GHz  and IRAS 100\micron\ data with a modified blackbody. The emission of the strongest point sources is subtracted from all maps before performing the fits, with gaps filled by the same method used for the dust intensity template.

For each pixel $p$, the spectral energy distribution is approximated with a modified blackbody, i.e.,
\begin{equation}
\label{Eq:greybody}
I_\nu(p) = I^F_{\rm 353\,GHz}(p) \left( \frac{\nu}{\rm 353\,GHz} \right)^{\beta_{{\rm FIR}}(p)} \left( \frac{B_\nu(T_{\rm dust}(p))}{B_{\rm 353\, GHz}(T_{\rm dust}(p))} \right),
\end{equation}
where the maps of the dust temperature $T_{\rm dust}$ and spectral index $\beta_{{\rm FIR}} $ are those presented in \citet{planck2013-p06b}.  The sky map
$I^F_{\rm 353\,GHz}$ is produced by applying the multi-dimensional GNILC of \cite{2011MNRAS.418..467R}.

Maps of the $Q/I$ and $U/I $ ratios between Stokes parameters were computed from the best versions of the \Planck\ maps available at the time of generation of the simulations in June 2014, smoothed to $30\arcmin$ resolution to reduce noise. At $353$\,GHz, the observed maps are used for polarization, while the intensity map is obtained from multi-dimensional GNILC processing.  All maps are corrected for the dust bandpass mismatch leakage.  Structure on small scales is added in order to have no break in the power spectra of dust polarization. These maps determine the polarization orientation and fraction. For each frequency, the Stokes $Q$ and $U$ maps of the model are then obtained by multiplying the $Q/I$ and $U/I$ maps by the total intensity map at full resolution.  The polarization orientation does not depend on frequency, but the difference between $\beta_{{\rm mm}}$ and $\beta_{{\rm FIR}}$ introduces a  spectral dependence of  the polarization fraction, which varies over the sky.

To extrapolate the model to frequencies below 353\,GHz we follow \citet{planck2013-XVII}, using a spectral index map $\beta_{{\rm mm}}$  slightly smaller than $\beta_{{\rm FIR}} $ for the total dust intensity, but not for the polarized emission. The difference between $\beta_{{\rm mm}}$ and $\beta_{{\rm FIR}} $ is a quadratic function of $T_{\rm dust}$,  which has been fitted comparing the \Planck{} maps at 143 and $353\,$GHz, after subtraction of the CMB anisotropies. It introduces a small spectral difference between dust intensity and polarization that varies over the sky, similar to that reported in \citet{planck2014-XXII}.

\subsubsection{Spinning dust}
\label{subsubsec:spin-dust}

Several observations indicate an excess of microwave emission in the $10--100$ GHz range \citep{1996ApJ...464L...5K, 1997ApJ...486L..23L, 1999ApJ...527L...9D, 2005ApJ...624L..89W}. This emission has been proposed to be the dipole radiation from rapidly spinning nano-scale dust grains \citep{1998ApJ...508..157D}. Detailed analyses of the data from various experiments show detections of a component that is consistent with the spinning dust model \citep{2006MNRAS.370.1125D, 2008A&A...490.1093M, 2011ApJ...734....4K, planck2011-7.2,planck2013-XV, planck2014-a12}.

\begin{figure}[!ht]
  \begin{center}
    \includegraphics[width=88mm]{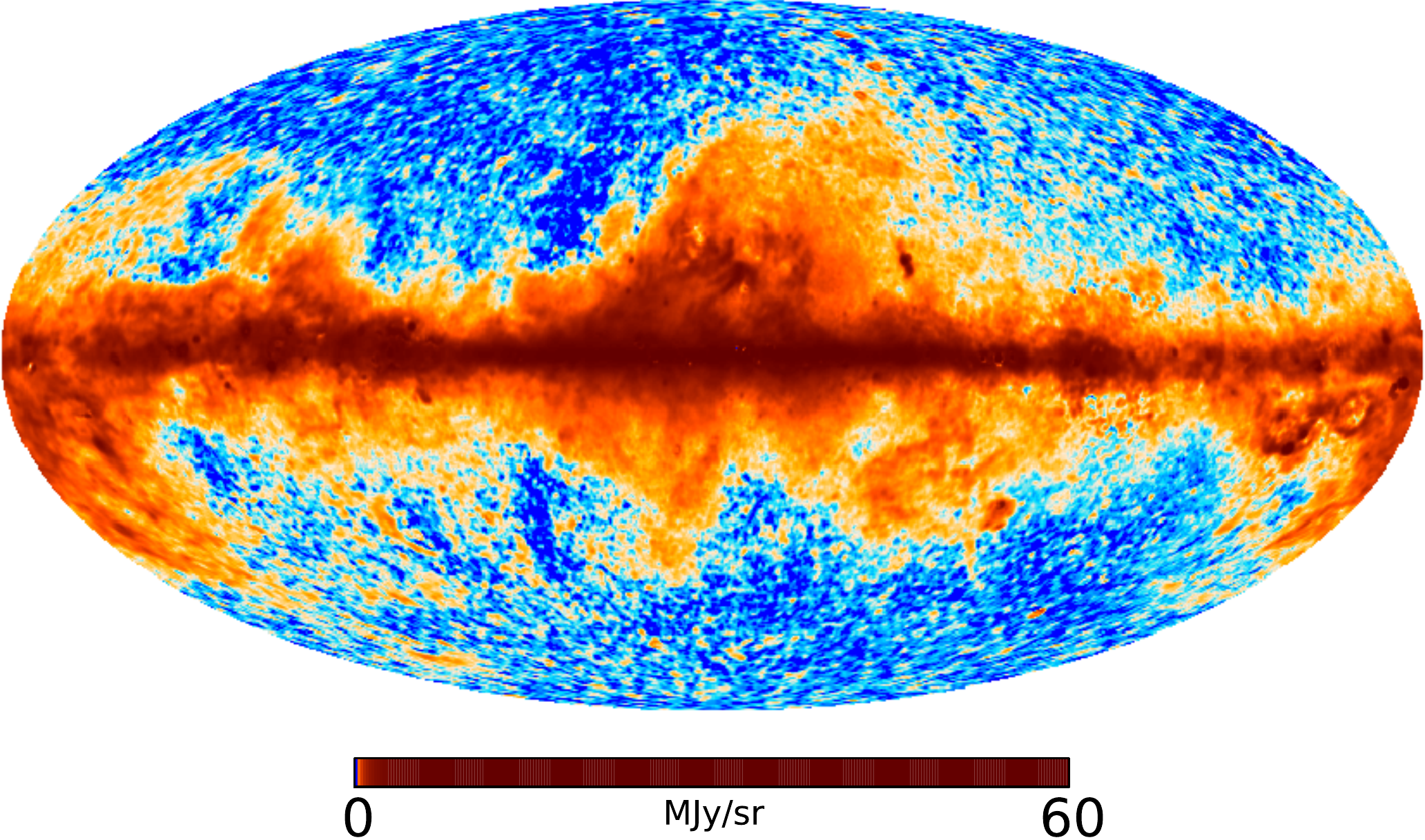}
    \caption{\label{fig:spin_temp} Template of spinning dust intensity.}
  \end{center}
\end{figure}

The spinning dust map used for FFP8 simulations is a simple realization of the spinning dust model, post-processed to remove negative values occurring in a few pixels because of the generation of small-scale fluctuations on top of the spinning dust template extracted from WMAP data by \cite{2007A&A...469..595M}. The model uses a single template map, illustrated in Fig. \ref{fig:spin_temp}, which is scaled by a single emission law parameterized following the model of \cite{1998ApJ...508..157D}.

\subsubsection{Synchrotron}
\label{subsubsec:synchrotron}

At low frequencies (below about $80$ GHz) the dominant contaminant of the polarized CMB signal is the Galactic synchrotron emission that arises from the acceleration of relativistic cosmic rays in the Galactic magnetic field. FFP8 synchrotron emission is modelled on the basis of the template emission map observed at 408 MHz by \cite{1982A&AS...47....1H}. This template synchrotron map is extrapolated in frequency using a spectral index map corresponding to a simple power law.

\subsubsection{Free-free}
\label{subsubsec:free-free}

Electron-ion interactions in the ionized phase of the ISM produce emission that is in general fainter than both the synchrotron and the thermal dust emission outside of the active star-forming regions in the Galactic plane. The free-free model uses a single template, which is scaled in frequency by a specific emission law. The free-free spectral index is a slowly varying function of frequency and depends only slightly on the local value of the electron temperature $T_{e}$ \citep{1992ApJ...396L...7B, 2003ApJS..148...97B, 2003MNRAS.341..369D}. In FFP8, the free-free spectral dependence is modelled by assuming a constant electron temperature $T_{e} = 7000$ K, as described in \cite{2003MNRAS.341..369D}.

\subsubsection{Cosmic infrared background}
\label{subsubsec:cib}

The CIB model relies on the distribution of individual galaxies in template maps based on the distribution of dark matter at a range of relevant redshifts. We assume the CIB galaxies can be grouped into three different populations (proto-spheroid, spiral, starburst) as described in \cite{2013ApJ...768...21C}. Within each population, galaxies have the
same SED, while the flux density is randomly distributed according to redshift-dependent number counts obtained from JCMT/SCUBA-2 observations \citep{2013ApJ...762...81C} and the \Planck\ ERCSC \citep{2013MNRAS.429.1309N}, as well as observations cited in \cite{2013ApJ...768...21C}, most notably {\it Herschel}/Spire (Hermes; \citealt{2013A&A...557A..66B}) and AzTEC/ASTE (Combined ACES; \citealt{2012MNRAS.423..575S}).

\begin{figure}[!ht]
\begin{center}
    \includegraphics[width=88mm]{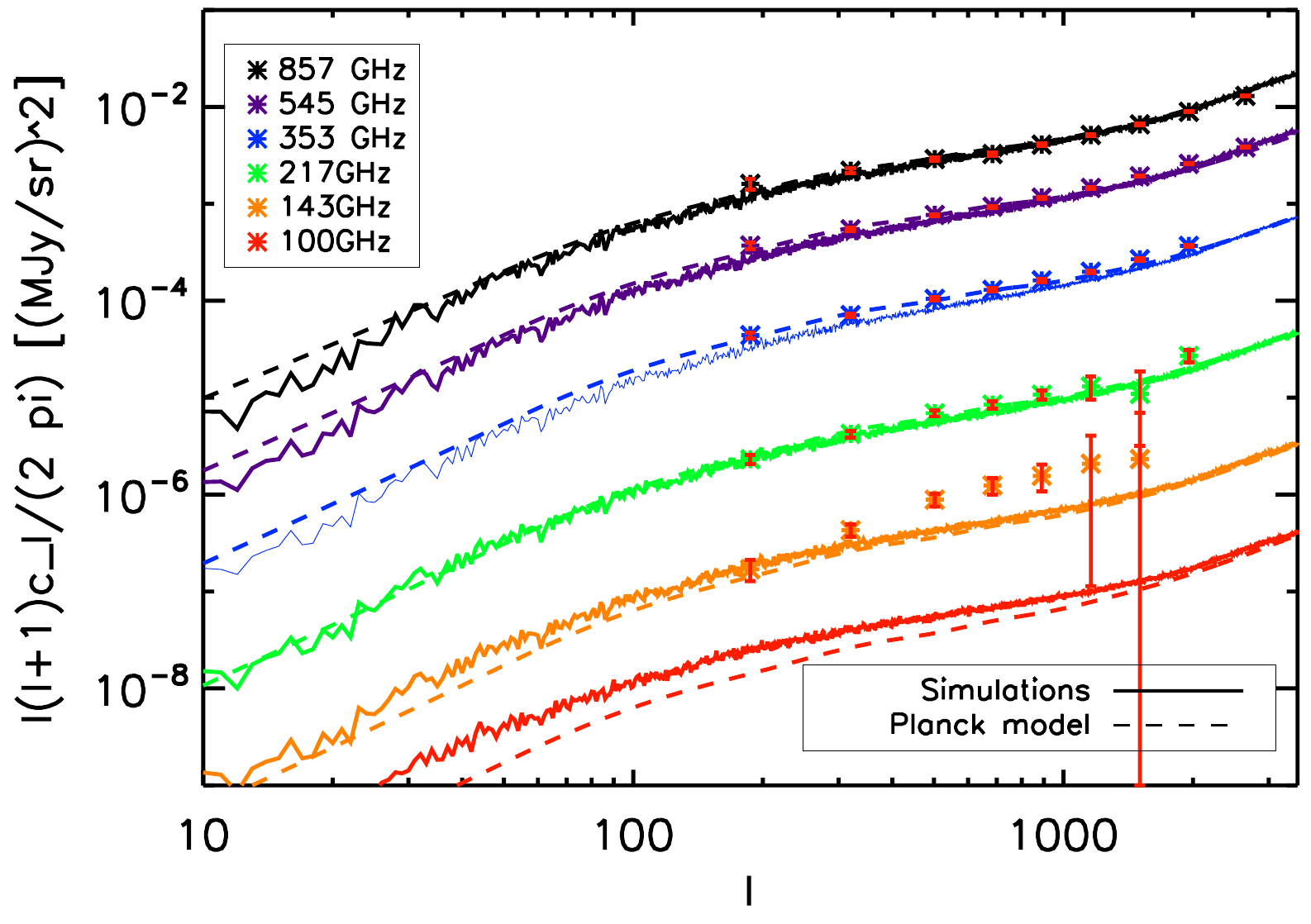}
    \caption{ \label{fig:cib_spec} Simulated and observed cosmic infrared background power spectra. Solid lines are from the PSM simulations; dashed lines are the \Planck\ ``extended halo model'' which includes the correlated and Poisson contributions to the CIB power spectrum from dusty galaxies. Data points are \Planck\ observations, corrected for the contribution of radio sources, all from \cite{planck2013-pip56}.}
\end{center}
\end{figure}

We use the {\tt Class} \citep{2011JCAP...07..034B} software to generate dark matter maps at $17$ different redshifts between $1$ and $5.5$. Since the galaxy distribution does not exactly follow the dark matter distribution, we modify the $a_{lm}^{\rm DM}$ coefficients of dark matter anisotropies given by  {\tt Class} at redshift $z_i$ such that
\begin{equation}
a_{\ell m, j} (z_i)=a_{\ell m}^{\rm DM}(z_i) \times b_j(z_i) \times \left( 1+ \left( \frac{\ell}{\beta} \right) ^\alpha \right),
\end{equation}
where $b_j(z)$ is the population-dependent component of the bias of the population $j$, and is identical for spirals and starbursts.
$\left( 1+ \left( \frac{\ell}{\beta} \right) ^\alpha \right)$ is the scale-dependent part of the bias. Values of $\alpha=1.44$ and $\beta=2\,654$ were obtained by fitting \Planck\ spectra.
Template maps generated from the $a_{\ell m, j} (z_i)$ coefficients are then exponentiated to avoid negative pixels.

Galaxies are randomly distributed with a probability of presence proportional to the pixel values of the template maps.  One map is generated for each population, at each redshift, and associated with a redshifted SED depending on the population.  The emission of these maps (initially at a reference frequency) can be extrapolated to any frequency using the associated redshifted SED. By summing the emission of all maps, we can generate CIB maps at any frequency in the range of validity of our model.

A comparison of CIB power spectra of the simulated maps and the maps obtained by \Planck\ observations at HFI frequencies is shown in Fig.~\ref{fig:cib_spec}; agreement is excellent at all but the lowest two HFI frequencies where the CIB is subdominant. See \cite{planck2013-pip56} for further discussion.

\subsubsection{Radio sources}
\label{subsubsec:pt-source}

Radio source modelling in the PSM was described in detail in \cite{2002ASPC..257...81D}. Here, we describe only the main changes since that pre-launch version.

For ``strong'' radio sources ($S_{30}>0.5$\,Jy), we use radio sources compiled by \cite{2008MNRAS.384..711G} at 0.84, 1.4, or 4.85\,GHz. For sources observed at two of these frequencies, we extrapolate or interpolate to the third frequency assuming the spectral index estimated from two observed. For sources observed at only one frequency, we use differential source counts from \cite{2010A&ARv..18....1D} to obtain the ratio of steep- to flat-spectrum sources in each interval of flux density considered. From this ratio, we assign spectral indices (randomly) to each source within each flux density interval. Fiducial Gaussian spectral index distributions as a function of spectral class are obtained from the literature \citep{2011MNRAS.415.1597M,2008MNRAS.385.1656S}. These are then adjusted slightly until there is reasonable agreement between the PSM differential counts and the model counts predicted by \cite{2010A&ARv..18....1D}. The final means and standard deviations of these spectral index distributions are $-0.7\pm0.2$ for steep-spectrum sources, $-0.3\pm0.2$ for flat-spectrum radio quasars (FSRQ), and $-0.1\pm0.2$ for BL Lacertae objects (BL~Lacs).

For ``faint'' radio sources ($S_{30}\leq 0.5$\,Jy), the pre-launch PSM showed a deficit of sources resulting from inhomogeneities in surveys at different depths. We address this issue by constructing a simulated catalogue of sources from the model differential counts of \cite{2010A&ARv..18....1D} at 1.4\,GHz. We replace the simulated sources by the observed ones, wherever possible. If, however, in any particular pixel, we have a shortfall of observed sources, we make up the deficit with the simulated sources. Every source in this new catalogue is given a model-derived spectral class. We thus assign a spectral index to each source based on the spectral class, and model the spectrum of each source using four power laws covering the ranges $\nu<5$\,GHz, $5<\nu<20$\,GHz, $20<\nu<100$\,GHz, and $\nu>100$\,GHz.  We also assume some steepening of the spectral index with frequency, with fiducial values of the steepening obtained from the literature \citep{2011MNRAS.415.1597M,2008MNRAS.385.1656S}. The steepening of the spectral index assumed for 5--20\,GHz is $-0.3$ for steep-spectrum sources and $-0.3$ for flat-spectrum sources (FRSQs and BL~Lacs).  For 20--100\,GHz, we assume no further steepening for steep-spectrum sources, and steepening of $-0.3$ for FRSQs and BL~Lacs.

\begin{figure}[!ht]
  \begin{center}
    \includegraphics[width=88mm]{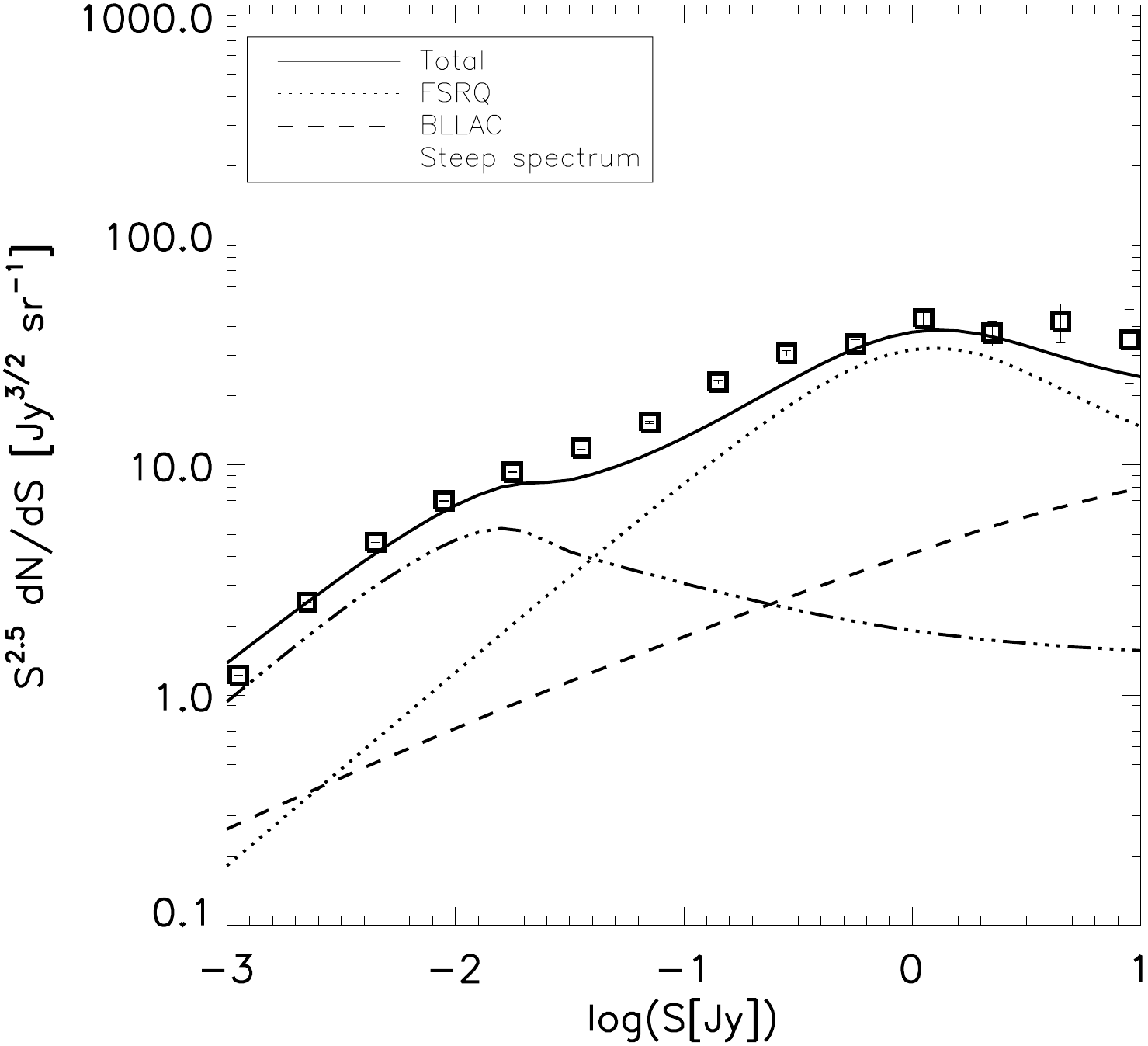}
    \includegraphics[width=88mm]{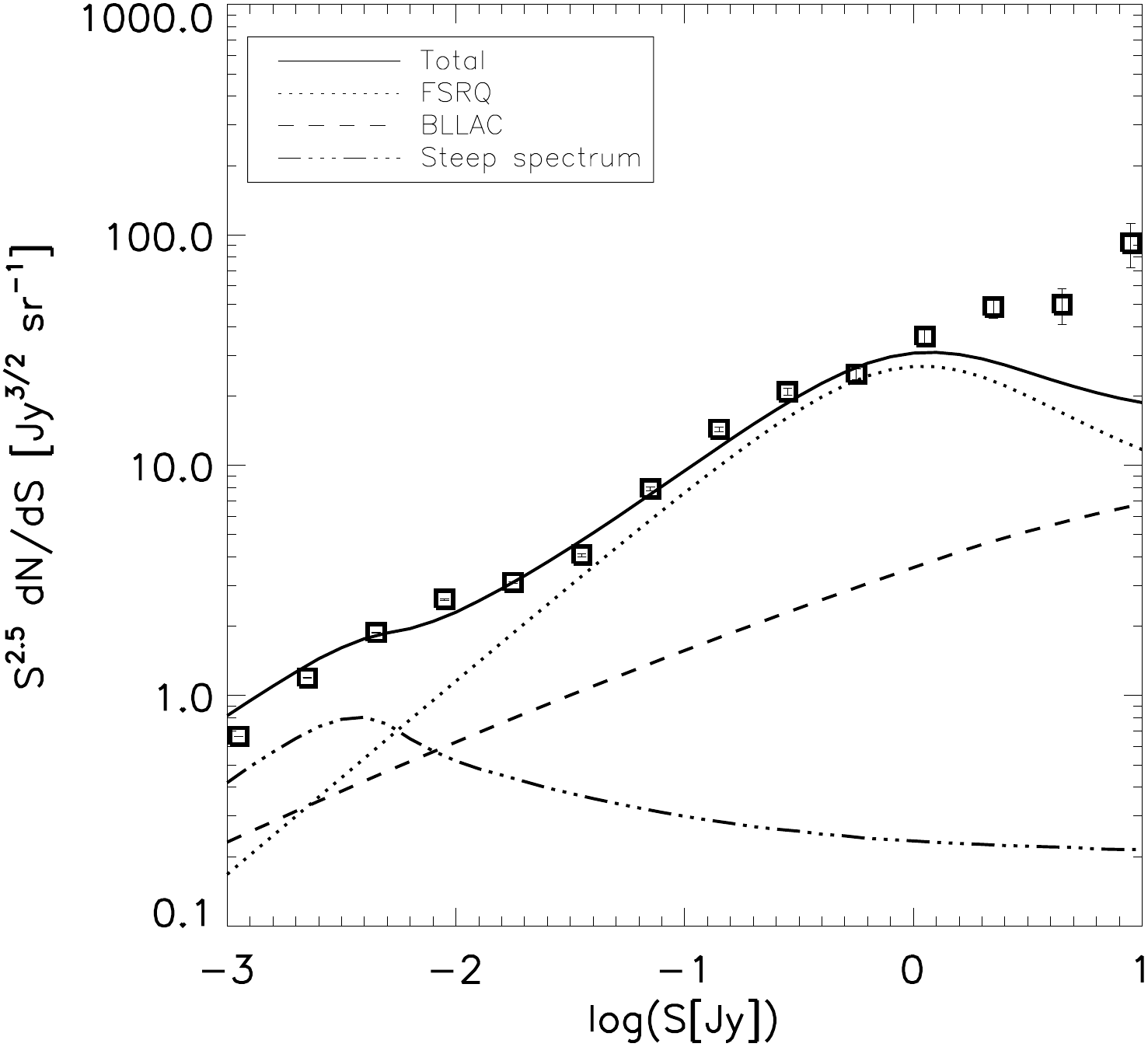}
    \caption{\label{fig:counts.ps} Counts of the faint and the strong radio sources at $30$\,GHz (top) and $100$\,GHz (bottom). The differential counts of the faint and the strong radio sources between $0.005$\,Jy and $1$\,Jy are shown by vertical bars.}
  \end{center}
\end{figure}

We combine the faint and strong radio source catalogues we constructed and compute the differential source counts on these sources between 0.005\,Jy and 1\,Jy (see Fig.~\ref{fig:counts.ps}). These show good agreement with the model differential counts predicted by \cite{2010A&ARv..18....1D}, particularly at the faintest end. Finally we also model the polarization of these radio sources using the measured polarization fractions from \cite{2004A&A...415..549R}, comprising 71 steep- and 126 flat-spectrum sources; for each simulated source we draw a polarization fraction at random from the list of real sources of the same spectral type.

\subsubsection{SZ clusters}
\label{subsubsec:sz-cluster}

SZ clusters are simulated following the model of Delabrouille, Melin, and Bartlett (DMB) as implemented in the PSM (\citealt{2002ASPC..257...81D}). A catalogue of halos is drawn from a Poisson distribution of the mass function from~\cite{2008ApJ...688..709T} with a limiting mass of $M_{500,\rm true}>2 \times 10^{13} M_\odot$.\footnote{$M_{500}$ is the total cluster mass within a sphere of radius $R_{500}$, defined as the radius within which the mean mass over-density of the cluster is 500 times the cosmic critical density at the redshift of the cluster.  $M_{500} = (4\pi/3) R_{500} 3 [500\rho_{\rm c}(z)]$, with $\rho_{\rm c}(z) = 3H^2(z)/(8\pi G)$, where $H(z)$ is the Hubble parameter with present-day value $H_0$.} We use the pressure profile from~\cite{2010A&A...517A..92A} to model the thermal SZ emission of each halo given its redshift and mass, $M_{500, \rm x}$. In order to match the observed counts in \Planck, we set $M_{500, \rm x}/M_{500,\rm true}=0.65$.  Fixing this ratio to unity would lead to twice as many clusters simulated as appear in the actual data, as shown in~\cite{planck2013-p15}. We determine the cluster temperature from the $M_{500, \rm x}$--$T$ relation of~\cite{2005A&A...441..893A}, and assume that the profiles are isothermal. These steps allow us to compute the first-order thermal relativistic correction~\citep{1998ApJ...502....7I} and the kinetic SZ effect for each cluster, both of which are included in the simulation. Finally, we inject MCXC~\citep{2011A&A...534A.109P} and MaxBCG~\citep{2007ApJ...660..239K} clusters following the same model, and remove from the simulation corresponding clusters in each redshift and mass range. Hence the SZ simulation features the majority of known X-ray and optical clusters, and is fully consistent with X-ray scaling laws and observed \Planck\ SZ counts.

\subsubsection{Total foreground sky}
\label{subsubsec:total-fg}

The sky model is simulated at a resolution common to all components by smoothing the maps with an ideal Gaussian beam of FWHM of 4\arcm.  The \healpix\footnote{\url{http://healpix.sourceforge.net}} pixelization in Galactic coordinates is used for all components, with $\Nside = 2048$ and $\ell_{max} = 6000$.  Sky emission maps are generated by numerically band-integrating the sky model maps (emission law of each component, in each pixel) over the frequency bands both of each detector in the focal plane and -- using an average over the detectors at a given frequency -- of each channel. The band-integrated maps are essentially observations of the model sky simulated by an ideal noiseless instrument with ideal Gaussian beams of FWHM equal to the resolution of the model sky.

\subsection{The CMB sky}
\label{subsec:cmb}

\begin{table*}[tb]
  \begingroup
  \newdimen\tblskip \tblskip=5pt
  \caption{Cosmological parameter values from the PR1-2013 and PR2-2015 data releases, used in FFP8 and FFP8.1, respectively}
  \label{table:params}
  \nointerlineskip
  \vskip -3mm
  \footnotesize
  \setbox\tablebox=\vbox{
    \newdimen\digitwidth
    \setbox0=\hbox{\rm 0}
    \digitwidth=\wd0
    \catcode`*=\active
    \def*{\kern\digitwidth}
    \newdimen\signwidth
    \setbox0=\hbox{-}
    \signwidth=\wd0
    \catcode`!=\active
    \def!{\kern\signwidth}
\halign{\hbox to 6.5cm{#\leaderfil}\tabskip 4em&
      \hfil#\hfil \tabskip 4em&
      \hfil#\hfil \tabskip 4em&
      \hfil#\hfil \tabskip 0pt\cr
      \noalign{\doubleline}
      \omit\hfil Parameter \hfil&\omit\hfil Symbol \hfil&\omit\hfil PR1-2013 (FFP8) \hfil&\omit\hfil PR2-2015 (FFP8.1) \hfil\cr
      \noalign{\vskip 3pt\hrule\vskip 4pt}
      Baryon density&$\omega_{\rm b} = \Omega_{\rm b}h^2$&$0.0222$&$0.0223$\cr
      Cold dark matter density&$\omega_{\rm c} = \Omega_{\rm c}h^2$&$0.1203$&$0.1184$\cr
      Neutrino energy density&$\omega_{\nu} = \Omega_{\nu}h^2$&$0.00064$&$0.00065$\cr
      Dark energy density&$\Omega_{\Lambda}$&$0.6823$&$0.6931$\cr
      Hubble parameter, $H_0 = 100 h \,\mathrm{km}\,\mathrm{s}^{-1}\,\mathrm{Mpc}^{-1}$&$h$&$0.6712$&$0.6787$\cr
      \omit Primordial curvature perturbation spectrum\hfil\cr
      {\hskip 10pt}amplitude (at $k=0.05\,\mathrm{Mpc}^{-1}$)&$A_{\rm s}$&$2.09\times 10^{-9}$&$2.14\times 10^{-9}$\cr
      {\hskip 10pt}spectral index&$n_{\rm s}$&$0.96$&$0.97$\cr
      Thomson optical depth through reionization&$\tau$&$0.065$&$0.067$\cr
      \noalign{\vskip 3pt\hrule\vskip 5pt}
    }
  }
  \endPlancktablewide                 
  \endgroup
\end{table*}

The CMB sky is simulated in three distinct components, namely lensed scalar, tensor, and non-Gaussian complement.  The total CMB sky is then the weighted sum with weights 1, $\sqrt{r}$, and $f_\mathrm{NL}$, respectively. For FFP8, all CMB sky components are produced as spherical harmonic representations of the $I$, $Q$, and $U$ skies.

The FFP8 CMB sky is derived from our best estimate of the cosmological parameters available at the time of its generation, namely those from the first \Planck\ data release \citep[PR1-2013;][]{planck2013-p01}, augmented with a judicious choice of reionization parameter $\tau$, as listed in Table~\ref{table:params}.

\subsubsection{The scalar CMB sky}

The scalar component of the CMB sky is generated including lensing, Rayleigh scattering, and Doppler boosting effects. Using the \camb\ code \citep{Lewis:1999bs}, we first calculate fiducial unlensed CMB power spectra $C_{\elt}^{TT}$, $C_{\elt}^{EE}$, $C_{\elt}^{TE}$, the lensing potential power spectrum $C_{\elt}^{\phi\phi}$, and the cross-correlations $C_{\elt}^{T\phi}$ and $C_{\elt}^{E\phi}$. We then generate Gaussian $T$, $E$, and $\phi$ multipoles with the appropriate covariances and cross-correlations using a Cholesky decomposition and three streams of random Gaussian phases. These fields are simulated up to $\elt_{\rm max}=5120$. In addition, we add a dipole component to $\phi$ to account for the Doppler aberration due to our motion with respect to the CMB \citep{Challinor:2002zh}:
\be
\phi_{\beta}(\vec{\hatn}) = \vec{\beta} \cdot \vec{\hatn},
\ee
where $\vec{\beta}$ is in the direction of Galactic coordinates $(l,b)=(264\pdeg4,48\pdeg4)$ with amplitude $|\vec{\beta}| = 0.00123$, corresponding to a velocity of $369\kms$ \citep{Hinshaw:2008kr}.

We we compute the effect of gravitational lensing on the temperature and polarization fields, using an algorithm similar to {\tt LensPix} \citep{Lewis:2005tp}. We use a fast spherical harmonic transform to compute the temperature, polarization, and deflection fields:
\begin{eqnarray}
[d_{\theta} + i d_{\varphi}] &=& \sum_{\elt m} -\sqrt{\elt (\elt+1)}\, \yslm{1}{\elt m}(\hatn) \, \phi_{\elt m}.
\end{eqnarray}
The unlensed CMB fields $T$, $Q$, and $U$ are evaluated on an equicylindrical pixelization (ECP) grid with $N_{\theta}=32\,768$ and $N_{\varphi} = 65\,536$, while the deflection field is evaluated on a $\healpix$ $\Nside=2048$ grid. We then calculate the ``lensed positions'' $\hatn'$ for each $\Nside=2048$ $\healpix$ pixel. In the flat-sky limit these are given schematically as $\hatn' = \hatn + \vec{d}(\hatn)$; in practice on the curved sky we parallel transport $\hatn$ a distance $| \vec{d}(\hatn) |$ along the great circle in the direction $d(\hatn)$. We then interpolate $T$, $Q$, $U$ at the lensed positions using 2-D cubic Lagrange interpolation on the ECP grid.

After lensing, we incorporate the frequency-dependent Doppler modulation effect~\citep{planck2013-pipaberration}, multiplying the temperature and polarization maps by ${[1+b_{\nu}\vec{\beta} \cdot \hatn]}$, with $b_{\nu}$ factors given in Table~\ref{table:boost_factors}.
\begin{table}[tb] 
  \begingroup
  \newdimen\tblskip \tblskip=5pt
  \caption{Boost factors}
  \label{table:boost_factors}
  \nointerlineskip
  \vskip -3mm
  \footnotesize
  \setbox\tablebox=\vbox{
    \newdimen\digitwidth
    \setbox0=\hbox{\rm 0}
    \digitwidth=\wd0
    \catcode`*=\active
    \def*{\kern\digitwidth}
    \newdimen\signwidth
    \setbox0=\hbox{-}
    \signwidth=\wd0
    \catcode`!=\active
    \def!{\kern\signwidth}
\halign{\hbox to 2cm{#\leaderfil}\tabskip 1em&
     \hfil#\hfil\tabskip 0pt\cr
      \noalign{\doubleline}
      \omit\hfil Frequency\hfil&\omit\hfil Boost factor\hfil\cr
      \omit\hfil [GHz]\hfil&\omit\hfil $b_\nu$\hfil\cr
      \noalign{\vskip 3pt\hrule\vskip 4pt}
      $*30$&$*1.05$\cr
      $*44$&$*1.1$\cr
      $*70$&$*1.25$\cr
      $100$&$*1.51$\cr
      $143$&$*1.96$\cr
      $217$&$*3.07$\cr
      $353$&$*5.38$\cr
      $545$&$*8.82$\cr
      $857$&$14.20$\cr
      \noalign{\vskip 3pt\hrule\vskip 5pt}
    }
  }
  \endPlancktablewide                 
  \endgroup
\end{table}
These boost factors incorporate bandpass corrections calculated given the frequency-dependence of the Doppler effect. We then evaluate lensed, Doppler boosted $T_{\elt m}$, $E_{\elt m}$, and $B_{\elt m}$ up to $\elt_{\rm max}=4\,096$ with a harmonic transform of the $\Nside=2\,048$ $\healpix$ map of these interpolated $T$, $Q$, and $U$ values.

We next add frequency-dependent Rayleigh scattering effects, using the ``R'' matrices defined in Eq.~(7) of \cite{Lewis:2013yia}. We use effective central frequencies for the $\nu^4$ terms of $30$, $44$, $70$, $105$, $148$, $229$, $372$, $577$, and $891$\,GHz for the $30$--$857$\,GHz channels, and effective central frequencies for the $\nu^6$ terms of $30$, $44$, $70$, $108$, $151$, $233$, $378$, $585$, and $905$\,GHz.

Finally we add a second-order temperature quadrupole following the equations in \cite{Kamionkowski:2002nd}. Since the main \Planck\ data processing removes the frequency-independent part~\citep{planck2014-a09}, we simulate only the residual frequency-dependent temperature quadrupole. After subtracting the frequency-independent part, the simulated quadrupole has frequency dependence $\propto (b_{\nu}-1)/2$, which we calculate using the  bandpass-integrated $b_{\nu}$ boost factors given in Table~\ref{table:boost_factors}. Defining the $z$-axis to be in the direction of the solar dipole, at each frequency we add
\begin{equation}
a^T_{20} = \frac{4 (b_\nu-1)}{6} \sqrt{\frac{\pi}{5}} \,\, \beta^2 \, T_0.
\end{equation}

\subsubsection{The tensor CMB sky}

In addition to the scalar CMB simulations, we also generate a set of CMB skies containing primordial tensor modes. Using the fiducial cosmological parameters of Table~\ref{table:params}, we calculate the tensor power spectra $C_{\ell}^{TT, {\rm tensor}}$, $C_{\ell}^{EE, {\rm tensor}}$, and $C_{\ell}^{BB, {\rm tensor}}$ using \camb\ with a primordial tensor-to-scalar power ratio $r=0.2$ at the pivot scale $k=0.05\,$Mpc$^{-1}$. We then simulate Gaussian $T$, $E$, and $B$-modes with these power spectra, and convert these to spherical harmonic representations of the corresponding $I$, $Q$ and $U$ maps. Note that the default $r=0.2$ means that building the FFP8a-d maps requires rescaling each CMB tensor map by $\sqrt{r/0.2}$ for each of the values of $r$ in Table~\ref{table:rfnl}.

\subsubsection{The non-Gaussian CMB sky}

We use a new algorithm to generate simulations of CMB temperature and polarization maps containing primordial non-Gaussianity. Non-Gaussian fields in general have a non-vanishing bispectrum contribution sourced by mode correlations. The bispectrum, the Fourier transform of the 3-point correlation function, can then be characterized as a function of three wavevectors, $F(k_1, k_2, k_3)$. Depending on the physical mechanism responsible for generating the non-Gaussian signal, it is possible to introduce broad classes of model that are categorized by the dependence of $F$ on the type of triangle formed by the three momenta $k_i$ \citep[see e.g.,][]{2004JCAP...08..009B}. Here, we focus on non-Gaussianity of local type, where the bulk of the signal comes from squeezed triangle configurations, $k_1 \ll k_2 \approx k_3$. This is typically predicted by multi-field inflationary models \citep[e.g.,][]{2001PhLB..522..215M, 2002NuPhB.626..395E, 2003PhRvD..67b3503L}. In this scenario, the primordial gravitational potential $\Phi$ is defined in terms of a Gaussian auxiliary field $\Phi_{\mathrm{L}}$,
\begin{equation}
\label{eq:fnl_def}
\Phi = \Phi_{\mathrm{L}} + f_{\mathrm{NL}} \left(\Phi_{\mathrm{L}}^2 -
\langle \Phi_{\mathrm{L}} \rangle^2 \right) \, ,
\end{equation}
where higher-order terms are neglected \citep{2000MNRAS.313..141V, 2000PhRvD..61f3504W}. Here, the non-Gaussian contribution to the primordial potential is parameterized by means of the scalar prefactor $f_{\mathrm{NL}}$. The spherical harmonic coefficients of the CMB anisotropies are then given by a linear equation, the line-of-sight integral \citep{2003ApJ...597...57L}
\begin{align}
\label{eq:fnl_los_integral}
a_{\ell m}^{X} &= \frac{2}{\pi} \int dr \, r^2 \Phi_{\ell m}(r)
\alpha_{\ell}^{X}(r) \nonumber \\
&= \tens{M}_{\ell}^{X} \Phi_{\ell m} \, ,
\end{align}
where we introduce the real space transfer function $\alpha_{\ell}^{X}(r)$ for temperature ($X = T$) and polarization ($X = E$). We also make use of a compact matrix notation for the integral that, for all practical purposes, is evaluated numerically on a finite number of grid points.

Our aim is to augment the Gaussian CMB simulations described above with non-Gaussian templates, defined by the second term in Eq.~(\ref{eq:fnl_def}). Given the noiseless Gaussian map, we first compute a Wiener filter reconstruction of the primordial potential $\Phi_{\mathrm{L}}$ on the full sky. Let $\tens{P}$ be the covariance matrix of $\Phi_{\mathrm{L}}$ on the radial grid \citep[e.g.,][]{2003ApJ...597...57L},
\begin{align}
  \tens{P} &= \left\langle \Phi_{\mathrm{L} \ \ell_1 m_1}(r_1) \,
  \Phi^{*}_{\mathrm{L} \ \ell_2 m_2}(r_2) \right\rangle \nonumber \\
  &= 4 \pi \ \delta_{\ell_2}^{\ell_1}\ \delta_{m_2}^{m_1}
  \int dk \ \frac{\Delta^2_{\Phi}(k)}{k} \ j_{\ell_1}(kr_1)
  \ j_{\ell_2}(kr_2) \, ,
\end{align}
where $\Delta^2_{\Phi}(k)$ is the primordial power spectrum and the $j_{\ell}$ are spherical Bessel functions. Further denoting $\tens{C}$ as the CMB covariance matrix given by the fiducial power spectrum of the Gaussian simulated map $s$, then
\begin{equation}
\Phi_{\mathrm{L}}^{\mathrm{WF}}(s) = \tens{P} \tens{M}^{\dagger} \tens{C}^{-1} s \, .
\end{equation}

Since the Wiener filter is a biased estimate, we now supplement it with a fluctuation term to yield a Gaussian constrained realization \citep{1987ApJ...323L.103B}. The covariance of the fluctuation term is completely determined by the kernel of $\tens{M}$,
\begin{equation}
\tens{P}^{\mathrm{fluc}} = \tens{P} - \tens{P} \tens{M}^{\dagger} \tens{C}^{-1} \tens{M} \tens{P} \, .
\end{equation}

We finally obtain an unbiased reconstruction of the linear gravitational potential for each simulation $s$,
\begin{equation}
\Phi_{\mathrm{L}}(s) = \Phi_{\mathrm{L}}^{\mathrm{WF}}(s) + \left( \tens{P}^{\mathrm{fluc.}} \right)^{1/2} g \, ,
\end{equation}
where $g$ is a vector of univariate Gaussian random numbers. With the reconstruction in hand, it is now straightforward to obtain the non-Gaussian template $s_\mathrm{NG}$,
\begin{equation}
s_\mathrm{NG} = \tens{M} \left(\Phi_{\mathrm{L}}(s)^2 - \langle \Phi_{\mathrm{L}} \rangle^2 \right) \, .
\end{equation}

We evaluate all line-of-sight integrals (Eq.~\ref{eq:fnl_los_integral}) on 70 radial shells, where quadrature weights and node positions are optimized as described in \citet{2009ApJS..184..264E}. The resulting non-Gaussian map is exact in the sense that all higher-order correlation functions are correctly represented.

\subsubsection{The FFP8.1 CMB skies}

The FFP8 simulations are an integral part of the analyses used to derive PR2-2015, and so were necessarily generated prior to determining that release's cosmological parameters. As such there is inevitably a mismatch between the FFP8 and the PR2-2015 cosmologies, which we address in two ways. The quick-and-dirty fix is to determine a single rescaling factor that minimizes the difference between the PR1-2013 and PR2-2015 TT power spectra and apply it to all of the FFP8 CMB maps; this number is determined to be 1.0134, and the rescaled maps have been used in several repeat analyses to confirm the robustness of various PR2-2015 results.

More rigorously though, we also generate a second set of CMB realizations based on the PR2-2015 cosmology, dubbed FFP8.1, and perform our reanalyses using these in place of the FFP8 CMB skies in both the fiducial and MC realizations. Table~\ref{table:params} lists the cosmological parameters used for FFP8.1 while Table~\ref{table:maps} enumerates the current status of the FFP8.1 CMB MCs.

\section{FFP8 pipelines}
\label{sec_pipelines}

The two \Planck\ DPCs have distinct internal data formats and access protocols, including elements that are either impossible or impractical to port to supercomputing facilities. Moreover, data access can be a significant computational bottleneck, especially in an HPC environment, so aggressive IO optimization is necessary for the most compute-intensive operations. To address these issues, we use the \Planck\ exchange format\footnote{The file format used in DPC exchanges is similar to the official released timelines \citep{planck2014-ES} except that detector pointing is not supplied but rather interpolated on-the-fly from $8\,$Hz satellite attitude files.} (EF), developed for inter-DPC exchanges and public releases, for all time-domain data, adding the necessary interfaces to EF files to the \toast\ data abstraction layer. \toast\ allows any analysis code to access any supported data set -- including \Planck\ -- without worrying about the details of the underlying file formats or data distributions on disk. Instead the path to, format of, and metadata about, each data file required in a particular analysis is stored in an XML document called a run configuration file, or ``runconfig''. Any request by a \toast-interfaced analysis code for data is then passed to \toast\ to resolve and satisfy, using the information in the runconfig. This data abstraction also enables \toast\ to cache heavily re-used data (such as the telescope pointing information) and to generate synthetic data on-the-fly (bypassing the crippling IO costs associated with performing distinct time-domain simulation and analysis phases), and to do both of those things transparently to the calling analysis code. Since the general analysis of large, correlated data sets involves global operations on all of the data, requiring very massive parallelism, \toast\ uses a hybrid MPI/OpenMP programming model to enable these operations to be efficiently executed on the architectures of the current largest-scale computing resources. All of the codes used in the various FFP8 pipelines are interfaced to \toast.

\subsection{Fiducial realizations}

\begin{figure}[!ht]
\begin{center}
\includegraphics[width=0.48\textwidth]{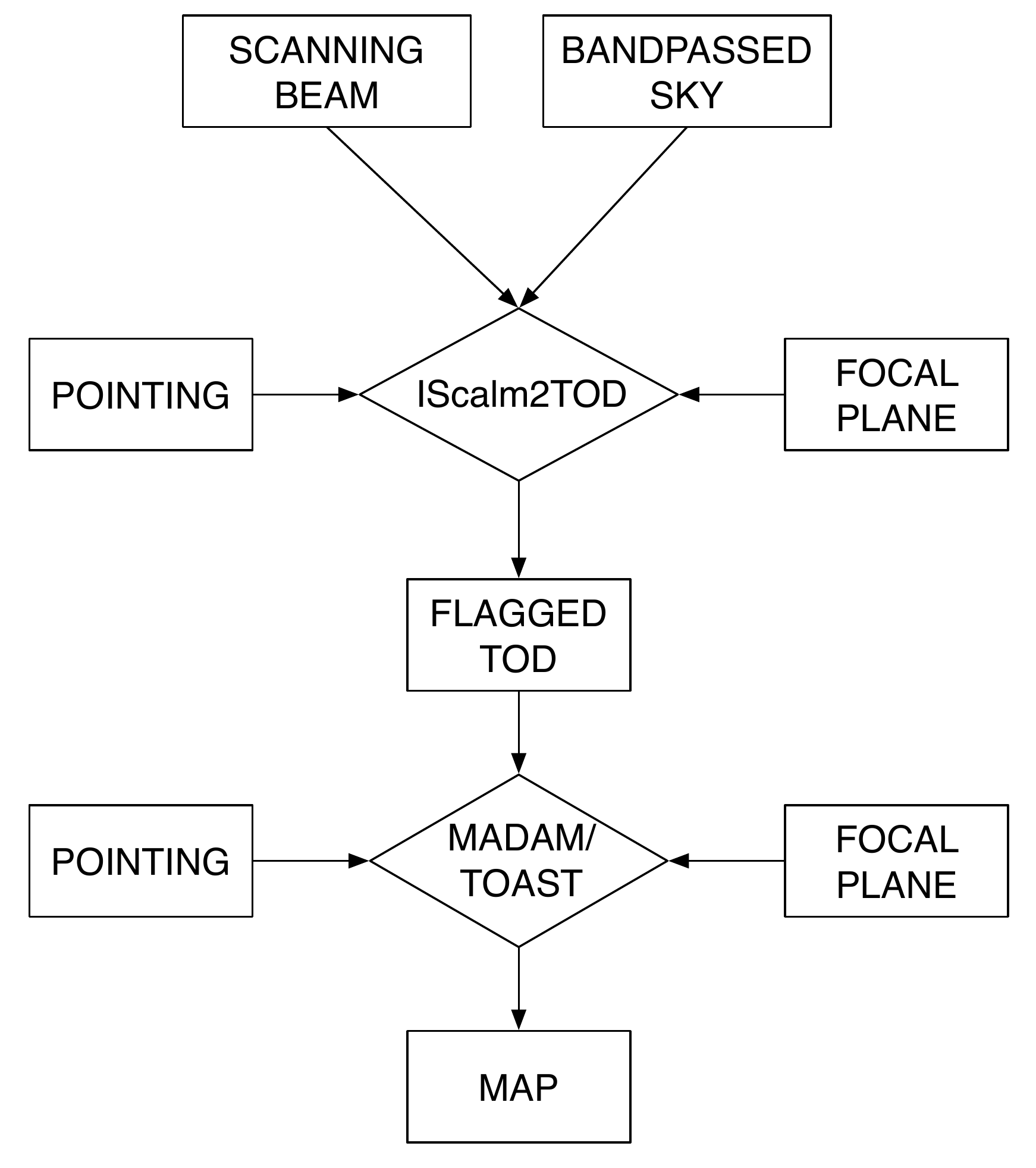}
\caption{Fiducial pipeline schematic.}
\label{fig:fiducial}
\end{center}
\end{figure}

The FFP8 fiducial realization is generated in two steps: (1)~simulation of the full mission TOD for every detector; and (2)~calculation of maps from the various detector subsets, intervals, and data cuts (Fig.~\ref{fig:fiducial}). Simulation of explicit TODs allows us to incorporate each detector's full beam (including its far sidelobes) and unique input sky (including its bandpass). As noted above, the fiducial realization is generated in six separate components -- the three CMB components (lensed scalar, tensor, and non-Gaussian complement), two foreground realizations (with and without bandpass mismatch), and noise. The first five of these are simulated as explicit TODs and then mapped, while the noise is generated using the on-the-fly approach described in the noise MC subsection below.

\begin{figure*}[ht!]
\begin{center}
\includegraphics[width=0.73\textwidth]{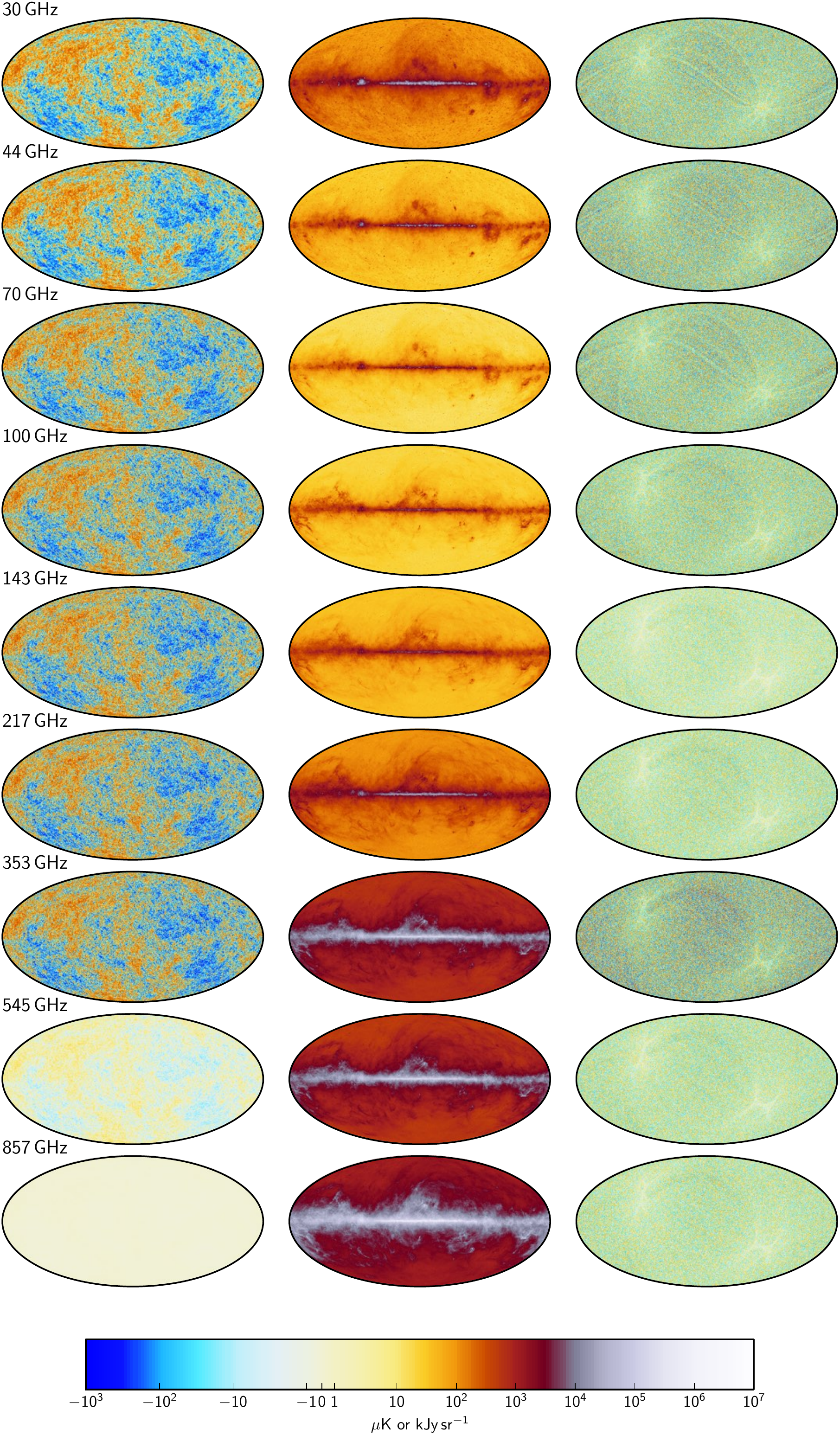}
\caption{Left to right, the baseline FFP8 fiducial CMB, foreground and noise component temperature channel/mission/full maps at each frequency. Frequencies $30$--$353\,$GHz are plotted in $\mu$K, while $545$ and $857\,$GHz are plotted in kJy\,sr$^{-1}$. See Fig.~\ref{fig:total_temperature} for the combined temperature maps.}
\label{fig:ftmaps}
\end{center}
\end{figure*}

\begin{figure*}[ht!]
\begin{center}
\includegraphics[width=0.73\textwidth]{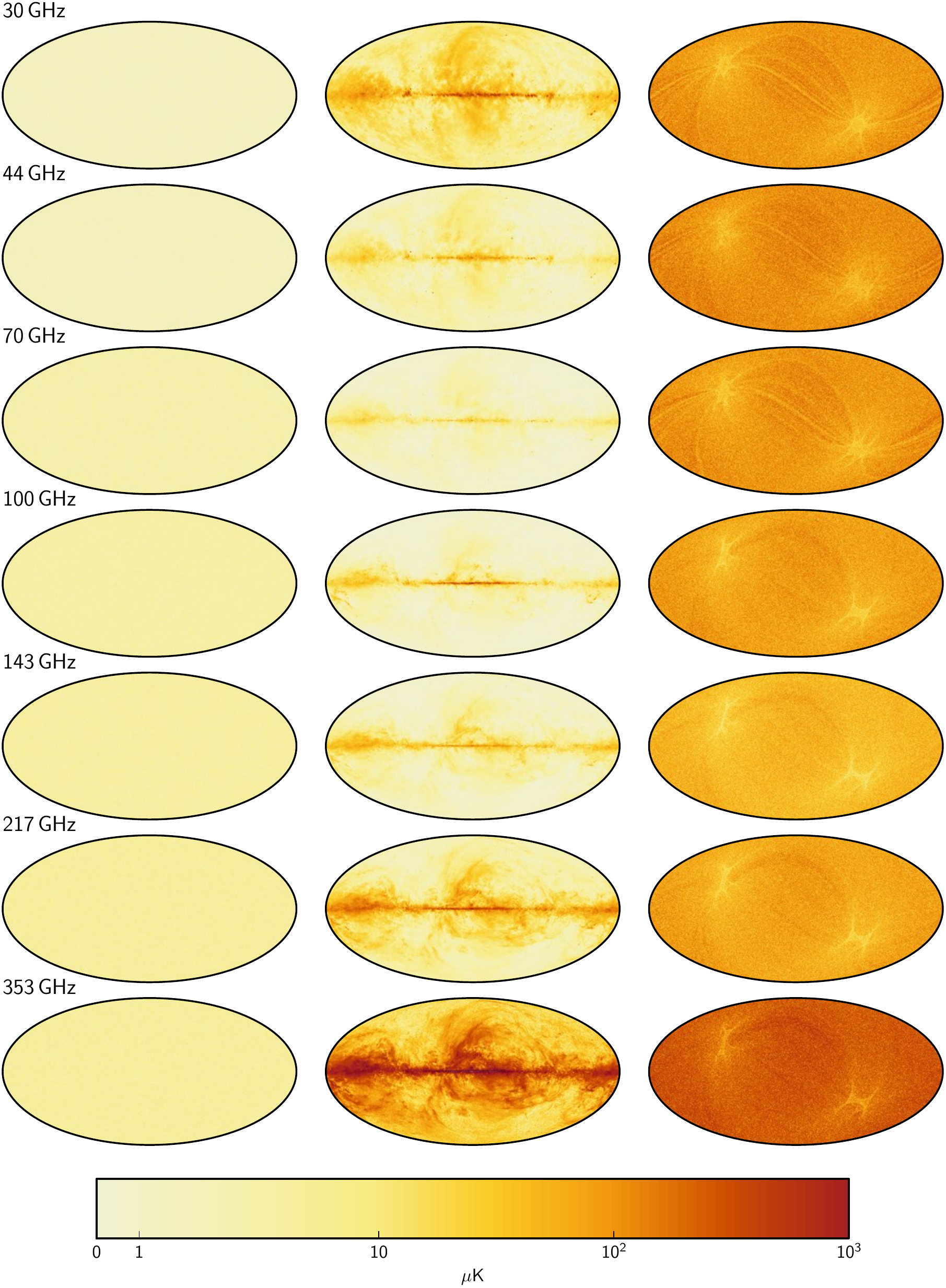}
\caption{Left to right, the baseline FFP8 fiducial CMB, foreground and noise component total polarization channel/mission/full maps at each frequency. See Fig.~\ref{fig:total_polarization} for the combined polarization maps.}
\label{fig:fpmaps}
\end{center}
\end{figure*}

TOD generation for any detector proceeds by first convolving the appropriate sky component with the beam at every point in a uniformly sampled data cube of Euler angle triplets (encoding the pointing and polarization orientation) to produce the ``beamskyset'', and then generating the time-ordered data by interpolating over the beamskyset data cube to the exact pointing and polarization orientation of each sample. Previous FFP simulations, including FFP6, accompanying the 2013 \Planck\ data release, used the \levels\ software package \citep{reinecke2006} to do this, using the \conviqt\ \citep{Prezeau:2010mx}  and \multimod\ tools, respectively. However, this required format conversions for the input pointing data (from exchange to \levels) and the output time-ordered data (from \levels\ to exchange), at significant IO and disk space costs. For FFP8 we have therefore embedded the critical parts of these routines into a new code, \iscalmtod, which uses \toast\ to interface directly with exchange format data. As an additional benefit, by combining the two steps into a single code the beamskyset never has to be written to disk, further reducing the IO and disk space costs. This approach also allows us trivially to handle the LFI multi-component beams (subsection \ref{subsec_beams}), where each beam is expressed at a different $(\ell_{\rm max}, m_{\rm max})$ resolution. Previously this meant generating distinct TODs and adding them only after they had been written, tripling the IO and disk space requirements, whereas now they are generated separately but combined within \iscalmtod\ prior to being written.

All of the FFP8 fiducial maps are produced using \madamtoast, a \toast\ port of the \madam\ generalized destriping code \citep{keihanen2005,keihanen2010}, which allows for destriping with an arbitrary baseline length, with or without a prior on the baseline distribution (or noise filter). \madam\ is used to produce the official LFI maps, and its destriping parameters can be chosen so that it reproduces the behaviour of \polkapix, the official HFI mapmaking code. Comparison of the official maps and \madamtoast\ maps run using exchange data show that mapmaker differences are negligible compared to small differences in pointing and (for HFI) dipole subtraction that do not impact the simulation. The sky components are mapped from the TODs generated by \iscalmtod, while the fiducial noise is taken to be realization 10\,000 of the noise MC (with realizations 0000-9999 reserved for the noise MC itself). Summarizing the key differences in the map making parameters for each \Planck\ frequency:

\begin{itemize}

\item $30$\,GHz is destriped with $0.25$\,s baselines; $44$ and $70$\,GHz are destriped using $1$\,s baselines; and $100$--$857$\,GHz are destriped using pointing-period baselines ($30$--$75$\,min).

\item $30$--$70$\,GHz are destriped with a $1/f$-shape noise prior, while $100$--$857$\,GHz are destriped without a noise prior.

\item $30$, $44$, and $70$\,GHz have separate destriping masks, while $100$--$857$\,GHz use the same 15\,\% galaxy+point source mask.

\item $30$--$70$\,GHz maps are destriped using baselines derived exclusively from the data going into the particular map, while $100-857$\,GHz maps are destriped using baselines derived from the full data set.

\end{itemize}

The baseline ($r = f_{\rm NL} = 0$) fiducial channel/mission/full maps for temperature and total polarization are shown in Figs.~\ref{fig:ftmaps} and \ref{fig:fpmaps}, respectively.

\subsection{Noise Monte Carlos}

\begin{figure}[!ht]
\begin{center}
\includegraphics[width=88mm]{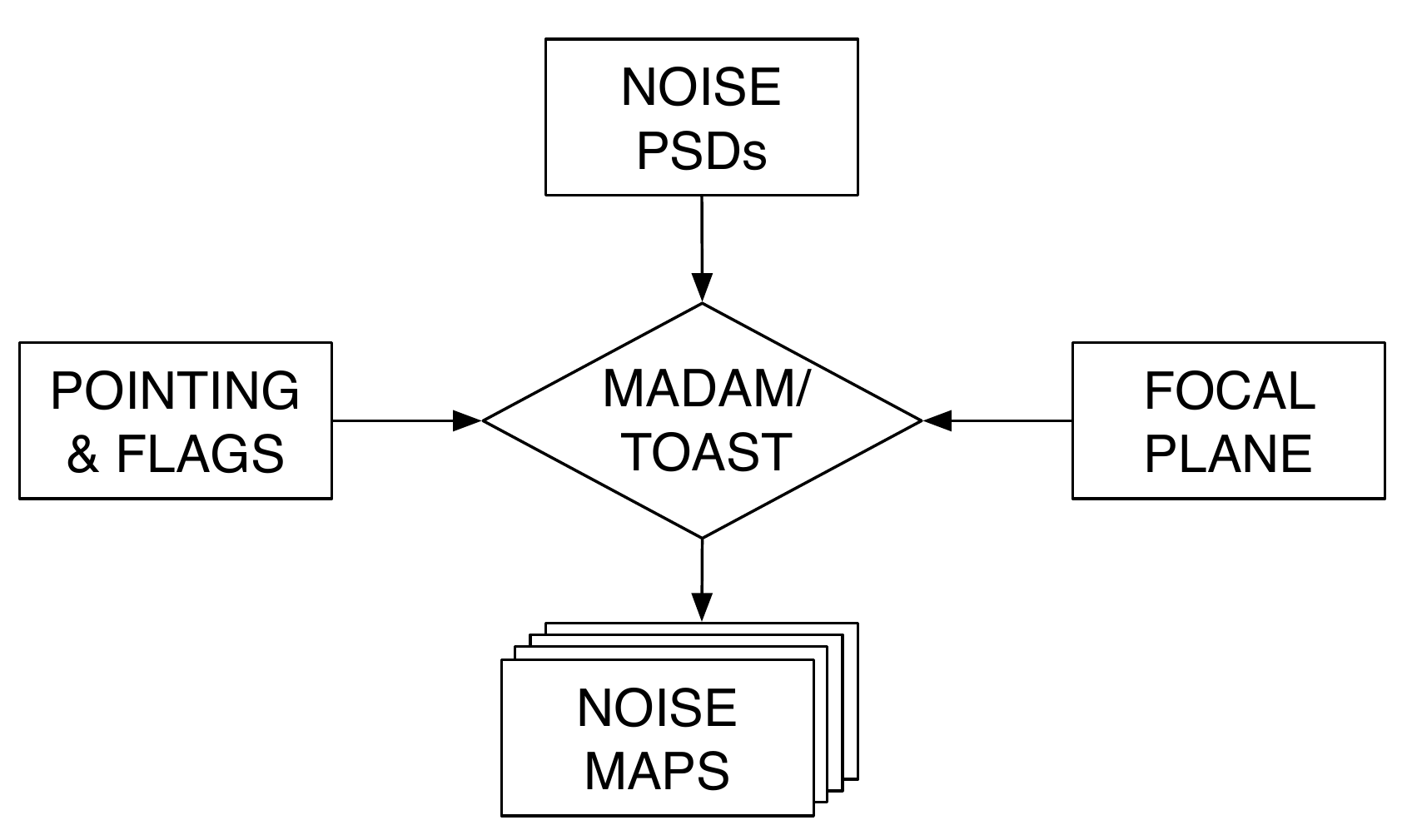}
\caption{Noise Monte Carlo pipeline schematic.}
\label{fig:noisemc}
\end{center}
\end{figure}

The FFP8 noise MCs are generated using \madamtoast, exploiting \toast's on-the-fly noise simulation capability to avoid the IO overhead of writing a simulated TOD to disk only to read it back in to map it (Fig.~\ref{fig:noisemc}). In this implementation, \madam\ runs exactly as it would with real data, but whenever it submits a request to \toast\ to provide it with the an interval of the noise TOD, that interval is simply simulated by \toast\ in accordance with the noise power spectral densities provided in the runconfig, and returned to \madam.

For a simulation set of this size and complexity, requiring of the order of ${10^{17}}$ random numbers over $10^{12}$ disjoint and uncorrelated intervals, care must be take with the pseudo-random number generation to ensure that it is fast, reliable (and specifically uncorrelated), and reproducible, in particular enabling any process to generate any element of any subsequence on demand. To achieve this \toast\ uses a Combined Multiple Recursive Generator (CMRG) \citep{L'Ecuyer} that provides more than sufficient period, excellent statistical robustness, and the ability to skip ahead to an arbitrary point in the pseudo-random sequence very quickly. Rather than seed each realization separately (with the associated risk of generating overlapping subsequences), we then set a single seed for the entire simulation set, divide the resulting sequence into $2^{64}$ disjoint subsequences, and assign a specific subset of subsequences to each pointing period of each detector for each realization, with the allocation of several rather than a single subsequence allowing us to support multiple independent timestream components, each of which depends on random number generation, such as the HFI detectors' uncorrelated and correlated noise components. Consequently when \madam\ requests the noise TOD for a particular interval of a particular realization, the process can skip ahead to the appropriate location in the overall sequence; furthermore the random number generation itself can be threaded, with each thread then skipping ahead to the appropriate place within the subsequence.

\madam\ supports MC runs by enabling the pointing and flag data common to all realizations to be read once, before looping over the realizations that share them. However, to support maximum MC throughput some changes were also made to the standard \madam\ implementation:

\begin{itemize}

\item once the full data for any time span are loaded, \madamtoast\ can independently destripe and map the full and two half ring data sets;

\item communication at high concurrency is more efficiently supported through use of the advanced MPI collective operation \texttt{alltoallv};

\item the map is distributed between the MPI processes based on pixels covered by the distributed TOD rather than using a simple round-robin scheme;

\item HFI full-mission destriping is supported on-the-fly so that once the full mission data have been destriped, all subset maps can be binned as part of the same run;

\item OpenMP/MPI hybridization allows for runs at higher concurrencies without suffering from MPI communication congestion.

\end{itemize}

\subsection{CMB Monte Carlos}

The FFP8 CMB MCs are generated using the \febecop\ software package~\citep{mitra2010}, which produces beam-convolved maps directly in the pixel domain rather than sample-by-sample, as is done for the fiducial maps. The goal of this approach is to reduce the computational cost by the ratio of time-samples to map-pixels (i.e., the number of hits per pixel).

As illustrated in Fig.~\ref{fig:cmbmc}, \febecop\ proceeds in 3 steps:

\begin{figure}[ht]
\begin{center}
\includegraphics[width=0.48\textwidth]{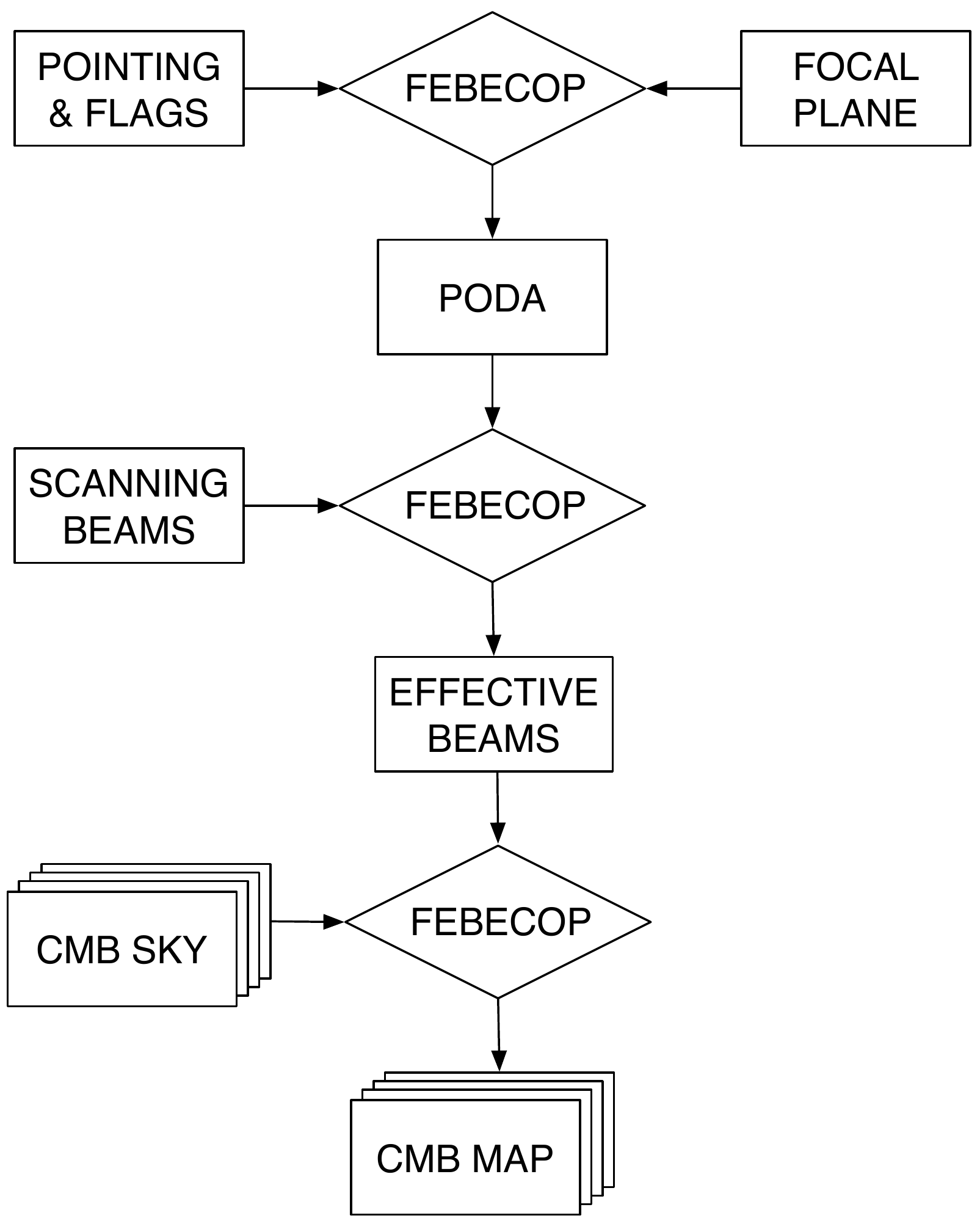}
\caption{CMB Monte Carlo pipeline schematic.}
\label{fig:cmbmc}
\end{center}
\end{figure}

\begin{enumerate}

\item Given the satellite pointing and flags and the focal plane (accessed through the \toast\ interface), for every channel \febecop\ first re-orders all of the samples in the mission by pixel instead of time, localizing all of the observations of each pixel, and writes the resulting pixel-ordered detector dngles (PODA) to disk. Note that since the PODA also contains the detector, time-stamp, and weight of each observation this is a one-time operation for each frequency, and does not need to be re-run for different time intervals or detector subsets, or for changes in the beam model or its chosen cut-off radius.

\item For every time interval and detector subset to be mapped, and for every pixel in the map, \febecop\ uses the PODA and the scanning beams to generate an effective-beam for that pixel which is essentially the weighted average of the discretized beam functions for every sample in the pixel included in the time interval and detector subset. The total effective-beam array is also written to disk. Given the PODA, this is a one-time operation for any beam definition.

\item Finally, \febecop\ applies the effective-beam pixel-by-pixel to every CMB sky realization in the MC set to generate the corresponding beam-convolved CMB map realization.

\end{enumerate}

The effective-beams provide a direct connection between the true and observed sky, explicitly incorporating the detailed pointing for every detector through a linear convolution. By providing the effective-beams at every pixel, \febecop\ enables precise control of systematic effects, e.g., the point-spread functions can be fitted at each pixel on the sky and used to determine point source fluxes \citep{planck2014-a35,planck2014-a36}.

\subsection{Validation}

\begin{figure*}
  \centerline{\includegraphics[width=1.0\textwidth]{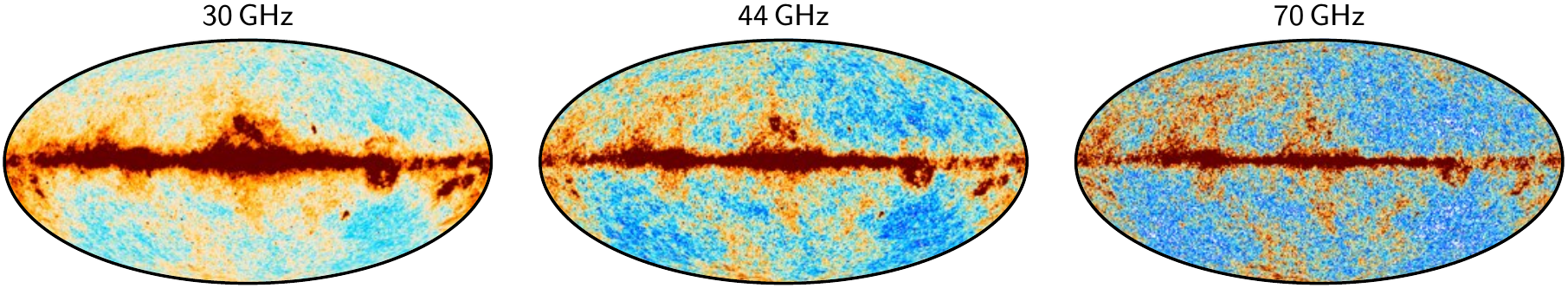}}
  \centerline{\includegraphics[width=1.0\textwidth]{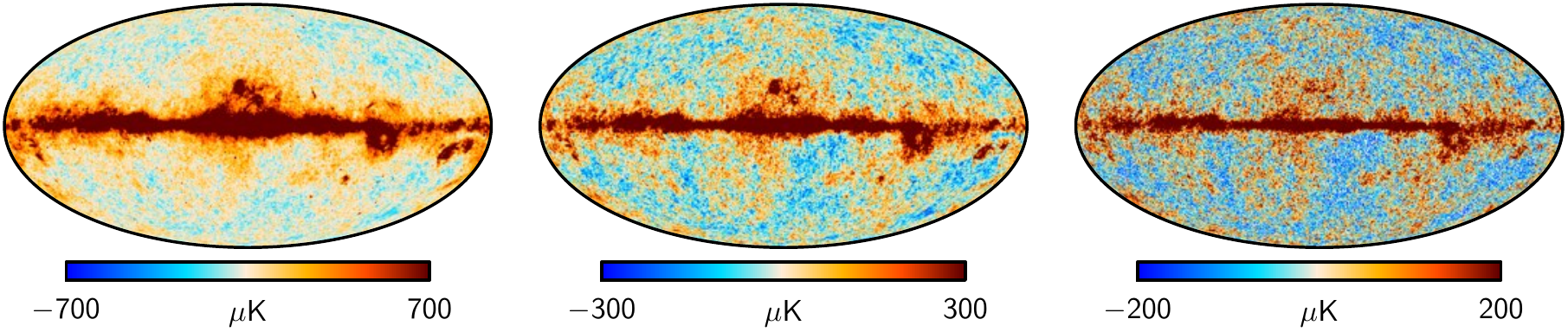}}
    \vspace*{0.2in}
  \centerline{\includegraphics[width=1.0\textwidth]{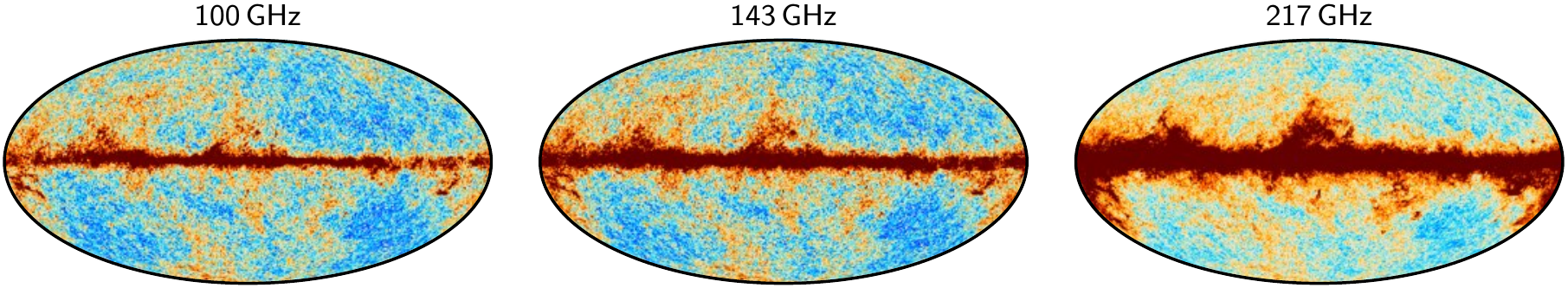}}
  \centerline{\includegraphics[width=1.0\textwidth]{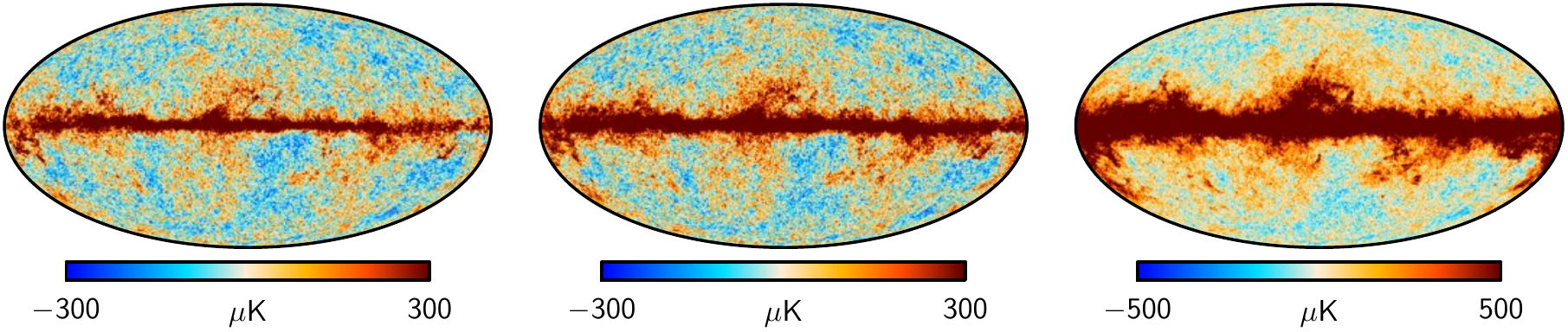}}
    \vspace*{0.2in}
  \centerline{\includegraphics[width=1.0\textwidth]{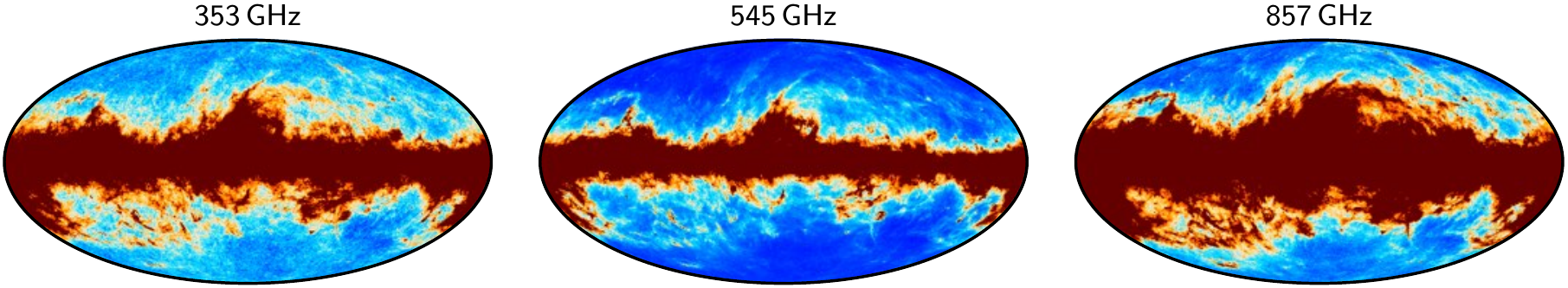}}
  \centerline{\includegraphics[width=1.0\textwidth]{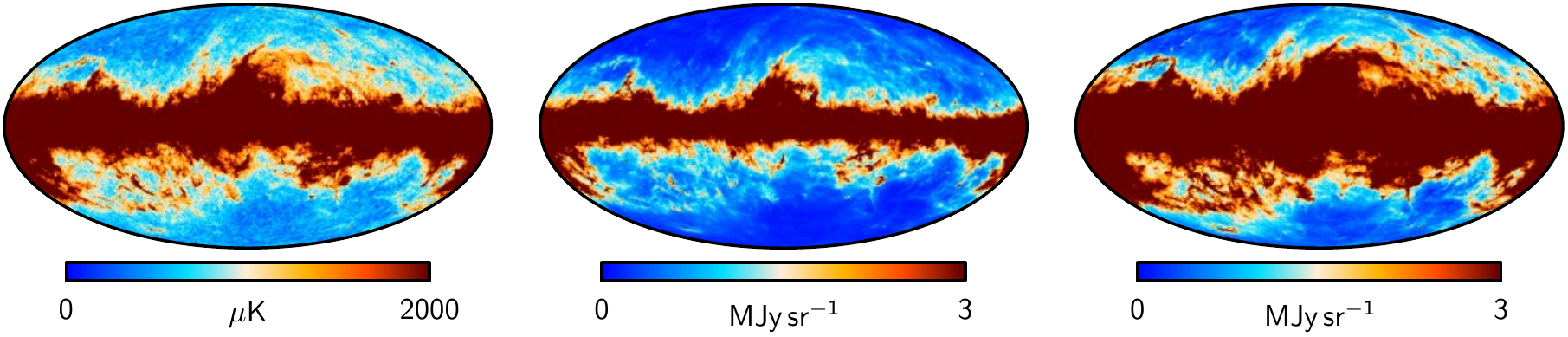}}
  \caption{\label{fig:total_temperature}
    Pairs of channel/mission/full temperature maps, comparing FFP8 simulations (upper) and 2015 \Planck\ data (lower).  All maps are downgraded to $\Nside=256$. Visible differences away from the galactic plane are between the actual and simulated CMB sky.
  }
  \label{fig:tempcomp}
\end{figure*}

\begin{figure*}
  \centerline{\includegraphics[width=1.0\textwidth]{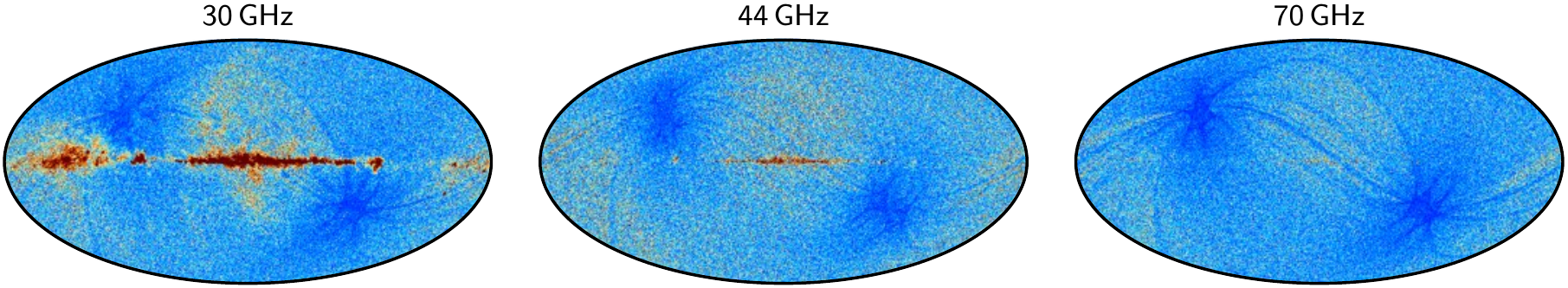}}
  \centerline{\includegraphics[width=1.0\textwidth]{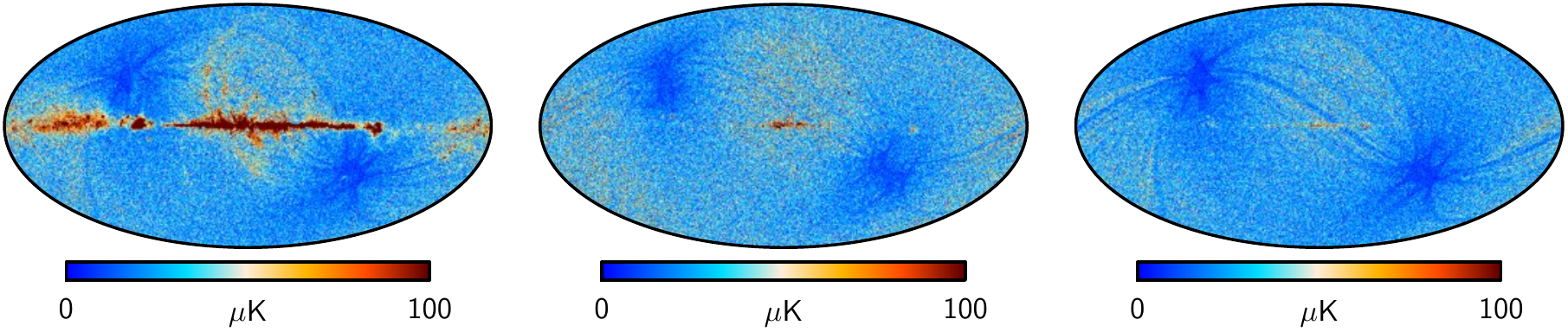}}
    \vspace*{0.2in}
  \centerline{\includegraphics[width=1.0\textwidth]{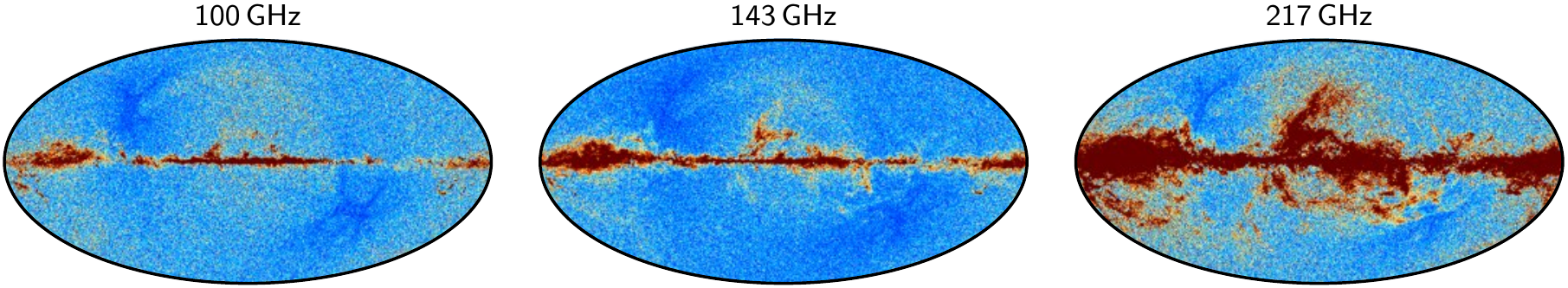}}
  \centerline{\includegraphics[width=1.0\textwidth]{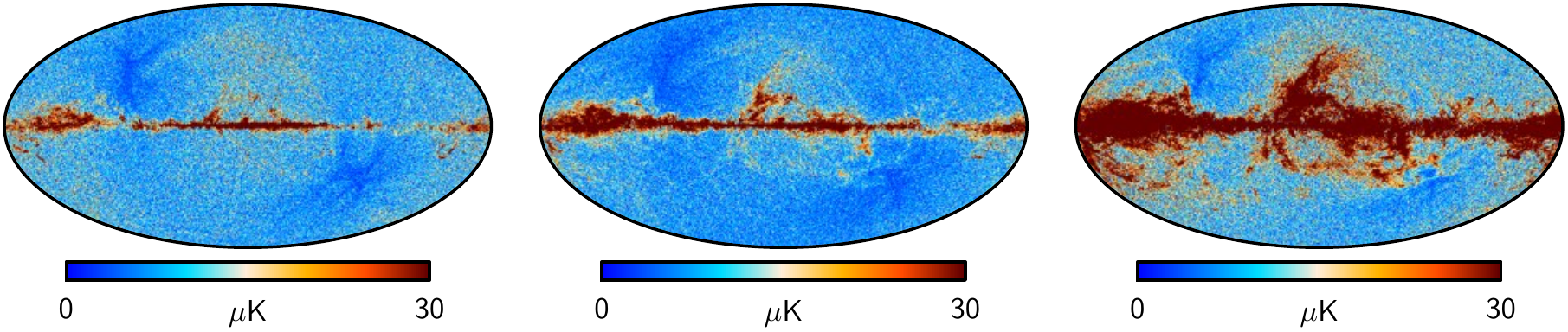}}
  \vspace*{0.2in}
    \includegraphics[width=0.315\textwidth]{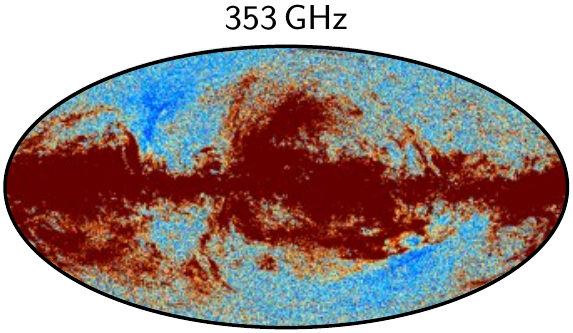}\\
    \includegraphics[width=0.315\textwidth]{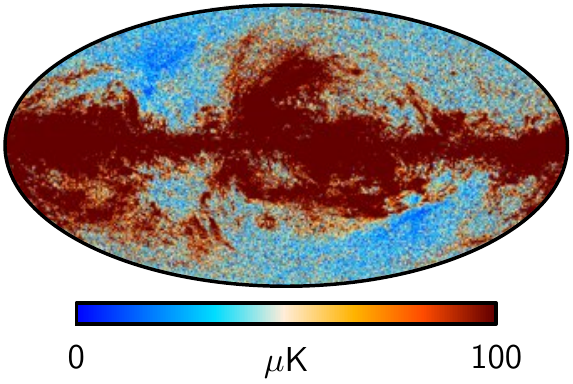}
  \caption{\label{fig:total_polarization}
    Pairs of channel/mission/full total polarization maps, comparing FFP8 simulations (upper) and 2015 \Planck\ (lower) data.  All maps are downgraded to $\Nside=256$. 
  }
  \label{fig:polcomp}
\end{figure*}

Our goal for the FFP8 simulation set is that it be not only internally self-consistent, but also a good representation of the real data. In addition to the validation steps carried out on all of the inputs individually and noted in their respective sections above, we must also validate the final outputs. A first crude level of validation is provided simply by visual inspection of the FFP8 and real \Planck\ maps -- Figs.~\ref{fig:tempcomp} (temperature) and \ref{fig:polcomp} (total polarization) -- where the only immediately apparent difference is the CMB realization.

\begin{figure*}[!ht]
  \centerline{\includegraphics[trim=0 0 260 0,clip,width=0.5\textwidth]{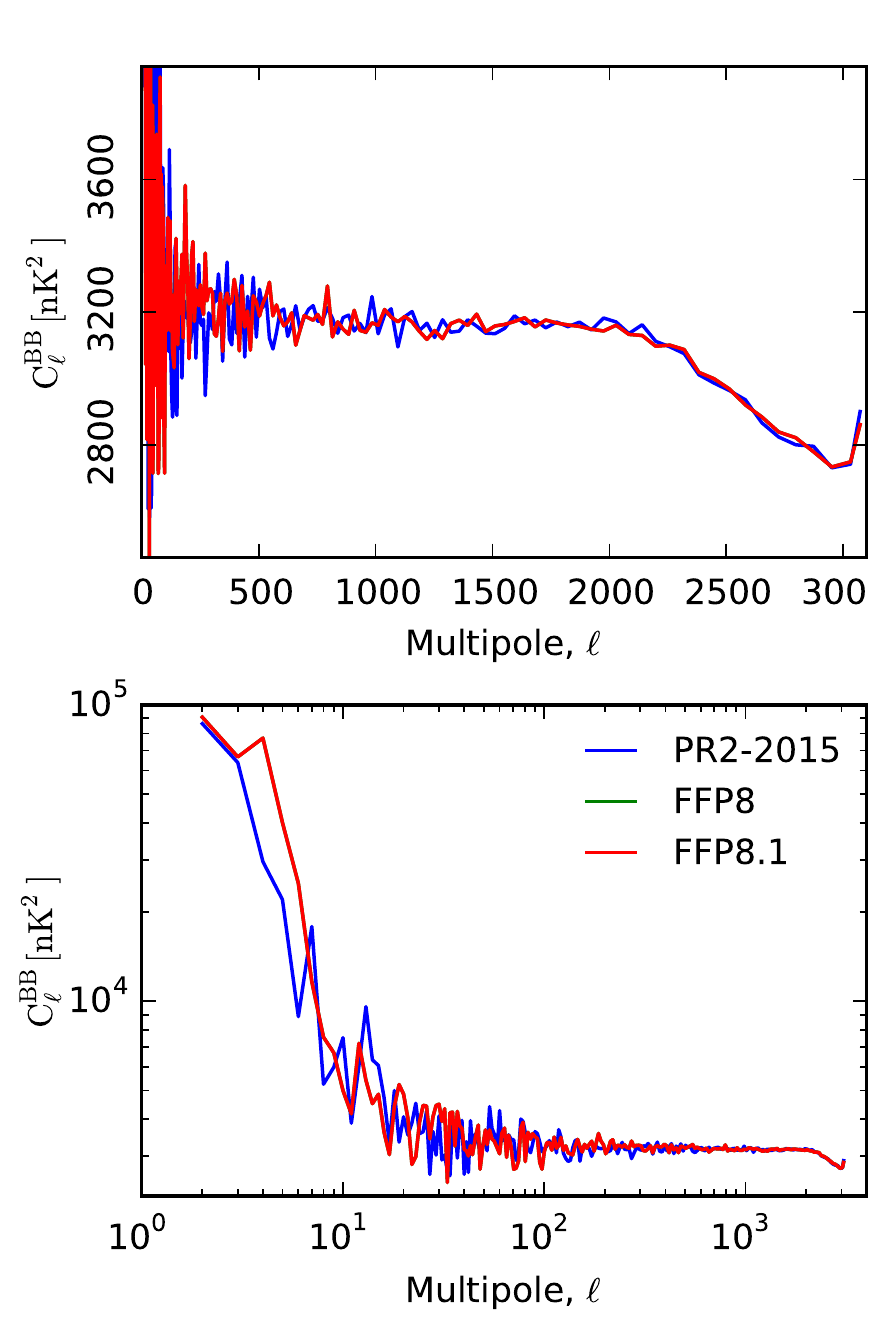}\includegraphics[trim=0 0 260 0,clip,width=0.5\textwidth]{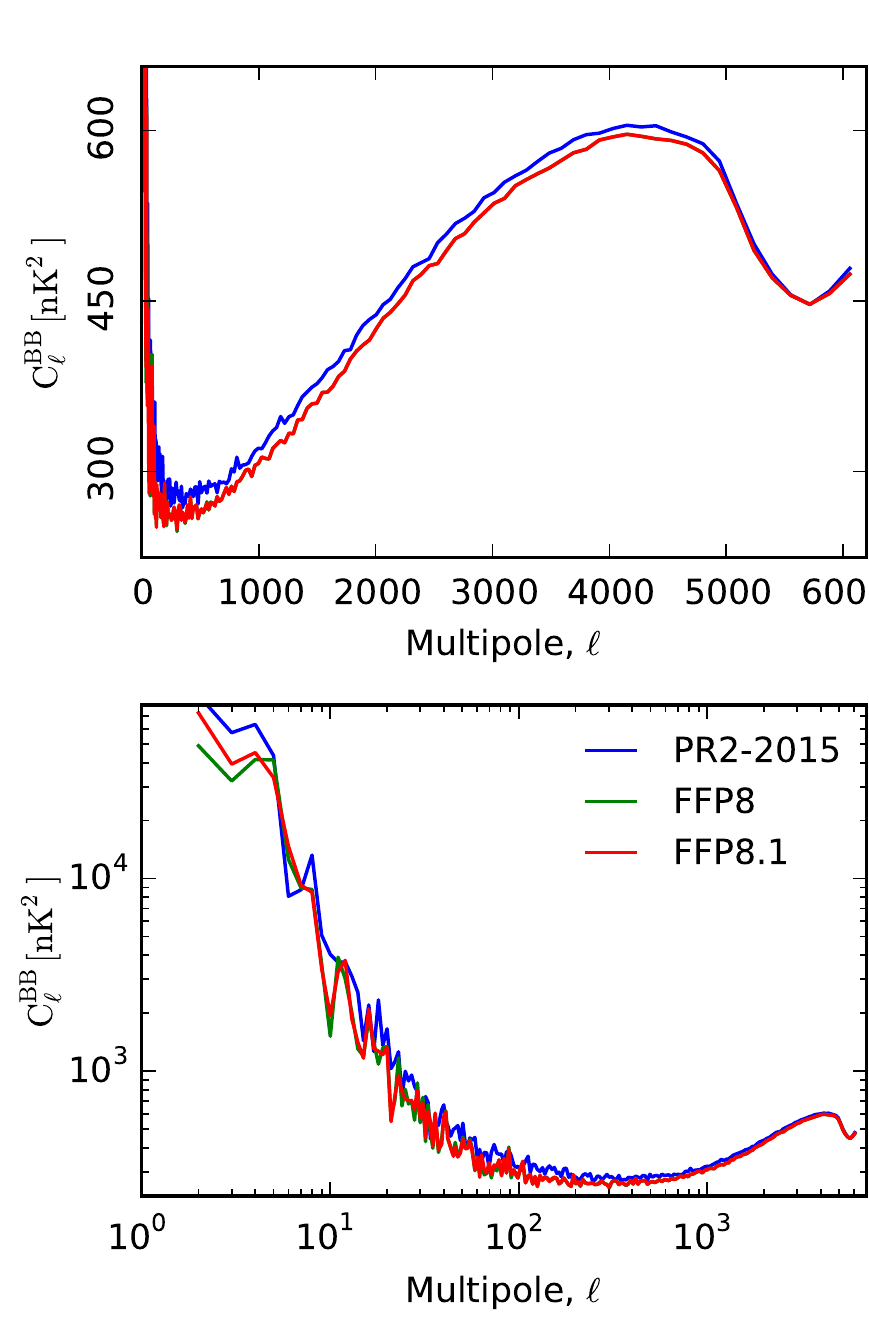}}
  \caption{\label{fig:full_data}
    Comparison of the $BB$ spectra of the channel/mission/full simulated and flight data maps using pseudo-spectra computed on $75\,$\% of the sky with the Galaxy and point sources masked, showing excellent agreement at $70\,$GHz but a few percent discrepancy at almost all angular scales at $100\,$GHz.
    \emph{Left}: 70\,GHz full map $BB$-spectra.
    \emph{Right}: 100\,GHz full map $BB$-spectra.
    \emph{Top}: Linear horizontal axis shows small scale behaviour.
    \emph{Bottom}: Logarithmic horizontal axis shows large scale behaviour.
  }
\end{figure*}

While this is a necessary test, it is hardly sufficient, and the next step is to compare the angular power spectra of the simulated and real channel/mission/full maps. As illustrated in Fig.~\ref{fig:full_data} and Table~\ref{table:ratios}, LFI channels show excellent agreement across all angular scales, while HFI channels show a significant power deficit at almost all angular scales. Since this missing HFI power is not picked up in the noise estimation, it must be sky-synchronous (frequency bins corresponding to sky-synchronous signals being discarded when fitting the noise PSDs due to their contamination by signal residuals). This is now understood to be a systematic effect introduced in the HFI pre-processing pipeline, and we are working both to incorporate it as a systematic component in existing simulations and to ameliorate if for future data releases.

Finally, the various analyses of the FFP8 maps in conjuction with the flight data provide powerful incidental validation. To date the only issues observed here are the known mismatch between the FFP8 and PR2-2015 cosmologies, and the missing systematic component in the HFI maps.  As noted above, the former is readily addressed by rescaling or using FFP8.1; however, the characterization and reproduction of the latter is an ongoing effort.  Specific details of the consequences of this as-yet unresolved issue, such as its impact on null-test failures and {\it p}-value stability in studies of non-Gaussianity, can be found in the relevant papers (e.g., \citealt{planck2014-a11}; \citealt{planck2014-a12}; \citealt{planck2014-a13}; \citealt{planck2014-a18}).

\begin{table}[htb] 
  \begingroup
  \newdimen\tblskip \tblskip=5pt
  \caption{$EE$ and $BB$ mission full map spectrum ratios.}
  \label{table:ratios}
  \nointerlineskip
  \vskip -3mm
  \footnotesize
  \setbox\tablebox=\vbox{
    \newdimen\digitwidth
    \setbox0=\hbox{\rm 0}
    \digitwidth=\wd0
    \catcode`*=\active
    \def*{\kern\digitwidth}
    \newdimen\signwidth
    \setbox0=\hbox{-}
    \signwidth=\wd0
    \catcode`!=\active
    \def!{\kern\signwidth}
\halign{\hbox to 2cm{#\leaderfil}\tabskip 2em&
    \hfil#\hfil \tabskip 2em&
    \hfil#\hfil \tabskip 0pt\cr
      \noalign{\doubleline}
      \omit\hfil Frequency\hfil\cr
      \omit\hfil [GHz]\hfil&\omit\hfil $EE$\hfil&\omit\hfil $BB$\hfil\cr
      \noalign{\vskip 3pt\hrule\vskip 4pt}
      $*30$&$0.998$&$1.000$\cr
      $*44$&$0.999$&$1.000$\cr
      $*70$&$0.998$&$1.000$\cr
      $100$&$1.019$&$1.020$\cr
      $143$&$1.033$&$1.029$\cr
      $217$&$1.017$&$1.007$\cr
      $353$&$1.042$&$1.039$\cr
      \noalign{\vskip 3pt\hrule\vskip 5pt}
    }
  }
  \endPlancktable                    
  \tablenote {{\mbox{}}} This is the full mission median $C_\ell$ ratio (PR2-2015 / FFP8) in the noise-dominated range $\ell\in[2000,3000]$ (LFI) and $\ell\in[3000,4000]$ (HFI); as expected, FFP8.1 exhibits the same values.\par
  \endgroup
\end{table}

\section{High performance computing considerations}
\label{sec_hpc}

\subsection{Simulations at NERSC}

NERSC is the US Department of Energy's general purpose supercomputing centre, with a new top-10 HPC system\footnote{\url{http://top500.org}}
being installed about every 3~years. For the last 15~years it has provided annual allocations of cycles and storage to the \Planck\ collaboration, with a unique agreement between NASA and DOE guaranteeing \Planck\ access to NERSC through the mission lifetime \citep{Borrill:2014}. The US \Planck\ collaboration has also purchased exceptional levels of service at NERSC, including 240\,TB of globally accesssible disk space and one cabinet (640 cores) of the 9\,984-core IBM iDataplex {\em Carver}. In 2014, \Planck\ was awarded about 75\,000\,000 NERSC CPU-hours, which were shared by the 170 \Planck\ data analysts who have requested NERSC accounts.

For FFP8, the fiducial realization, all HFI noise MCs, the first 1\,000 LFI channel noise MCs, and all CMB MCs were performed on NERSC's 133\,824-core Cray XC30 {\em Edison}. Subsequently, all of the FFP8.1 CMB MCs were generated there too.

\subsubsection{The fiducial realization}

The FFP8 fiducial foreground and CMB components were generated via explicit timelines and the noise via an implicit timeline, as detailed in the noise MC section below. For the explicit cases each frequency used a separate job script, allowing multiple frequencies to be executed simultaneously without concerns about load-balancing or synchronization. Within each frequency-specific job each component was run in turn, allowing the same target TOD files to be re-used (overwritten) to save disk space. Within each component, each detector was run in turn to prevent overloading the file system (since the exchange format TODs are stored by frequency). Once a full set of TODs had been generated for a given component at a given frequency, the full set of 1134 maps was produced before moving on to the next component. The concurrencies and runtimes for each frequency are listed in Table~\ref{tbl:fidrun}.  With five components generated via explicit timelines and one via on-the-fly simulation within the mapmaking, the total cost was approximately 250\,000 CPU-hours. This same set-up also allowed us to generate each individual foreground component to cross-check the component separation pipelines.

\begin{table}[h]
  \begingroup
  \newdimen\tblskip \tblskip=5pt
  \caption{Computational cost of generating one component of the FFP8 fiducial realization for each frequency. TOD generation includes all of the detectors, and mapmaking all of the maps, at that frequency. Note the much higher cost of LFI mapmaking, due to its shorter baselines and run-baseline destriping.}
  \label{tbl:fidrun}
  \nointerlineskip
  \vskip -3mm
  \footnotesize
  \setbox\tablebox=\vbox{
    \newdimen\digitwidth
    \setbox0=\hbox{\rm 0}
    \digitwidth=\wd0
    \catcode`*=\active
    \def*{\kern\digitwidth}
    \newdimen\signwidth
    \setbox0=\hbox{-}
    \signwidth=\wd0
    \catcode`!=\active
    \def!{\kern\signwidth}
\halign{\hbox to 1.7cm{#\leaderfil}\tabskip 1.5em&
    \hfil#\hfil&
    \hfil#\hfil&
    \hfil#\hfil&
    \hfil#\hfil\tabskip 0pt\cr
      \noalign{\doubleline}
      \omit\hfil Frequency\hfil&Concurrency&TODs&Maps&Cost\cr
      \omit\hfil [GHz]\hfil&(Cores)&[min]&[min]&[CPU-hrs]\cr
      \noalign{\vskip 4pt\hrule\vskip 6pt}
      *30&3\,600&*12&20&1\,920\cr
      *44&3\,600&*36&15&3\,060\cr
      *70&3\,600&108&48&9\,360\cr
      100&4\,800&*64&*3&5\,360\cr
      143&4\,800&*88&*5&7\,440\cr
      217&4\,800&*96&*5&8\,080\cr
      353&4\,800&*96&*5&8\,080\cr
      545&4\,800&*18&*2&1\,600\cr
      857&4\,800&*24&*2&2\,080\cr
      \noalign{\vskip 3pt\hrule\vskip 4pt}
    }
  }
  \endPlancktable                    
  \endgroup
\end{table}

\subsubsection{The noise Monte Carlos}

For the noise MCs, we noted that mapmaking runs most efficiently at the lowest concurrency with sufficient memory to hold all of the data at a given frequency. Additional parallelism was then achieved by running multiple instances of a single frequency simultaneously. In practice, for each instrument and frequency we first ran a job containing 10 instances, each generating $10^2$ realizations of all map-types, to produce the first $10^3$ realizations, and then, for HFI only (for LFI, see Sect.~\ref{subsection:CSC}), ran a second job running nine instances each generating $10^3$ realizations of only the channel/mission/full maps  to fill out the full $10^4$ realizations. For LFI, where every data subset is individually destriped, the instances were further subdivided to keep the data per core constant, so that the same number of cores would process one channel/mission, two channel/half-mission, four channel/year, or eight channel/survey maps, or (for the 70\,GHz data) three detset/mission, six detset/half-mission, twelve detset/year or twenty-four detset/survey maps. Table~\ref{tbl:nmcrun} shows the cost per realization of all the maps at one frequency (for realizations 0--999) and of the HFI channel/mission/full maps alone (for realizations 1000 - 9999). The overall cost of these runs was approximately 3\,000\,000 CPU-hours for the $10^3$ realizations of all maps and 5M CPU-hours for the $9 \times 10^3$ additional HFI channel/mission/full maps.

\begin{table}[h]
  \begingroup
  \newdimen\tblskip \tblskip=5pt
  \caption{Computational cost of generating one realization of the FFP8 noise MC for each frequency, both for the full set of maps and for the HFI channel/mission/full (CMF) maps alone.}
  \label{tbl:nmcrun}
  \nointerlineskip
  \vskip -3mm
  \footnotesize
  \setbox\tablebox=\vbox{
    \newdimen\digitwidth
    \setbox0=\hbox{\rm 0}
    \digitwidth=\wd0
    \catcode`*=\active
    \def*{\kern\digitwidth}
    \newdimen\signwidth
    \setbox0=\hbox{-}
    \signwidth=\wd0
    \catcode`!=\active
    \def!{\kern\signwidth}
\halign{\hbox to 1.7cm{#\leaderfil}\tabskip 1em&
    \hfil#\hfil&
    \hfil#\hfil\tabskip 0.5em&
    \hfil#\hfil\tabskip 1em&
    \hfil#\hfil\tabskip 0.5em&
    \hfil#\hfil \tabskip 0pt\cr
      \noalign{\doubleline}
      \omit&&\multispan2\hfil All maps\hfil&\multispan2\hfil CMF maps\hfil\cr
      \noalign{\vskip -3pt}
      \omit\hfil Frequency\hfil&Concurrency&\multispan2\hrulefill&\multispan2\hrulefill\cr
      \omit\hfil [GHz]\hfil&(Cores)&[min]&[CPU-hrs]&[min]&[CPU-hrs]\cr
      \noalign{\vskip 4pt\hrule\vskip 6pt}
      *30&2\,304&10&384&\dots&\dots\cr
      *44&3\,456&*8&460&\dots&\dots\cr
      *70&4\,608&13&949&\dots&\dots\cr
      100&2\,640&*3&176&2&*88\cr
      143&3\,600&*3&180&2&120\cr
      217&3\,960&*3&198&2&132\cr
      353&3\,960&*3&198&2&132\cr
      545&*\,600&*4&*40&2&*20\cr
      857&*\,600&*6&*60&3&*30\cr
      \noalign{\vskip 3pt\hrule\vskip 4pt}
    }
  }
  \endPlancktable                    
  \endgroup
\end{table}

\subsubsection{The CMB Monte Carlos}

For the CMB MCs, the computational cost depends on the numbers of time-samples ${\cal N}_{\rm t}$, map-pixels ${\cal N}_{\rm p}$, beam-pixels ${\cal N}_{\rm b}$, and MC realizations ${\cal N}_{\rm mc}$. Generating the PODA is essentially a re-ordering of the time-domain data, scaling as ${\cal N}_{\rm t}$; generating the effective-beam is a calculation of beam-pixel weights for every observation, scaling as ${\cal N}_{\rm t} \, {\cal N}_{\rm b}$; and generating the beam-convolved maps is an effective-beam-matrix/map-vector multiplication for each realization, scaling as ${\cal N}_{\rm mc} \, {\cal N}_{\rm p} \, {\cal N}_{\rm b}$. Table~\ref{table:effbeam} shows the values of these parameters for each \Planck\ channel, indicating that for FFP8 the effective-beam convolution over the CMB skies has, for the first time, become the dominant cost.  By comparison, for FFP6, although the numbers of samples were a factor of 2 (HFI) to 4 (LFI) smaller, the smaller MC set (with only $10^3$ realizations) and smaller numbers of pixels (temperature only) made the effective-beam calculation dominate. Running the 46 distinct $10^4$-realization CMB MC map sets included in FFP8 then required some 8M CPU-hours in total.

\begin{table}[h]
  \begingroup
  \newdimen\tblskip \tblskip=5pt
  \caption{The numbers of samples (${\cal N}_{\rm t}$), map-pixels (${\cal N}_{\rm p}$), beam-pixels (${\cal N}_{\rm b}$) and the derived computational costs of the \febecop\ effective-beam calculation (${\cal N}_{\rm t} \, {\cal N}_{\rm b}$) and MC convolution (${\cal N}_{\rm mc} \, {\cal N}_{\rm p} \, {\cal N}_{\rm b}$) for each channel in FFP8. The number of effective-beam pixels at each frequency depends on the \healpix\ resolution of the map ($\Nside = 2048$ and $1024$ for HFI and LFI, respectively) and the cut-off radius of the scanning beam (113.6\arcm, 79\arcm, 52\arcm\ and 100\arcm\ for 30\GHz, 44\GHz, 70\GHz\ and HFI, respectively). }
  \label{table:effbeam}
  \nointerlineskip
  \vskip -3mm
  \footnotesize
  \setbox\tablebox=\vbox{
    \newdimen\digitwidth
    \setbox0=\hbox{\rm 0}
    \digitwidth=\wd0
    \catcode`*=\active
    \def*{\kern\digitwidth}
    \newdimen\signwidth
    \setbox0=\hbox{-}
    \signwidth=\wd0
    \catcode`!=\active
    \def!{\kern\signwidth}
    \halign{\hbox to 1.25cm{#\leaderfil}\tabskip 0.9em&\hfil#\hfil&\hfil#\hfil&\hfil#\hfil&\hfil#\hfil&\hfil#\hfil \tabskip 0pt\cr
      \noalign{\doubleline}
      \omit\hfil Frequency\hfil& ${\cal N}_{\rm t}$&${\cal N}_{\rm p}$&${\cal N}_{\rm b}$ &${\cal N}_{\rm t} \, {\cal N}_{\rm b}$&${\cal N}_{\rm mc} \, {\cal N}_{\rm p} \, {\cal N}_{\rm b}$\cr
      \noalign{\vskip 4pt\hrule\vskip 6pt}
      *30&$1.5 \! \times \! 10^{10}$&$3.8 \! \times \! 10^{6}$&$3.5 \! \times \! 10^3$&$5.2 \! \times \! 10^{13}$&$1.3 \! \times \! 10^{15}$\cr
      *44&$3.2 \! \times \! 10^{10}$&$3.8 \! \times \! 10^{6}$&$1.7 \! \times \! 10^3$&$5.4 \! \times \! 10^{13}$&$6.4 \! \times \! 10^{14}$\cr
      *70&$1.1 \! \times \! 10^{11}$&$3.8 \! \times \! 10^{6}$&$7.3 \! \times \! 10^2$&$7.9 \! \times \! 10^{13}$&$2.8 \! \times \! 10^{14}$\cr
      100&$7.5 \! \times \! 10^{10}$&$1.5 \! \times \! 10^{7}$&$1.1 \! \times \! 10^4$&$8.3 \! \times \! 10^{14}$&$1.7 \! \times \! 10^{16}$\cr
      143&$1.0 \! \times \! 10^{11}$&$1.5 \! \times \! 10^{7}$&$1.1 \! \times \! 10^4$&$1.1 \! \times \! 10^{15}$&$1.7 \! \times \! 10^{16}$\cr
      217&$1.1 \! \times \! 10^{11}$&$1.5 \! \times \! 10^{7}$&$1.1 \! \times \! 10^4$&$1.2 \! \times \! 10^{15}$&$1.7 \! \times \! 10^{16}$\cr
      353& $1.1 \! \times \! 10^{11}$&$1.5 \! \times \! 10^{7}$&$1.1 \! \times \! 10^4$&$1.2 \! \times \! 10^{15}$&$1.7 \! \times \! 10^{16}$\cr
      545& $3.2 \! \times \! 10^{10}$&$5.0 \! \times \! 10^{6}$&$1.1 \! \times \! 10^4$&$3.6 \! \times \! 10^{14}$&$5.5 \! \times \! 10^{15}$\cr
      857& $4.3 \! \times \! 10^{10}$&$5.0 \! \times \! 10^{6}$&$1.1 \! \times \! 10^4$&$4.8 \! \times \! 10^{14}$&$5.5 \! \times \! 10^{15}$\cr
      \noalign{\vskip 3pt\hrule\vskip 4pt}
    }
  }
  \endPlancktable                    
  \endgroup
\end{table}

Originally \febecop\ was designed to optimize MC simulations with well-localized beams (and hence small ${\cal N}_{\rm b}$), but it is clear that with FFP8 we have moved out of that regime and it is time to re-evaluate the comparison of the computational cost and accuracy of the effective-beam and explicit time-domain approaches. Since these scale as ${\cal N}_{\rm p} \, {\cal N}_{\rm b}$ and ${\cal N}_{\rm t} $, respectively, the cross-over point is when the number of pixels in the effective-beam is comparable to the number of hits per map-pixel. Possible ways to reduce ${\cal N}_{\rm b}$ for \febecop\ would be to use multi-component beams with each component pixelized at a different resolution, or to restrict the main beam to the pixel domain and process the intermediate and far sidelobes in the spherical harmonic domain. Looking forward, this will be an active area of research and development. We note, though, that next-generation $B$-mode experiments are specifically designed to have very many more hits per pixel than \Planck\ in order to achieve the necessary signal-to-noise ratio to meet their science goals, shifting the balance back towards pixel-domain calculations.

\subsection{Simulations at CSC}
\label{subsection:CSC}

A significant fraction of the LFI noise MCs were performed at the CSC-IT Center for Science in Finland, using their Cray XC30 supercomputer {\em Sisu} with 40\,512 computing cores. These runs were performed using the same software stack as was used at NERSC, with cross-checks confirming that the two systems produce results identical to numerical precision.

The CSC runs included the channel/mission/full and half-ring maps for realizations 1\,000--9\,999, totalling 81\,000 maps and using a total of $4$ million CPU-hours. These were run in about a dozen different jobs, gradually increasing the number of instances of the code running simultaneously and the number of cores used.  The largest jobs for each frequency were as follows:

\begin{itemize}

\item for $30$\,GHz, $10$ simultaneous instances generating $5\,000$ realizations in all, running on $23\,040$ cores for $17.8$ hours for a total of $410\,000$ CPU-hours;

\item for $44$\,GHz, $10$ simultaneous instances generating $5\,000$ realizations in all, running on $11\,520$ cores for $21.8$ hours for a total of $251\,000$ CPU-hours;

\item for $70$\,GHz, $10$ simultaneous instances generating $2\,757$ realizations in all, running on $36\,000$ cores for $24$ hours\footnote{The maximum runtime allowed at CSC for jobs of this size.}
for a total of $864\,000$ core-hours.

\end{itemize}

In addition, we produced a smaller number of realizations for the 70\,GHz detector subsets, including 1\,000 realizations of the mission and 300 realizations each of the two half-missions, four years, and eight surveys. With three detector subsets and three maps (full and half-ring) for every realization, this resulted in 46\,800 maps and took a total of 200\,000 CPU-hours.

In total, $127\,800$ maps were produced at CSC, using 4\,200\,000 CPU-hours and requiring a total of $17.7$\,TB of disk space; these were then transferred to NERSC for wider distribution.

\section{Results}
\label{sec_res}

The core FFP8 simulation set consists of:

\begin{itemize}

\item 90\,720 fiducial maps, comprising 1\,134 maps of each of six components (three CMB, two foreground, one noise), then combined to produce two sets of total maps (with and without bandpass mismatch), all for five different values of the $(r, f_{NL})$ parameter pair.

\item 671\,400 noise MC maps, comprising $10^4$ realizations of all channel/mission/full maps, at least $10^3$ realizations of all other full maps, and at least $10^2$ realizations of all half-ring maps.

\item 460\,000 CMB MC maps, comprising $10^4$ channel/mission/full and channel/half-mission/full maps and selected detset/mission/full maps.

\end{itemize}
while the FFP8.1 simulation set currently consists of:

\begin{itemize}

\item 18\,144 fiducial maps, comprising 1\,134 maps of each of three CMB components, then combined with the FFP8 foreground and noise maps to produce two sets of total maps (with and without bandpass mismatch), restricted the the baseline ($r = f_{NL} = 0$) case.

\item 27\, 000 CMB MC maps, comprising 1\, 000 channel/mission/full and channel/half-mission/full maps.

\end{itemize}

\subsection{Uses of FFP8}

As well as supporting the validation and verification of all of the \Planck\ analysis codes, FFP8 is widely used in the analysis of \Planck\ data.\\

\noindent{\bf Fiducial realizations} are used to:

  \begin{itemize}

  \item test how various choices of the smoothing kernel affect signal aliasing in low resolution maps (LFI) \citep{planck2014-a07};

  \item demonstrate inaccuracies in the measurements of the instrumental bandpasses \citep{planck2014-a12};

  \item measure and subtract the LFI sidelobe pick-up \citep{planck2014-a03};

  \item estimate calibration and bandpass measurement uncertainties in component separation \citep{planck2014-a12};

  \item de-bias and measure uncertainty in lensing reconstruction \citep{planck2014-a17};

  \item assess SZ catalog reliability \citep{planck2014-a36};

  \item estimate noise for mapping the thermal Sunyaev-Zeldovich (tSZ) effect \citep{planck2014-a28}.

  \end{itemize}

\noindent{\bf Noise MCs} are used to:

  \begin{itemize}

  \item measure the accuracy of the LFI pixel-pixel noise covariance matrix construction \citep{planck2014-a07}; and

  \item inform low-$\ell$ data selection by measuring the distribution of low-multipole amplitudes due to noise as well as confirming the anomalous nature of Surveys 2 and 4 \citep{planck2014-a03};

  \end{itemize}

\noindent{\bf CMB MCs} are used to:

  \begin{itemize}

  \item produce beam transfer functions for each map type listed in Table~\ref{table:maps} by averaging over the
        corresponding MC realizations \citep{planck2014-a05,planck2014-a08}.

  \end{itemize}

In addition, the noise and CMB MC maps are combined using the same component separation pipelines as the real and fiducial data.  The resulting MC sets are then used to support the various CMB isotropy and statistics analyses \citep{planck2014-a18} and non-Gaussianity estimators \citep{planck2014-a19}.

Various ancilliary data sets were also produced to address the specific requirements of particular analyses. The most notable of these were:

\begin{itemize}

\item nine channel/mission/full maps for each of the ten individual foreground components, to validate the component separation analyses; 

\item $10^3$ channel/mission/full of the scalar CMB MCs with no Doppler boosting applied, to validate the \biposh\ reconstruction of Doppler boosting \citep{planck2014-a18}.

\end{itemize}

The capabilities developed for FFP8 also allowed for many other tests of data and analysis systematics, including running the mapmaking:

\begin{itemize}

\item with both mission- and run-baseline destriping to measure correlations induced by mission-baseline destriping in maps spanning different time intervals or including disjoint detector subsets, leading to the acceptance of this approach by HFI (where these are generally small) and its rejection by LFI (where they are significant);

\item with both temperature-only and temperature+polarization destriping to demonstrate the loss of polarization power induced by the former, leading to the adoption of the latter by both HFI and LFI;

\item using pointing that includes the expected uncertainty in its reconstruction to test the impact of pointing reconstruction error for LFI maps; 

\item using polarization angles that include the expected uncertainty in their determination to test the impact of polarization angle error for LFI maps.

\end{itemize}

\subsection{Known issues with FFP8}

Generating a simulation set of the size and complexity of FFP8 requires the coordination of a large number of people and processes. In addition to the cosmology mismatch and HFI systematic residual issues noted above, there are a few errors that are known to have been inadvertently  introduced into FFP8 during its execution. The most significant of these derives from an approximation in the way the PSM applies the detector bandpasses to the input skies, which results in effective bandpasses that differ from the inputs by up to 2\,\% in the submillimetre channels. This effect has been incorporated into the fiducial maps, but when the submillimetre CMB MC maps are used with the fiducial realization they must be rescaled by factors of 0.9976 at 545\,GHz and 0.9828 at 857\,GHz.

The other known issues are not expected to have any significant impact, but we list them here for completeness.

\begin{itemize}

\item During the export of the TOD from the HFI DPC the sample time stamps were inadvertently truncated, resulting in pointing errors of up to 0\parcm3.

\item The channel-average bandpasses generated by the PSM used preliminary detector weights, resulting in a tiny discrepancy between the bandpass-mismatch and no-bandpass-mismatch cases.

\item The FFP8 mapmaking used a slightly more rigorous pixel-rejection criterion than the HFI DPC, resulting in some FFP8 maps having somewhat more missing pixels than the real data.  Since these are the noisiest pixels, their naive inclusion in any map-based analysis can have a disproportionate effect at high $\ell$.

\end{itemize}

\subsection{Community access to FFP8}

As was the case with FFP6 for the first \Planck\ data release, FFP8/FFP8.1 is being made available to the scientific community. A simulation set of this size requires significant computational resources to store and analyse, so once again we are making it available at NERSC, where there is also an allocation of compute cycles available to support its exploitation. Full details on this, including how to obtain a NERSC account, can be found at \url{http://crd.lbl.gov/cmb-data}.

\section{Conclusions}
\label{sec_conclusions}

Simulations play a critical role in the analysis of CMB data sets, most particularly in the validation and verification of analysis pipelines and the debiasing and uncertainty quantification of analysis results. The former requires a simulation that is as true to the mission as possible, while the latter requires enormous Monte Carlo sets of mission realizations. FFP8 provides the 2015 \Planck\ data release with both. The fiducial realization takes the most realistic model of the microwave sky available to us and simulates the full mission observation of that sky for every \Planck\ detector, incorporating our best estimates of that detector's beam, bandpass, and time-varying noise properties. The noise and CMB MC sets include $10^4$ realizations of the most important maps (namely those that incorporate all of the data at a given frequency), together with $10^2$--$10^4$ realizations of all of the other maps that include subsets of the observations and detectors at each frequency. The FFP8.1 CMB maps provide a smaller sample of fiducial and MC CMB skies consistent with the PR2-2015 cosmology. With around 1.25 million maps in total, and requiring some 25 million CPU-hours at multiple HPC centres, this is by far the largest CMB simulation set ever produced.

\begin{acknowledgements}
The Planck Collaboration acknowledges the support of: ESA; CNES and CNRS/INSU-IN2P3-INP (France); ASI, CNR, and INAF (Italy); NASA and DoE (USA); STFC and UKSA (UK); CSIC, MINECO, JA, and RES (Spain); Tekes, AoF, and CSC (Finland); DLR and MPG (Germany); CSA (Canada); DTU Space (Denmark); SER/SSO (Switzerland); RCN (Norway); SFI (Ireland); FCT/MCTES (Portugal); ERC and PRACE (EU). A description of the Planck Collaboration and a list of its members, indicating which technical or scientific activities they have been involved in, can be found at http://www.cosmos.esa.int/web/planck/planck-collaboration. Particular thanks are due to the extraordinary NERSC staff, who have supported the \Planck\ mission for over a decade and who facilitated the production and distribution of FFP8 in numerous ways.

\end{acknowledgements}

\bibliographystyle{aat}
\bibliography{Planck_bib,A14_Simulations}

\appendix

\section{Measuring band-pass mismatch residuals through spurious-component maps}
\label{appendix:iqus}

\begin{table*}[tb]
  \begingroup
  \newdimen\tblskip \tblskip=5pt
  \caption{Spurious map pointing weights}
  \label{table:iqus_weights}
  \nointerlineskip
  \vskip -3mm
  \footnotesize
  \setbox\tablebox=\vbox{
    \newdimen\digitwidth
    \setbox0=\hbox{\rm 0}
    \digitwidth=\wd0
    \catcode`*=\active
    \def*{\kern\digitwidth}
    \newdimen\signwidth
    \setbox0=\hbox{-}
    \signwidth=\wd0
    \catcode`!=\active
    \def!{\kern\signwidth}
    \halign{\hbox to 2.5cm{#\leaderfil}\tabskip 1em&\hfil#\hfil&\hfil#\hfil&\hfil#\hfil&\hfil#\hfil&\hfil#\hfil&\hfil#\hfil&\hfil#\hfil&\hfil#\hfil&\hfil#\hfil&\hfil#\hfil&\hfil#\hfil \tabskip 0pt\cr
      \noalign{\doubleline}
      \omit\hfil Detector\hfil&Horn 1&Horn 2&Horn 3&Horn 4&Horn 5&Horn 6&Horn 7&Horn 8&Horn 9&Horn 10&Horn 11\cr
      \noalign{\vskip 4pt\hrule\vskip 6pt}
      LFI18M&$+1$&$!0$&$!0$&$!0$&$!0$&$!0$&$+1$&$!0$&$!0$&$+1$&$+1$\cr
      LFI18S&$-1$&$!0$&$!0$&$!0$&$!0$&$!0$&$+1$&$!0$&$!0$&$+1$&$+1$\cr
      LFI19M&$!0$&$+1$&$!0$&$!0$&$!0$&$!0$&$-1$&$!0$&$!0$&$+1$&$+1$\cr
      LFI19S&$!0$&$-1$&$!0$&$!0$&$!0$&$!0$&$-1$&$!0$&$!0$&$+1$&$+1$\cr
      LFI20M&$!0$&$!0$&$+1$&$!0$&$!0$&$!0$&$!0$&$+1$&$!0$&$-1$&$+1$\cr
      LFI20S&$!0$&$!0$&$-1$&$!0$&$!0$&$!0$&$!0$&$+1$&$!0$&$-1$&$+1$\cr
      LFI21M&$!0$&$!0$&$!0$&$+1$&$!0$&$!0$&$!0$&$-1$&$!0$&$-1$&$+1$\cr
      LFI21S&$!0$&$!0$&$!0$&$-1$&$!0$&$!0$&$!0$&$-1$&$!0$&$!0$&$+1$\cr
      LFI22M&$!0$&$!0$&$!0$&$!0$&$+1$&$!0$&$!0$&$!0$&$+1$&$!0$&$-1$\cr
      LFI22S&$!0$&$!0$&$!0$&$!0$&$-1$&$!0$&$!0$&$!0$&$+1$&$!0$&$-1$\cr
      LFI23M&$!0$&$!0$&$!0$&$!0$&$!0$&$+1$&$!0$&$!0$&$-1$&$!0$&$-1$\cr
      LFI23S&$!0$&$!0$&$!0$&$!0$&$!0$&$-1$&$!0$&$!0$&$-1$&$!0$&$-1$\cr
      \noalign{\vskip 3pt\hrule\vskip 4pt}
    }
  }
  \endPlancktablewide                 
  \tablenote {{\mbox{}}} 
  Horns 1--6 are the physical horns that comprise two polarization-orthogonal arms. \\
  Horns 7--9 are the ``virtual horns''; each consists of two physical horns, LFI18+LFI19, LFI20+LFI21, and LFI22+LFI23. \\
  Horns 10--11 expand the hierarchy to include mismatch between the virtual horns.\par
  \endgroup
\end{table*}

We measure bandpass mismatch by projecting the mismatch into extra degrees of freedom called ``spurious'' maps. The mismatch has a specific signature: it is a sky-synchronous difference between the detectors at a given frequency that does not modulate under rotation (unlike polarization). We have extended the method to measure the mismatch between individual detectors and the noise-weighted frequency average. Conventionally, the mismatch is only resolved between two polarization-orthogonal detectors that have common pointing (i.e., are co-located on the focal plane).

Mathematically we can formulate this as an extension of the pointing model. We start from the conventional polarized detector response:
\begin{equation}
  \label{eq:iqu}
  d_t^D = I_p + \eta^D\left( Q_p\cos 2(\alpha_t+\psi^D) +U_p\sin 2(\alpha_t+\psi^D) \right) + n_t,
\end{equation}
where detector $D$ TOD sample, $d_t$, at time $t$ is a linear combination of the observed intensity sky, $I$, at pixel $p$ and linearly polarized $(Q,U)$ sky that is modulated by the polarization efficiency, $\eta^D$, the polarization sensitive angle of the detector, $\psi^D$ and the orientation of the detector $\alpha_t$. In this model, the sky signal includes the azimuthally symmetric part of the beam, and all instrumental noise and systematics are contained in the noise, $n$. The full data vector can be written concisely as
\begin{equation}
  \label{eq:iqu2}
  \vec d = \tens P \vec m + \vec n,
\end{equation}
where $\tens P$ is the pointing matrix that scans the smoothed sky map, $\vec m$, into the time domain according to the instrument scanning pattern.

We extend the model by assuming that the detectors are actually observing sky emission that is convolved with the individual detector bandpass, causing each detector to view a slightly different sky, $I^D$. The difference between an effective frequency average, $I$, and the sky seen by a particular detector, $I^D$, is the spurious (or bandpass mismatch) map, $S^D=I-I^D$. We can solve for the frequency-averaged $(I,Q,U)$ sky by introducing a hierarchy of spurious maps that captures the mismatch first within pairs of detectors (horns), then between pairs of horns (quadruplets), and ultimately between larger aggregates. We call all these combinations ``virtual horns''. We show in Table~\ref{table:iqus_weights} how the spurious maps can be assigned for the \Planck\ $70\,$GHz channel. The specific hierarchy of virtual horns is not important: it is only required that for each pair of detectors there exists exactly one spurious map (virtual or physical horn) for which they have opposite pointing weights. The total bandpass mismatch map between any detector and the frequency average can be produced as a linear combination of the spurious maps with the appropriate pointing weights.

The spurious maps extend the pointing model in Eq.~\ref{eq:iqu} to
\begin{equation}
  \label{eq:iqus}
  \widetilde{d_t^D} = d_t^D + \sum_i w_i S_i,
\end{equation}
where the sum is over all spurious maps that correspond to physical and virtual horns listed in Table~\ref{table:iqus_weights} and the associated weights are $+1$, $-1$, or $0$. These extra pointing weights are included in additional columns in the pointing matrix, $\tens P$, and a full map with all the spurious components can be binned using the standard mapmaking equation:
\begin{equation}
  \label{eq:mapmaking}
  \widetilde{\vec m} = \left( \tens P^\mathrm T \tens N^{-1} \tens P \right)^{-1}  \tens P^\mathrm T \tens N ^{-1} \vec d,
\end{equation}
where $\tens N$ is some model of the detector noise covariance, frequently a diagonal matrix with the detector white noise variances. In practice, the additional degrees of freedom in the spurious maps can only be resolved for pixels that have sufficient coverage in crossing angles to break the degeneracy between polarization and bandpass mismatch. Even then, the resulting noise levels in the $I$, $Q$, and $U$ maps are much higher than in the standard mapmaking case, making the $IQUS$ solutions only useful for low-resolution studies and checking other bandpass leakage correction schemes that rely on prior information such as detector bandpasses and models of the sky emission.

\end{document}